\documentclass[useAMS,usenatbib]{mn2e} \usepackage{graphicx}

\title[Methanol maser catalogue: 186$^{\circ}$ to 330$^{\circ}$]{ The 6-GHz methanol multibeam maser catalogue IV: Galactic longitudes 186$^{\circ}$ to 330$^{\circ}$ including the Orion--Monoceros region}

\author[Green et al.]
       {J. A. Green,$^1$\thanks{E-mail:james.green@csiro.au} J. L.
Caswell,$^1$ G. A. Fuller,$^2$ A. Avison,$^{2}$ S. L. Breen,$^{1,3}$  \newauthor
S. P. Ellingsen,$^3$ M. D. Gray,$^{2}$  M. Pestalozzi,$^{5}$  L. Quinn,$^2$ M. A. Thompson,$^4$ \newauthor
and M. A. Voronkov$^1$ \\ 
$^{1}$ CSIRO Astronomy and Space Science, Australia Telescope
National Facility, PO Box 76, Epping, NSW 2121, Australia; \\ $^{2}$
Jodrell Bank Centre for Astrophysics, Alan Turing Building, School of Physics and Astronomy, University of
Manchester,\\ Manchester, M13 9PL, UK; \\ 
$^{3}$ School of Mathematics and Physics, University of Tasmania, Private
Bag 37, Hobart, TAS 7001, Australia; \\ $^{4}$
Centre for Astrophysics Research, Science and Technology Research
Institute, University of Hertfordshire, College Lane, \\Hatfield, AL10
9AB, UK; \\ $^{5}$ INAF -- Istituto Fisica Spazio Interplanetario, via del Fosso del Cavaliere 100, I-00133 Roma, Italy \\}

\date{Accepted XXXX . Received XXX; in original form XXXX}

\pagerange{\pageref{firstpage}--\pageref{lastpage}} \pubyear{XXXX}
\begin{document} \maketitle

\label{firstpage}

\begin{abstract}
We present the fourth portion of a Galactic Plane survey of methanol masers at 6668\,MHz, spanning the longitude range 186$^{\circ}$  to 330$^{\circ}$.  We report 207 maser detections, 89 new to the survey. This completes the southern sky part of the Methanol Multibeam survey and includes a large proportion of new sources, 43\%. We also include results from blind observations of the Orion--Monoceros star forming region, formally outside the latitude range of the Methanol Multibeam survey; only the four previously known methanol emitting sites were detected, of which we present new positions and spectra for masers at Orion-A (south) and Orion-B, obtained with the MERLIN array.
\end{abstract}

\begin{keywords} 
masers - surveys - stars: formation
\end{keywords}

\section{Introduction}
The methanol multibeam (MMB) survey is a project to observe, for the first time,  the entire Galactic plane for 6668-MHz methanol masers \citep{green09a}. This maser transition of methanol is not only bright and widespread throughout the Galaxy, but has been found to exclusively trace one of the early stages of high-mass star formation \citep{pestalozzi02b, minier03, xu08}. The MMB survey, through its unbiased observations with uniform sensitivity, will establish a definitive catalogue for future studies. This catalogue is being released sequentially, with the first region covering Galactic longitudes 345$^{\circ}$ to 6$^{\circ}$ \citep{caswell10mmb1}, the second Galactic longitudes 6$^{\circ}$ to 20$^{\circ}$ \citep{green10mmb2} and the third Galactic longitudes 330$^{\circ}$ to 345$^{\circ}$ \citep{caswell11mmb3}. The current paper concludes the southern hemisphere portion of the catalogue, presenting detections within Galactic longitudes 186$^{\circ}$ and 330$^{\circ}$. 

A number of targeted searches have been made in the 186$^{\circ}$ to 330$^{\circ}$ longitude region including those by \citet{menten91,macleod92a,macleod92b,gaylard93,schutte93,caswell95a,walt95,walt96,walsh97,walsh98, slysh99, szymczak00}. The only previous blind survey observations in this longitude range were those of \citet{ellingsen96}, who observed 325$^{\circ}$ to  335$^{\circ}$ with a latitude range of  $\pm$0.53$^{\circ}$, detecting 21 sources in the overlapping longitude region of 5$^{\circ}$ with the current survey catalogue; and those of \citet{ellingsen96thesis}, who observed: 282$^{\circ}$ to 286$^{\circ}$ and 291$^{\circ}$ to 296$^{\circ}$ with a latitude range of --1.03$^{\circ}$ to 0.03$^{\circ}$, detecting 6 sources. High-resolution positions for 100 sources were determined by \citet[][and references therein]{caswell09a}. This region of the Galactic plane is particularly notable for tangents of the Crux-Scutum and Sagittarius spiral arms together with an unconfused portion of the Perseus arm. 

In this paper we present details of detections made by the MMB survey, high signal to noise spectra of the detected sources and highlight interesting characteristics of the sources. We also provide details of a blind search of the Orion-Monoceros region, which lies outside the standard survey $\pm$2$^{\circ}$ latitude range, but is a prominent nearby, extended site of high-mass star formation, already known to contain maser emission.

\section{Methanol multibeam survey parameters and equipment}
The techniques of the MMB survey are detailed in full by \citet{green09a}, hence we only present a summary of the pertinent points here. The observations involved the initial blind survey with the Parkes Radio Telescope using a purpose-built 7-beam receiver and covering a latitude range of $\pm$2$^{\circ}$ (termed `survey' observations from here onwards). These were followed by high-resolution observations with the Australia Telescope Compact Array in order to position (to within 0.4 arcsec) any new sources or sources without previous high-resolution positions. Finally, all detections were re-observed with the Parkes Telescope again, at the precise positions, to obtain high signal--to--noise spectra (termed `MX' observations from here onwards). The main survey observations were made between 2006 January and 2007 December, whilst the MX observations were made between 2007 January and 2009 April.  The correlator setup allowed for a simultaneous velocity coverage of $\sim$180\,km\,s$^{-1}$, and for the longitude range presented here we required only one central velocity setting in order to fully encompass the CO emission seen by \citet{dame01}. The central velocities are given in Table\,\ref{centralvels}.

\begin{table} \centering \caption{\small Central velocities for survey observations.} 
\begin{tabular}{cc}
\hline
\multicolumn{1}{c}{Longitude Range} & \multicolumn{1}{c}{Central Velocity}\\
 ($^\circ$) & (km\,s$^{-1}$ )\\
\hline
186$^{\circ}$ -- 190$^{\circ}$ &0\\
190$^{\circ}$ -- 236$^{\circ}$ &+20\\
236$^{\circ}$ -- 270$^{\circ}$ &+40\\
270$^{\circ}$ -- 276$^{\circ}$ &+30\\
276$^{\circ}$ -- 290$^{\circ}$ &+15\\
290$^{\circ}$ -- 296$^{\circ}$ &+10\\
296$^{\circ}$ -- 300$^{\circ}$ &+5\\
300$^{\circ}$ -- 306$^{\circ}$ &0\\
306$^{\circ}$ -- 310$^{\circ}$ & --5\\
310$^{\circ}$ -- 316$^{\circ}$ & --10\\ 
316$^{\circ}$ -- 320$^{\circ}$ & --20\\ 
320$^{\circ}$ -- 326$^{\circ}$ &--30\\
326$^{\circ}$ -- 330$^{\circ}$ &--40\\
\hline
\end{tabular} 
\label{centralvels}
\end{table}

\subsection{Observations of the Orion-Monoceros complex}
\begin{figure*}
 \begin{center}
 \renewcommand{\baselinestretch}{1.1}
\includegraphics[width=15cm]{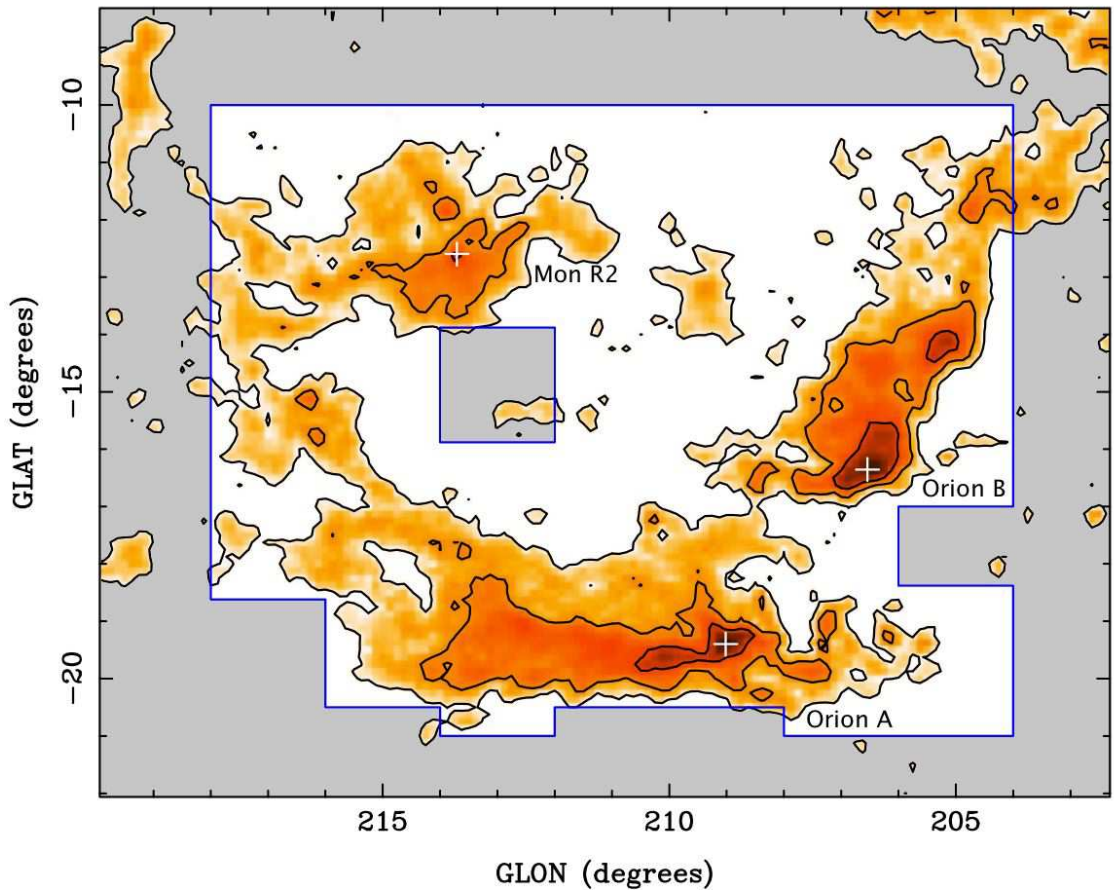}  
\caption{\small Map of the Orion-Monoceros complex with blue lines outlining the region surveyed - grey areas (including the small rectangular region centred at $l$=213$^{\circ}$ and $b$=$-$14.88$^{\circ}$) were not observed. The background shows the velocity integrated $^{12}$CO emission from \citet{wilson05} with the contours corresponding to 3.3, 16.4 and 49.1~K~km~s$^{-1}$. The methanol masers detected in the area are depicted by three white crosses.}
\label{orionmap}
\end{center}
\end{figure*}

In addition to the Galactic plane survey observations we also observed the Orion-Monoceros complex using the time when other regions of the main survey were below the telescope horizon. The complex contains three giant molecular clouds (GMCs), Orion~A, Orion~B and Mon~R2, which are known sites of high-mass star-formation \citep[see][and references therein]{wilson05}. Our survey covered all three GMCs in the region as shown in Figure\,\ref{orionmap}. The observations used the MMB receiver and shared the same scanning technique with the main survey \citep{green09a}. In particular, the region was also binned into blocks spanning 2$^{\circ}$ of Galactic longitude and the scanning was done along the lines of constant Galactic latitude. This allowed us to adjust the coverage area according to the amount of available observing time. Most of the data were taken towards the end of the project when the MMB receiver suffered a failure of one of its beams. This had little impact on the main survey which was at the stage of the follow-up MX observations (apart from the need to increase the observing time), but introduced a slight under sampling and corresponding sensitivity  variations in the raster maps of the Orion-Monoceros complex. To minimise these detrimental effects we changed the latitude offset between the forward and reverse (odd and even) scans to 7.5~arcmin (as opposed to alternating the offset between 1.07 and 15~arcmin, as was done for the main survey). To ensure even sensitivity across the region additional scans were taken towards regions with strong continuum emission. We adopted a central velocity of 10\,km\,s$^{-1}$ for the whole region (thus a velocity coverage of --80\,km\,s$^{-1}$ to 100\,km\,s$^{-1}$). We also observed at a limited range of hour angles to minimise rotation of the MMB receiver, required to compensate for changing parallactic angle. Overall, this strategy is equivalent to the strategy of the main survey in the middle of the map, but the variable sensitivity pattern at the latitude edges of the survey area extends over 15~arcmin rather than 1.07~arcmin. The small variations of the  sensitivity pattern result in a slightly higher detection threshold in the Orion-Monoceros region than for the main Galactic plane survey. We estimate our completeness to be greater than 85 per cent for 1.0~Jy sources.

The known sources associated with Orion-A (south) (209.016--19.398) and Orion-B (206.542--16.355) were also observed with the Multi-Element Radio Linked Interferometer Network (MERLIN) array. These observations were made in 2006 January with the broadband 4--8 GHz e-MERLIN receivers on 6 telescopes (the Lovell, MKII, Darnhall, Tabley, Knockin and Cambridge). The observations were centred at 6668.518 MHz, corrected for source velocities of --1.1\,km\,s$^{-1}$ for Orion-A (south) and 5.5\,km\,s$^{-1}$ for Orion-B, with 256 spectral channels across a bandwidth of 0.5-MHz (a velocity width of $\sim$22\,km\,s$^{-1}$). The longest baseline of MERLIN is 217\,km, giving a synthesized beam size of 43 mas at 6668-MHz. Primary flux and bandpass calibration was achieved through observations of 3C84 and 3C286, and phase calibration with observations of 0539-057 (in the broadband 16 channels across 32-MHz mode). The data were reduced using the local MERLIN software \citep{merlin03} and the Astronomical Image Processing Software ({\sc AIPS}).

\section{Survey results}
Results are presented in Table\,\ref{resotable}, which follows the layout of \citet{caswell10mmb1}: columns one and two are Galactic longitude
and latitude; columns three and four are Right Ascension and Declination in J2000 co-ordinates from ATCA observations; columns five and six are the minimum and maximum velocities of emission (the most extreme values observed in any observations); columns 7 and 8 are the MX peak velocities and flux densities; columns 9 and 10 are the peak velocities and flux densities observed in the survey cubes; and the final column provides extra information on the date of the ATCA observation (if it was observed by us) or a reference for comparable observations made previously. All velocities are with respect to the radio convention of the local standard of rest (LSR). For sources in close proximity, individual source velocity ranges have been determined from inspection of the higher spatial resolution ATCA data. The Parkes MX spectra for the sources are given in Figure\,\ref{spectra}. 
291.642--0.546, 294.977--1.734 and 321.704+1.168 have an asterisk to highlight that the survey cube spectra have been shown instead (the sources were confirmed and positioned with the ATCA observations, but were undetectable in the MX observations). 
Again, following the practice of  \citet{caswell10mmb1}, maser sites separated by more than 2 arcsec are considered different sites and listed as such. For completeness, we include in the table 290.411--2.915, a source known to exist outside the latitude range of the MMB survey \citep{walsh98, pestalozzi05} and give measurements from our targeted MX observations.

For the 186$^{\circ}$  to 330$^{\circ}$  longitude region we detected 207 sources, corresponding to an average of 1.4 sources per degree of longitude. However, unlike the already published longitude regions closer to the Galactic centre, the current results show extensive portions of the plane with no detections. There are 89 detections new to the 186$^{\circ}$  to 330$^{\circ}$ longitude region. 

\subsection{Results from the Orion--Monoceros complex}
Our blind survey revealed only the four previously known sources in this large region, three confirmed masers (213.705--12.597, 206.542--16.355 and 209.016--19.398) and one variable maser or thermal source (208.996--19.386, as discussed in the following source notes). In this portion of the survey, we achieved an 85\% completeness limit of 1\,Jy. 
The map and spectrum of Orion-A (south) (209.016--19.398) and Orion-B (206.542--16.355) were obtained with the MERLIN array and are shown in Figures\,\ref{orionmerlin} and \ref{orionBmerlin}. The MERLIN positions for the two sources listed in Table\,\ref{resotable} agree within the errors of the previously published ATCA positions \citep{minier03,voronkov05}.   

\subsection{Remarks on sites of maser emission.}\label{srcnotes}
Here we provide additional details on notable sources and sources with confused spectral structure. This includes: 11 sources with relatively large latitudes (including 2 sources associated with Orion A, one with Orion B and one with Mon R2); 20 sources demonstrating variability; and 7 with velocity ranges wider than 16\,km\,s$^{-1}$. 
The majority of sources within longitudes 270$^{\circ}$ to 330$^{\circ}$ are located within the solar circle (inferred from their negative LSR velocities) and therefore have a near-far ambiguity for kinematic distance estimates. Where possible we remark on sources which have been allocated a near or far distance in the literature.  

\subparagraph{188.794+1.031} This source was originally detected in 1999 by \citet{szymczak00} from a targeted observation of the IRAS source 06061+2151, with a position of RA (J2000) 06h09m06.5s, Dec +21$^{\circ}$50$'$26$''$. Although we were unable to obtain a high-resolution position with the ATCA (due to scheduling constraints), we obtained a position (2009 January) with the MERLIN array of RA (J2000) 06h09m06.96s, Dec +21$^{\circ}$50$'$41.3$''$. The intensity of a feature at $-$5.5\,km\,s$^{-1}$ has doubled to nearly 10\,Jy since the \citet{szymczak00} observations; a feature which they detected at $-$8.7\,km\,s$^{-1}$ with a peak flux density of $\sim$4\,Jy at their offset position, has faded to below the detection threshold of the MMB, but is included in the velocity range of Table\,\ref{resotable} for completeness.

\subparagraph{188.946+0.886} This source has a peak flux density varying between $\sim$400 and $\sim$600 Jy and is periodically variable with a period of 416 days \citep{goedhart04}. A weak feature near $-$5\,km\,s$^{-1}$, originally observed with a peak flux density of $\sim$0.4\,Jy \citep{caswell95a}, was detected in the survey cube (observed 2006 April--June) with a peak flux density of $\sim$0.6\,Jy, but was not detected in the subsequent MX observation taken in 2008 March. The spectrum in Figure\,\ref{spectra} is centred to show the full range of velocities (17\,km\,s$^{-1}$) over which emission has been seen.

\subparagraph{189.471--1.216} This is a new source for which we were able to obtain a high--resolution position with the MERLIN array (2009 January).

\subparagraph{189.778+0.345} Two roughly similar peaks of 14 to 15 Jy at 4\,km\,s$^{-1}$ and 5.6\,km\,s$^{-1}$ in 1992 \citep{caswell95a} have both declined in flux density. The 5.6\,km\,s$^{-1}$ feature decreased to a peak of 6.2\,Jy in 1999 \citep{caswell09a}, it was then observed to be 8\,Jy in the survey cube (2006 April--June), before decreasing further to 2.5\,Jy in the 2008 March MX observation. However it then increased in peak flux density to 4.5\,Jy in the 2009 March MX observation. The 4\,km\,s$^{-1}$ feature was 1.5\,Jy in the survey cube and $\sim$1\,Jy in the subsequent 2008 and 2009 MX observations.

\subparagraph{196.454--1.677} Intensity variations in this source show an unconfirmed periodicity of 668 days \citep{goedhart04}. The peak flux density of almost 60\,Jy at 15.2\,km\,s$^{-1}$ in 1992 \citep{caswell95a} decreased to 26\,Jy in 1999 as reported by \citet{caswell09a}. Our observations found it had faded further to 18\,Jy in the survey cube observations taken 2007 March/April before brightening slightly to 22\,Jy in the 2008 March MX observation.

\begin{figure*}
 \begin{center}
 \renewcommand{\baselinestretch}{1.1}
\includegraphics[width=12cm]{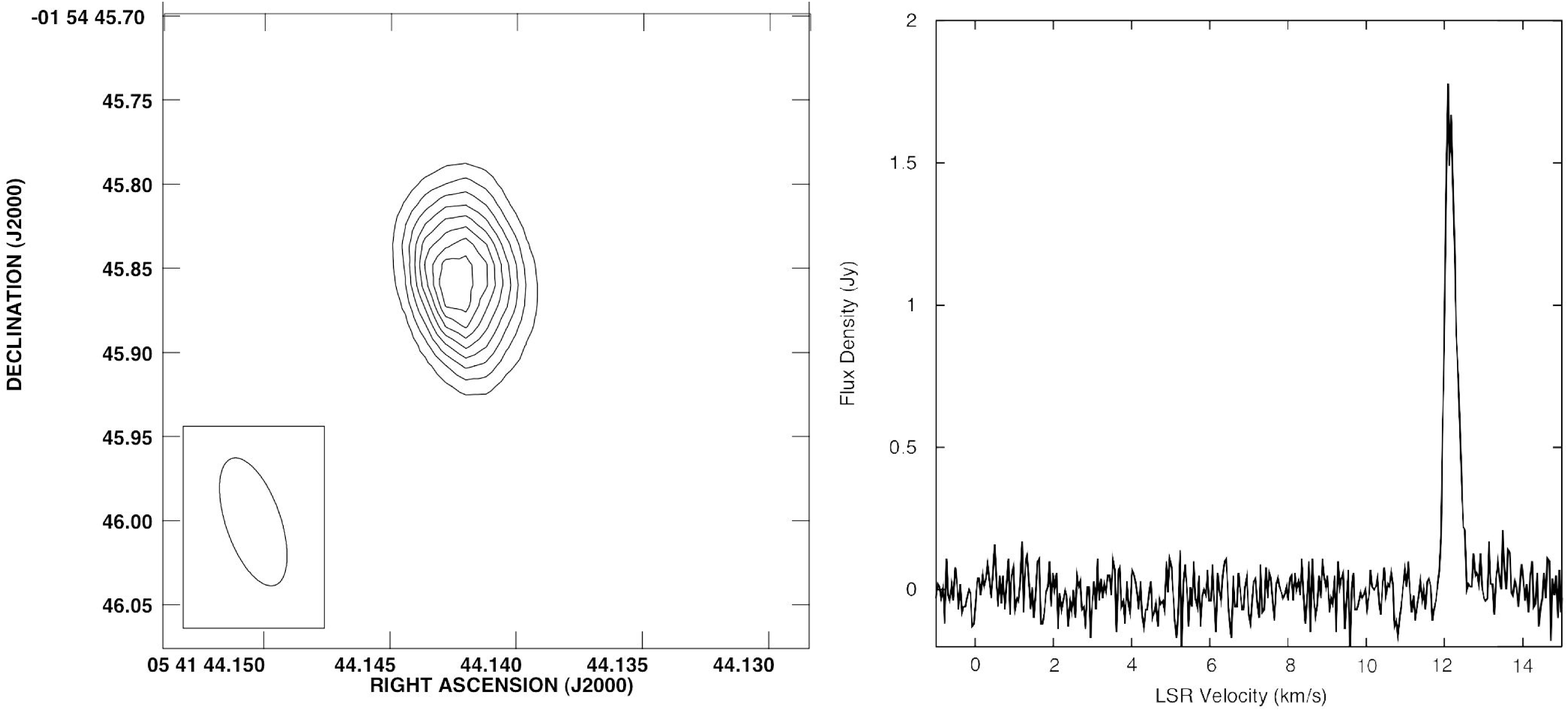}  
\caption{\small Map (left) and spectrum (right) of the Orion-B maser source (206.542--16.355) as observed with the MERLIN array in 2006 (with the same observational setup as Orion-A (south), see text). The map displays contours at 20 to 100\% (in steps of 10\%) of 1.137 Jy beam$^{-1}$ for the average of the channels with emission. }
\label{orionBmerlin}
\end{center}
\end{figure*}

\subparagraph{206.542--16.355} This source was first detected by \citet{minier03}. Its association with the Orion B cloud (also called OMC2) at an estimated distance of 0.4\,kpc \citep{lombardi11}, makes it one of the closest methanol masers in our Galaxy. We obtained observations with the MERLIN array in 2006, from which we were able to detect and position the source  (see Figure\,\ref{orionBmerlin}).

\begin{figure*}
 \begin{center}
 \renewcommand{\baselinestretch}{1.1}
\includegraphics[width=12cm]{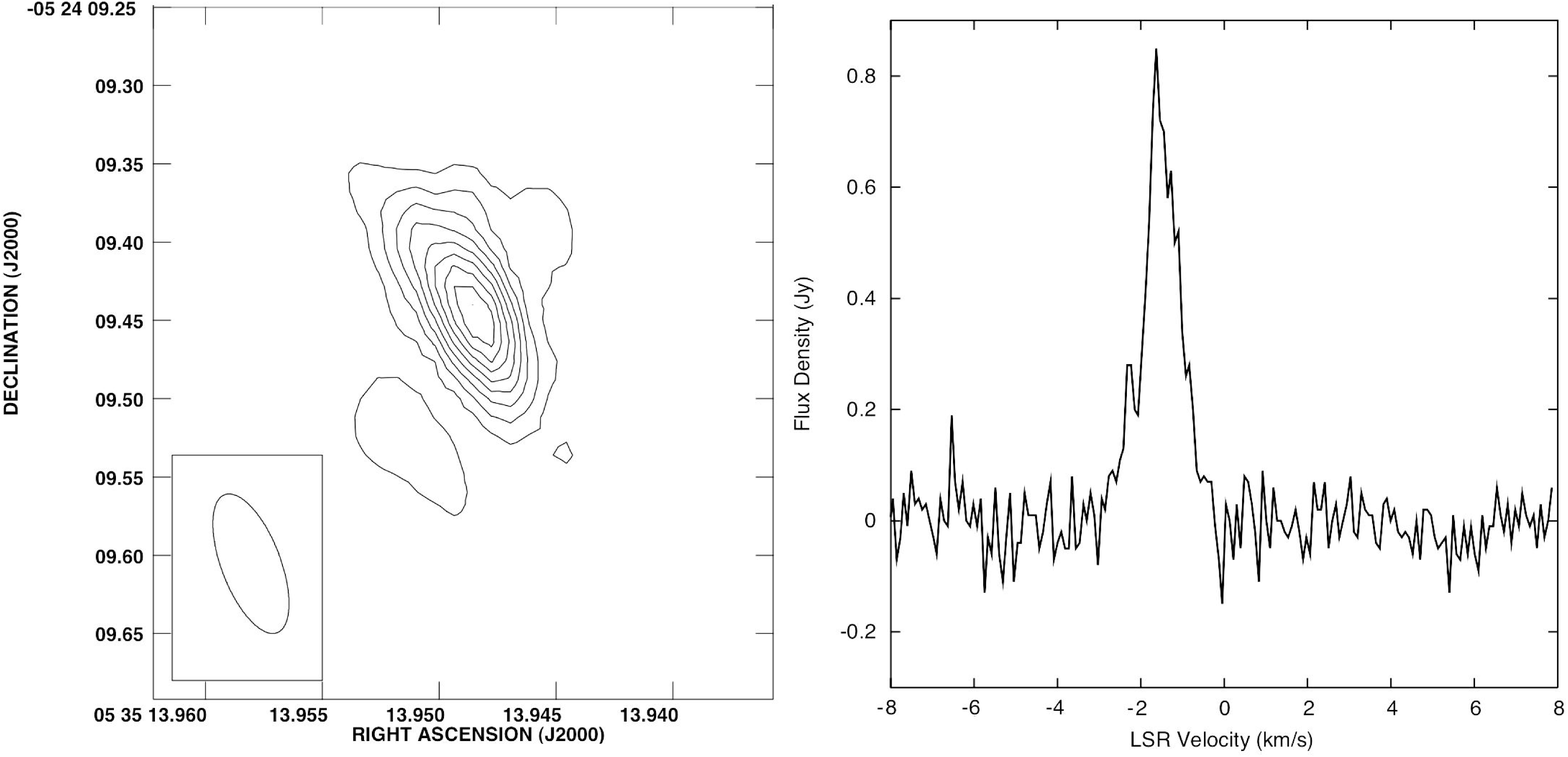}  
\caption{\small Map (left) and spectrum (right) of the Orion-A (south) maser source (209.016--19.398) as observed with the MERLIN array in 2006 (with the same observational setup as Orion-B, see text). The map displays contours at 20 to 100\% (in steps of 10\%) of 0.495 Jy beam$^{-1}$ for the average of the channels with emission. }
\label{orionmerlin}
\end{center}
\end{figure*}

\subparagraph{208.996--19.386 and 209.016--19.398} These two sources, one with a peak velocity of $\sim$7\,km\,s$^{-1}$, the other with a peak velocity of $-$1.5\,km\,s$^{-1}$, are associated with the Orion A cloud (also called OMC1), in the direction of very strong H{\sc ii} continuum, giving rise to an irregular baseline as well as prominent absorption. Both masers were positioned in 2003/2004 with the ATCA, as Orion-A (south) and Orion-KL respectively, by \citet{voronkov05}. The feature at $\sim$7\,km\,s$^{-1}$ was the first to be discovered towards Orion A, observed in 1992 by \citet{caswell95a}, with a peak flux density of $\sim$0.5\,Jy. The observations by  \citet{voronkov05} found the $-$1.5\,km\,s$^{-1}$ feature to be the brightest, at $\sim$1\,Jy, and partially resolved the $\sim$7\,km\,s$^{-1}$ feature (again with a flux density of $\sim$0.5\,Jy), leading to speculation that this feature may be thermal in nature. We obtained observations with the MERLIN array in 2006, from which we were able to detect and position the $-$1.5\,km\,s$^{-1}$ feature (see Figure\,\ref{orionmerlin}), but we did not separately observe the 7\,km\,s$^{-1}$ feature as it was resolved with the ATCA. MX observations obtained in 2009 March could not detect anything above $\sim$0.2\,Jy for the $\sim$7\,km\,s$^{-1}$ feature, although the feature lies in an absorption dip and taking this into account the upper limit may be consistent with 0.5\,Jy. We show in Figure\,\ref{spectra} the spectra at both positions, with the spectrum of 208.996--19.386 including emission at $\sim$7\,km\,s$^{-1}$, but below the 1$\sigma$ level. The peak flux density of 209.016--19.398 varied between the Orion-Monoceros survey observations and the targeted MX observations in 2009 March (2.8\,Jy and 0.8\,Jy respectively).

\subparagraph{212.06$-$0.74} This source was discovered by \citet{xu08} through observations with the Effelsberg telescope in 2006 February/March as follow-up of water maser detections towards IRAS sources. We, too, detected this source in 2007 August. The position listed in Table\,\ref{resotable} is that used for the Effelsberg observations (the IRAS source position), as we were unable to obtain a high-resolution position with the ATCA (due to scheduling constraints). The source velocity, $\sim$45\,km\,s$^{-1}$, is significantly positive, more than 20\,km\,s$^{-1}$ more positive than any other source in our survey within 20$^{\circ}$ of longitude. Also the peak flux density of this source varied significantly between the survey cube observations ($\sim$1\,Jy) and the MX observation ($\sim$0.3\,Jy).

\subparagraph{213.705--12.597} This source is associated with the well studied Mon R2 cloud \citep{white79,willson81,massi85,natta86,wolf90,kohno02} and known to be variable, both in the methanol \citep{goedhart04} and the associated 4.765-GHz OH line \citep{smits98}. Our Orion-Monoceros survey observations in 2008 March detected a peak flux density of 278\,Jy whilst the targeted MX observation obtained in 2009 March detected a peak flux density of 96\,Jy. 

\subparagraph{254.880+0.451, 259.939$-$0.041, 263.250+0.514, 264.140+2.018, 264.289+1.469, 269.153$-$1.128, 269.456$-$1.467, 269.658$-$1.270 and 270.255+0.835} With the exception of 269.456$-$1.467 (with a high positive velocity), these maser sites are loosely associated with the well studied Vela molecular cloud complex \citep[e.g.][Lo et al. in prep, Lowe et al. in prep, Giannini et al. in prep, ]{murphy91,olmi09,netterfield09,hill11}. 
264.140+2.018 is a new source detected at the edge of the survey cube region, with an initial position of 264.133+1.984 and peak flux density of 1.4\,Jy. With the ATCA observations this was found to be offset (by $\sim$2 arcmin) from the true position of 264.140+2.018 and therefore the measured survey cube flux density is less than half the true value, accounting for at least some of the apparent discrepancy with the survey MX detection.  
There has been some variation in the emission of the faint source 264.289+1.469. It was initially detected with a peak flux density of 0.2\,Jy \citep{caswell95a}, before being positioned in 1999 with a peak flux density of 0.4\,Jy \citep{caswell09a}. In our survey cube observations the peak flux density was 0.9\,Jy and then the 2008 March MX found a peak flux density of 0.5\,Jy. The secondary feature at 6.5\,km\,s$^{-1}$ flared in the MX observation (it was undetectable against the noise in the survey cube, but was 0.4\,Jy in the MX).
Finally, the known source 269.153--1.128 had faded in the survey cube observations (from 1.4\,Jy in 1992, as reported by \citet{caswell95a}, to only 0.65\,Jy), but flared in the MX to 4.1\,Jy.

\subparagraph{281.710--1.104} This new source had a peak flux density exceeding 1\,Jy in both the survey cube and ATCA observations, but had faded to 0.35\,Jy in the 2008 March MX observation. 

\subparagraph{285.337--0.002} This known source,  first detected by \citet{caswell95a}, has been allocated to the far kinematic distance by both \citet{busfield06} and \citet{green11b}.

\subparagraph{287.371+0.644} This source had slightly faded in peak intensity between the survey cube and MX observations (observed 2006 February and 2008 March respectively). This source was allocated to the near kinematic distance by \citet{green11b}.

\subparagraph{290.374+1.661} This source has a history of variability. It was originally detected at 2.4\,Jy by  \citet{caswell95a} before being positioned with the ATCA in 1999 with a peak flux density of 0.5\,Jy \citep{caswell09a}. It was then found with our initial survey observations to have a peak flux density of 0.85\,Jy before brightening to 2.1\,Jy in the later MX observation taken 2008 March.

\subparagraph{290.411--2.915} This source was detected by \citet{walsh97} and is formally outside our survey region, but we have included it in the catalogue, and presented a newly measured spectrum, for completeness.

\subparagraph{291.270--0.719 and 291.274--0.709} The second source is the brighter of the two, consisting only of the features between $-$31\,km\,s$^{-1}$ and $-$28\,km\,s$^{-1}$, peaking at $-$29.7\,km\,s$^{-1}$. The first source, with a peak of 7\,Jy at $-$31.2\,km\,s$^{-1}$,  now appears to have an extended velocity range relative to the previous reported observations \citep[e.g.][]{caswell09a}, due to the presence of an additional feature at $-$19\,km\,s$^{-1}$ in the 2008 March MX observation (which was not present in the survey cube observation); we have assigned the feature as most probably part of 291.270--0.719, but we were unable to position it precisely. 

\subparagraph{291.579--0.431 and 291.582--0.435} The first source of this pair is the weaker feature peaking at 15.2\,km\,s$^{-1}$, with a wider range of velocities; the second source is the brighter feature peaking at 10.4\,km\,s$^{-1}$, with a narrow range of velocities. There are some weak features with velocities of $\sim$19\,km\,s$^{-1}$ which we include in the listed range of the first source. These features have been seen before \citep{caswell95a}, but previously without a high degree of confidence in their authenticity.  

\subparagraph{291.642--0.546} This new source has demonstrated variability over the course of our observations. It was detected in the survey cube observations (2006 June-August) with a peak flux density of 1.1\,Jy at 12.1\,km\,s$^{-1}$. It was positioned in 2007 February with a peak flux density of 0.6\,Jy. Then two MX observations obtained in 2009 March could only put upper limits on the peak flux density of 0.3\,Jy and 0.2\,Jy. The survey cube spectrum is presented in Figure\,\ref{spectra}.

\subparagraph{292.074--1.131} This source varied between the survey cube observation (2006 June-August) and the later MX observation (2009 March), with respective peak flux densities of 1.76\,Jy and 0.85\,Jy.
 
\subparagraph{292.468+0.168} This new source has a feature at 23.5\,km\,s$^{-1}$, which was seen in the survey cube in excess of 4\,Jy (observed 2006 June-August), positioned with a flux density of only $\sim$0.5\,Jy with the ATCA (2007 February), and then was barely detectable in the 2009 March MX observation ($<$0.2\,Jy).   

\subparagraph{294.337--1.706} This new source was discovered in the survey cube observations (obtained 2006 August and 2007 January) with a peak flux density of 14.5\,Jy at $-$11.5\,km\,s$^{-1}$. It was positioned in 2007 February with a peak flux density of 30\,Jy. It then faded to just 0.8\,Jy in an MX taken in 2008 March. Two new MX observations during 2009 March, both failed to detect the source (with a limit of $<$0.1\,Jy). The survey cube spectrum is presented in Figure\,\ref{spectra} as well as the 2008 March MX spectrum.

\begin{table*} \centering \caption{\small Methanol maser positions. The references are: CVF95: \citet{caswell95ab}, C+95: \citet{caswell95a}, C97: \citet{caswell97}, W97: \citet{walsh97}, W98: \citet{walsh98}, P98: \citet{phillips98}, M01: \citet{minier01}, M03: \citet{minier03}, V05: \citet{voronkov05}, X08: \citet{xu08}, C2009: \citet{caswell09a}. Position is from non-bracketed reference. Sources with a $^{\dagger}$ have positions (in italics) from single--dish observations only. The source with an $^{\alpha}$ is possibly thermal, see main text for details. The MX flux densities with a $^{\beta}$ signify these were specific epoch MX observation values as the sources were undetected at other epochs (see individual source notes for details).} 
\begin{tabular}{lcrcccrcrl}
\hline

\multicolumn{1}{c}{Source Name} & \multicolumn{2}{c}{Equatorial
Coordinates} & \multicolumn{2}{c}{Velocity range} & \multicolumn{2}{c}{MX
data} & \multicolumn{2}{c}{Survey Cube data} & \multicolumn{1}{l}{Position}\\

\ (~~~l,~~~~~~~b~~~)    &       RA(2000)        &       Dec(2000)       &
$\rm V_{L}$&$\rm V_{H}$ &  $\rm V_{pk}$(MX)     &  $\rm S_{pk}$(MX)  &
$\rm V_{pk}$(SC) & $\rm S_{pk}$(SC)     &   Refs,
epoch    \\
\ (~~~$^\circ$~~~~~~~$^\circ$~~~) & (h~~m~~~s) & (~$^\circ$~~ '~~~~") &
\multicolumn{2}{c}{(km\,s$^{-1}$ )} & (km\,s$^{-1}$ ) &  (Jy) & (km\,s$^{-1}$ ) & (Jy) & \\

\hline
188.794+1.031& 06 09 06.96& +21 50 41.3&$-$6.0&$-$4.0&$-$5.5&9.50&$-$5.5&7.82&MERLIN\\
188.946+0.886&06 08 53.32& +21 38 29.1&$-$5.0&12.0&10.8&607.17&10.8&517.30&C2009\\
189.030+0.783&06 08 40.65& +21 31 07.0&8.5&10.0&8.9&16.97&8.9&11.05&C2009\\
189.471$-$1.216& 06 02 08.37& +20 09 20.1&18.5&19.5&18.8&1.87&18.8&2.08&MERLIN\\
189.778+0.345&06 08 35.28& +20 39 06.7&2.0&6.0&5.5&2.51&5.7&7.69&C2009\\
192.600$-$0.048&06 12 53.99& +17 59 23.7&1.0&6.0&4.6&83.00&4.6&67.85&C2009 (M01)\\
196.454$-$1.677&06 14 37.03& +13 49 36.6&13.5&16.0&15.2&21.58&15.2&18.13&C2009 (M01)\\
206.542$-$16.355&05 41 44.14& $-$01 54 45.9&11.5&13.0&12.3&1.97&12.2&3.65&MERLIN (M03)\\
208.996$-$19.386$^{\alpha}$&05 35 14.50& $-$05 22 45.0&5.0&10.0&7.3&$<$0.20&7.3&$<$0.50&V05\\
209.016$-$19.398&05 35 13.95& $-$05 24 09.4&$-$2.4&0.5&$-$1.5&1.3&$-$1.6&$2.78$&MERLIN (V05)\\
212.06$-$0.74$^{\dagger}$&{\it 06 47 12.9}&{\it +00 26 07}&42.5&49.0&43.3&0.32&44.4&1.03&X08\\
213.705$-$12.597&06 07 47.85& $-$06 22 55.2&8.0&14.0&10.7&96.34&10.8&278.40&C2009\\
232.620+0.996&07 32 09.79& $-$16 58 12.4&18.5&24.5&22.9&164.96&22.9&178.20&C2009 (W98)\\
254.880+0.451&08 22 25.25& $-$36 15 00.1&29.5&31.0&30.1&1.59&30.1&1.76&2007nov24\\
259.939$-$0.041&08 35 31.07& $-$40 38 23.9&$-$1.5&$-$0.5&$-$1.0&1.74&$-$1.0&2.15&2008aug23 (W97,W98)\\
263.250+0.514&08 48 47.84& $-$42 54 28.3&11.0&17.5&12.3&67.26&12.3&68.93&C2009 (W98)\\
264.140+2.018&08 58 11.66& $-$42 37 30.6&7.5&9.0&8.1&7.60&8.1&1.40&2008aug23\\
264.289+1.469&08 56 26.80& $-$43 05 42.1&5.5&10.0&8.7&0.51&8.7&0.90&C2009\\
269.153$-$1.128&09 03 33.46& $-$48 28 02.6&7.5&16.5&16.0&4.10&16.0&0.65&C2009\\
269.456$-$1.467&09 03 14.78& $-$48 55 11.2&53.5&56.5&56.1&5.23&56.1&5.96&2007feb05 (W98)\\
269.658$-$1.270&09 04 57.72& $-$48 56 18.4&13.5&17.0&16.2&5.39&16.2&4.52&2007feb05\\
270.255+0.835&09 16 41.51& $-$47 56 12.1&2.5&5.5&3.9&0.42&3.9&0.50&C2009\\
281.710$-$1.104&10 05 05.63& $-$56 57 24.7&0.5&3.0&0.9&0.35&0.9&1.32&2006apr01\\
284.352$-$0.419&10 24 10.89& $-$57 52 38.8&3.0&11.0&3.9&2.47&3.9&2.71&C97\\
284.694$-$0.361&10 26 36.29& $-$58 00 34.3&12.0&14.0&13.3&3.28&13.3&5.01&2006apr01\\
285.337$-$0.002&10 32 09.62& $-$58 02 04.6&$-$8.5&3.0&0.7&11.37&0.7&17.95&C2009 (W98)\\
286.383$-$1.834&10 31 55.12& $-$60 08 38.6&8.0&10.5&9.6&15.49&9.6&17.58&2006apr01\\
287.371+0.644&10 48 04.44& $-$58 27 01.0&$-$3.5&0.0&$-$1.9&104.25&$-$1.9&115.70&C2009 (W98)\\
290.374+1.661&11 12 18.10& $-$58 46 21.5&$-$28.0&$-$22.0&$-$24.2&2.07&$-$24.2&0.85&C2009\\
290.411$-$2.915&10 57 33.89& $-$62 59 03.5&$-$17.5&$-$14.5&$-$16.0&4.56&$-$&$-$&2008oct20 (W98)\\
291.270$-$0.719&11 11 49.44& $-$61 18 51.9&$-$32.0&$-$19.0&$-$31.2&8.00&$-$31.2&7.00&C2009 (W98)\\
291.274$-$0.709&11 11 53.35& $-$61 18 23.7&$-$31.0&$-$28.0&$-$29.7&69.28&$-$29.7&71.17&C2009 (W98)\\
291.579$-$0.431&11 15 05.76& $-$61 09 40.8&11.0&19.0&15.2&1.00&15.2&1.10&C2009\\
291.582$-$0.435&11 15 06.61& $-$61 09 58.3&8.0&11.0&10.4&2.75&10.4&2.14&C2009\\
291.642$-$0.546&11 15 14.32& $-$61 17 26.7&11.5&12.5&12.1&$<$0.2&12.1&1.10&2007feb05\\
291.879$-$0.810&11 16 17.35& $-$61 37 20.7&30.0&34.5&33.5&1.34&33.5&1.34&2007feb05\\
292.074$-$1.131&11 16 51.24& $-$61 59 32.6&$-$19.5&$-$18.5&$-$19.1&0.85&$-$19.1&1.76&2007nov24\\
292.468+0.168&11 23 42.17& $-$60 54 33.5&8.0&24.0&10.9&4.40&10.9&6.23&2007feb05\\
293.723$-$1.742&11 28 32.97& $-$63 07 18.6&23.0&26.0&24.2&0.55&24.2&0.82&2007nov24\\
293.827$-$0.746&11 32 05.56& $-$62 12 25.3&35.0&39.5&37.0&2.50&37.0&2.93&2008aug23 (W98)\\
293.942$-$0.874&11 32 42.09& $-$62 21 47.5&37.0&41.5&41.1&4.30&41.1&3.66&2007feb05 (W97,W98)\\
294.337$-$1.706&11 33 49.91& $-$63 16 32.5&$-$12.5&$-$11.0&$-$11.7&0.80$^{\beta}$&$-$11.7&14.51&2007feb05\\
294.511$-$1.621&11 35 32.25& $-$63 14 43.2&$-$13.5&$-$4.5&$-$11.9&7.14&$-$11.9&9.00&C97\\
294.977$-$1.734&11 39 13.94& $-$63 29 04.6&$-$7.0&$-$5.0&$-$5.3&$<$0.10&$-$5.3&1.99&2007feb05\\
294.990$-$1.719&11 39 22.88& $-$63 28 26.4&$-$13.0&$-$11.5&$-$12.3&11.00&$-$12.3&9.08&C2009\\
296.893$-$1.305&11 56 50.07& $-$63 32 05.5&20.5&23.0&22.2&1.20&22.2&2.32&C2009\\
297.406$-$0.622&12 02 30.42& $-$62 58 03.8&25.5&28.5&27.8&1.42&27.8&1.65&2007feb05\\
298.177$-$0.795&12 08 57.56& $-$63 16 25.8&22.5&28.0&23.5&2.62&23.5&1.74&2008jan21\\
298.213$-$0.343&12 09 55.18& $-$62 50 01.1&32.5&38.0&33.3&1.30&33.3&1.17&C2009\\
298.262+0.739&12 11 47.65& $-$61 46 20.9&$-$31.5&$-$29.0&$-$30.1&14.64&$-$30.1&14.95&2008aug23 (W98)\\
298.632$-$0.362&12 13 31.63& $-$62 55 01.0&36.5&45.5&37.2&1.35&38.7&1.35&2008jan21\\
298.723$-$0.086&12 14 39.50& $-$62 39 25.9&13.5&25.5&23.5&1.12&23.5&1.83&2008jan21\\
299.013+0.128&12 17 24.60& $-$62 29 03.7&18.0&19.5&18.3&8.23&18.3&7.85&C2009\\
299.772$-$0.005&12 23 48.97& $-$62 42 25.3&$-$10.0&0.0&$-$6.8&15.65&$-$6.8&13.38&2008jan21\\
300.504$-$0.176&12 30 03.58& $-$62 56 48.7&2.5&10.5&7.5&4.11&7.5&7.04&C2009\\
300.969+1.148&12 34 53.29& $-$61 39 40.0&$-$40.0&$-$34.5&$-$37.2&4.84&$-$37.2&4.36&C2009 (C97)\\
301.136$-$0.226&12 35 35.14& $-$63 02 32.6&$-$41.0&$-$37.0&$-$39.6&1.69&$-$39.6&1.57&C2009\\
\hline
\end{tabular} 
\label{resotable}
\end{table*}

\begin{table*} \addtocounter{table}{-1} \centering \caption{\small cont.} 
\begin{tabular}{lcrcccrcrl}
\hline
\multicolumn{1}{c}{Source Name} & \multicolumn{2}{c}{Equatorial
Coordinates} & \multicolumn{2}{c}{Velocity range} & \multicolumn{2}{c}{MX
data} & \multicolumn{2}{c}{Survey Cube data} & \multicolumn{1}{l}{Position}\\
\ (~~~l,~~~~~~~b~~~)    &       RA(2000)        &       Dec(2000)       &
$\rm V_{L}$&$\rm V_{H}$ &  $\rm V_{pk}$(MX)     &  $\rm S_{pk}$(MX)  &
$\rm V_{pk}$(SC) & $\rm S_{pk}$(SC)     &   Refs,
epoch    \\
\ (~~~$^\circ$~~~~~~~$^\circ$~~~) & (h~~m~~~s) & (~$^\circ$~~ '~~~~") &
\multicolumn{2}{c}{(km\,s$^{-1}$ )} & (km\,s$^{-1}$ ) &  (Jy) & (km\,s$^{-1}$ ) & (Jy) & \\
\hline
302.032$-$0.061&12 43 31.92& $-$62 55 06.7&$-$43.0&$-$32.5&$-$35.7&10.65&$-$35.7&8.27&C2009 (W98)\\
302.034+0.625&12 43 43.44& $-$62 13 58.5&$-$50.0&$-$38.5&$-$39.1&2.70&$-$39.1&1.90&2008aug23\\
302.455$-$0.741&12 47 08.65& $-$63 36 30.2&32.0&38.0&32.6&1.16&32.6&1.29&2008aug23 (2008jan21)\\
303.507$-$0.721&12 56 36.66& $-$63 35 14.5&10.5&18.5&14.2&2.06&14.2&1.62&2008aug23 (2008jan21)\\
303.846$-$0.363&12 59 33.37& $-$63 13 14.7&22.5&33.0&24.8&7.40&25.4&6.85&2008jan21\\
303.869+0.194&12 59 35.84& $-$62 39 47.6&$-$37.5&$-$36.0&$-$36.9&0.90&$-$36.9&1.00&2008jan21\\
304.367$-$0.336&13 04 09.82& $-$63 10 20.2&31.0&33.0&32.7&0.94&32.7&1.49&2008jan21\\
304.887+0.635&13 08 11.95& $-$62 10 21.9&$-$35.5&$-$34.5&$-$35.1&1.01&$-$35.1&0.81&2008jan21\\
305.199+0.005&13 11 17.20& $-$62 46 46.0&$-$45.0&$-$40.0&$-$42.8&6.00&$-$42.8&4.25&C2009 (C97, W98)\\
305.200+0.019&13 11 16.93& $-$62 45 55.1&$-$38.0&$-$29.5&$-$33.1&46.00&$-$33.1&44.98&C2009 (C97, W98)\\
305.202+0.208&13 11 10.49& $-$62 34 38.8&$-$46.5&$-$42.0&$-$44.0&92.00&$-$44.0&75.00&C2009 (CVF95)\\
305.208+0.206&13 11 13.71& $-$62 34 41.4&$-$42.0&$-$34.0&$-$38.3&447.00&$-$38.3&430.00&C2009 (CVF95)\\
305.248+0.245&13 11 32.47& $-$62 32 09.1&$-$35.5&$-$28.0&$-$32.1&7.09&$-$32.1&6.39&C2009 (P98)\\
305.362+0.150&13 12 35.86& $-$62 37 17.9&$-$38.0&$-$35.0&$-$36.5&5.00&$-$36.5&4.97&C2009 (W98)\\
305.366+0.184&13 12 36.74& $-$62 35 14.7&$-$35.0&$-$29.5&$-$33.7&3.30&$-$33.7&2.89&C2009 (W98)\\
305.475$-$0.096&13 13 45.75& $-$62 51 28.3&$-$55.0&$-$26.5&$-$35.3&2.90&$-$39.0&2.57&2008jan21\\
305.563+0.013&13 14 26.90& $-$62 44 29.4&$-$42.0&$-$32.0&$-$37.3&4.94&$-$37.3&4.58&2008aug23 (W97,W98)\\
305.573$-$0.342&13 14 49.12& $-$63 05 38.4&$-$55.0&$-$50.0&$-$51.0&1.02&$-$51.0&0.60&2008aug23\\
305.615$-$0.344&13 15 11.31& $-$63 05 29.8&$-$36.0&$-$26.5&$-$34.9&3.57&$-$34.9&2.83&2008aug23 (2008jan21)\\
305.634+1.645&13 13 49.43& $-$61 06 32.4&$-$58.0&$-$54.0&$-$54.8&7.17&$-$54.8&7.30&2008jan21\\
305.646+1.589&13 13 58.05& $-$61 09 51.7&$-$60.0&$-$55.0&$-$58.1&3.86&$-$58.1&3.05&2008jan21\\
305.799$-$0.245&13 16 43.23& $-$62 58 32.9&$-$40.0&$-$27.0&$-$39.3&0.48&$-$39.3&0.63&C2009\\
305.822$-$0.115&13 16 48.71& $-$62 50 38.1&$-$47.0&$-$38.5&$-$42.2&2.69&$-$42.2&2.92&2008jan21\\
305.887+0.017&13 17 15.53& $-$62 42 23.0&$-$35.0&$-$29.0&$-$34.2&9.54&$-$34.2&7.93&C2009\\
305.940$-$0.164&13 17 53.05& $-$62 52 50.5&$-$51.0&$-$50.5&$-$50.9&0.77&$-$50.9&0.88&2008jan21\\
306.322$-$0.334&13 21 23.01& $-$63 00 29.5&$-$25.0&$-$21.0&$-$24.7&0.56&$-$24.7&0.50&C2009\\
307.132$-$0.476&13 28 37.94& $-$63 02 38.8&$-$36.0&$-$33.5&$-$34.0&1.20&$-$34.0&1.30&2007nov26 (2008jan22)\\
307.133$-$0.477&13 28 38.29& $-$63 02 40.4&$-$40.0&$-$36.0&$-$38.7&2.36&$-$38.7&1.82&2007nov26 (2008jan22)\\
308.056$-$0.396&13 36 32.16& $-$62 49 05.6&$-$12.5&$-$10.0&$-$11.8&1.50&$-$11.8&1.26&2007nov24\\
308.075$-$0.411&13 36 43.51&$-$62 49 45.2&$-$8.5&$-$6.5&$-$7.5&0.90&$-$7.5&0.80&2007nov24\\
308.651$-$0.507&13 41 50.19& $-$62 49 05.2&0.5&3.5&3.2&5.67&3.2&3.99&2007nov24\\
308.686+0.530&13 40 25.33& $-$61 47 37.6&$-$54.0&$-$44.0&$-$53.1&1.76&$-$53.1&1.81&2007nov26\\
308.715$-$0.216&13 41 53.59& $-$62 31 11.7&$-$17.0&$-$11.0&$-$12.5&1.04&$-$12.5&1.07&2007nov24\\
308.754+0.549&13 40 57.60& $-$61 45 43.4&$-$52.0&$-$38.5&$-$45.3&11.92&$-$45.3&11.63&C2009 (2007nov26)\\
308.918+0.123&13 43 01.85& $-$62 08 52.2&$-$56.0&$-$52.5&$-$54.8&43.00&$-$54.8&45.24&C2009\\
309.384$-$0.135&13 47 23.98& $-$62 18 12.0&$-$50.5&$-$49.0&$-$49.6&1.36&$-$49.6&1.35&C2009\\
309.901+0.231&13 51 01.05& $-$61 49 56.0&$-$57.0&$-$52.0&$-$54.6&20.98&$-$54.6&16.70&2008aug23\\
309.921+0.479&13 50 41.78& $-$61 35 10.2&$-$65.0&$-$53.5&$-$59.7&884.67&$-$59.7&885.30&C97 (P98,W98)\\
310.144+0.760&13 51 58.43& $-$61 15 41.3&$-$59.0&$-$53.5&$-$55.7&78.23&$-$55.7&93.87&C2009\\
310.180$-$0.122&13 54 01.78& $-$62 06 34.6&2.5&4.5&3.6&1.07&3.6&1.13&2007nov24 (W97)\\
311.230$-$0.032&14 02 27.47& $-$61 45 14.4&24.0&27.0&24.8&2.27&24.8&1.92&2007nov24\\
311.551$-$0.055&14 05 07.06& $-$61 41 15.9&$-$57.0&$-$55.5&$-$56.3&1.00&$-$56.3&1.13&2007nov26\\
311.628+0.266&14 04 59.20& $-$61 21 29.4&$-$60.0&$-$55.0&$-$57.8&4.33&$-$57.8&4.31&2007nov26 (W97)\\
311.643$-$0.380&14 06 38.77& $-$61 58 23.1&31.0&36.5&32.6&10.29&32.6&11.13&C2009 (W98)\\
311.729$-$0.735&14 08 13.97& $-$62 17 16.5&25.5&31.5&30.9&0.46&30.9&0.76&2007nov24\\
311.947+0.142&14 07 49.72& $-$61 23 08.3&$-$48.0&$-$38.0&$-$44.0&0.28&$-$38.7&0.25&C2009 (C+95)\\
312.071+0.082&14 08 58.20& $-$61 24 23.8&$-$30.5&$-$29.5&$-$34.8&67.86&$-$34.8&83.20&2007jul19\\
312.108+0.262&14 08 49.31& $-$61 13 25.1&$-$54.5&$-$48.5&$-$49.9&19.74&$-$49.9&22.89&C2009 (W98)\\
312.307+0.661&14 09 24.95& $-$60 47 00.5&$-$13.0&$-$11.5&$-$12.3&4.10&$-$12.3&3.83&2007jul18\\
312.501$-$0.084&14 12 48.95& $-$61 26 03.2&21.0&25.0&21.8&1.19&21.8&1.51&2007jul18\\
312.597+0.045&14 13 14.35& $-$61 16 57.7&$-$60.5&$-$59.0&$-$59.8&0.90&$-$59.8&1.00&C2009\\
312.598+0.045&14 13 15.03& $-$61 16 53.6&$-$68.5&$-$64.0&$-$67.9&21.45&$-$67.9&10.35&C2009 (W98)\\
312.698+0.126&14 13 49.85& $-$61 10 24.1&28.0&34.5&29.5&1.65&29.5&0.99&2007jul18\\
312.702$-$0.087&14 14 25.12& $-$61 22 29.0&$-$59.5&$-$55.0&$-$58.2&0.81&$-$58.2&1.39&2007jul19\\
313.469+0.190&14 19 40.94& $-$60 51 47.3&$-$15.5&$-$3.0&$-$9.3&29.94&$-$9.3&26.29&C2009 (W98)\\
313.577+0.325&14 20 08.58& $-$60 42 00.8&$-$53.5&$-$46.0&$-$47.8&100.51&$-$47.8&104.10&C2009\\
313.705$-$0.190&14 22 34.74& $-$61 08 26.8&$-$47.5&$-$41.0&$-$41.5&1.65&$-$41.5&1.74&C2009 (2007nov26)\\
313.767$-$0.863&14 25 01.73& $-$61 44 58.1&$-$57.0&$-$52.5&$-$56.3&9.50&$-$56.3&9.00&C2009 (W98)\\
313.774$-$0.863&14 25 04.80& $-$61 44 50.3&$-$46.0&$-$40.5&$-$44.8&14.30&$-$44.8&22.34&C2009 (W98)\\
313.994$-$0.084&14 24 30.78& $-$60 56 28.3&$-$9.0&$-$2.0&$-$4.9&15.03&$-$4.9&15.16&2007jul18\\
314.221+0.273&14 25 12.89& $-$60 31 38.4&$-$62.5&$-$60.5&$-$61.6&2.76&$-$61.6&2.70&2007jul19\\
314.320+0.112&14 26 26.20& $-$60 38 31.3&$-$59.0&$-$42.5&$-$43.5&36.94&$-$43.5&37.23&C2009\\
315.803$-$0.575&14 39 46.46& $-$60 42 39.6&7.5&9.0&8.2&8.89&8.2&3.85&2007jul18\\

\hline
\end{tabular} 
\label{resotable}
\end{table*}

\begin{table*} \addtocounter{table}{-1} \centering \caption{\small cont.} 
\begin{tabular}{lcrcccrcrl}
\hline
\multicolumn{1}{c}{Source Name} & \multicolumn{2}{c}{Equatorial
Coordinates} & \multicolumn{2}{c}{Velocity range} & \multicolumn{2}{c}{MX
data} & \multicolumn{2}{c}{Survey Cube data} & \multicolumn{1}{l}{Position}\\
\ (~~~l,~~~~~~~b~~~)    &       RA(2000)        &       Dec(2000)       &
$\rm V_{L}$&$\rm V_{H}$ &  $\rm V_{pk}$(MX)     &  $\rm S_{pk}$(MX)  &
$\rm V_{pk}$(SC) & $\rm S_{pk}$(SC)     &   Refs,
epoch    \\
\ (~~~$^\circ$~~~~~~~$^\circ$~~~) & (h~~m~~~s) & (~$^\circ$~~ '~~~~") &
\multicolumn{2}{c}{(km\,s$^{-1}$ )} & (km\,s$^{-1}$ ) &  (Jy) & (km\,s$^{-1}$ ) & (Jy) & \\
\hline
316.359$-$0.362&14 43 11.20& $-$60 17 13.3&1.5&8.5&3.4&98.39&3.4&107.30&C2009\\
316.381$-$0.379&14 43 24.21& $-$60 17 37.4&$-$6.0&1.5&$-$0.5&18.12&$-$0.5&22.48&C2009\\
316.412$-$0.308&14 43 23.34& $-$60 13 00.9&$-$7.0&2.0&$-$5.6&10.98&$-$5.6&9.00&C2009\\
316.484$-$0.310&14 43 55.37& $-$60 11 18.8&$-$16.0&$-$11.0&$-$11.5&0.72&$-$15.6&2.30&2007jul18\\
316.640$-$0.087&14 44 18.45& $-$59 55 11.5&$-$25.5&$-$14.5&$-$20.4&96.29&$-$20.4&95.28&C2009\\
316.811$-$0.057&14 45 26.43& $-$59 49 16.3&$-$49.0&$-$37.0&$-$45.8&52.00&$-$45.8&75.32&C2009\\
317.029+0.361&14 45 35.54& $-$59 20 58.6&$-$51.5&$-$46.0&$-$47.9&0.89&$-$47.9&1.24&2007jul19\\
317.061+0.256&14 46 10.31& $-$59 25 50.5&$-$45.0&$-$43.0&$-$43.8&0.66&$-$43.7&1.04&2007jul19\\
317.466$-$0.402&14 51 19.69& $-$59 50 50.7&$-$51.0&$-$35.0&$-$37.7&51.35&$-$37.7&51.54&2007jul19\\
317.701+0.110&14 51 11.69& $-$59 17 02.1&$-$48.0&$-$40.0&$-$42.3&23.10&$-$42.3&26.51&C2009\\
318.043$-$1.404&14 59 08.61& $-$60 28 25.5&44.0&47.0&46.3&6.07&46.3&7.78&C2009\\
318.050+0.087&14 53 42.67& $-$59 08 52.4&$-$59.0&$-$46.0&$-$51.8&11.36&$-$51.8&12.60&C2009\\
318.472$-$0.214&14 57 42.86& $-$59 13 15.9&$-$20.4&$-$18.5&$-$19.1&1.05&$-$19.1&1.26&2007feb05 (2007jul18)\\
318.948$-$0.196&15 00 55.40& $-$58 58 52.1&$-$39.0&$-$31.0&$-$34.6&569.20&$-$34.6&616.50&2007feb05 \\
319.163$-$0.421&15 03 13.74& $-$59 04 30.5&$-$22.5&$-$12.5&$-$21.1&7.03&$-$21.1&6.93&2007feb05\\
319.836$-$0.197&15 06 54.65& $-$58 33 00.0&$-$14.0&$-$8.5&$-$9.2&0.23&$-$9.2&0.65&C2009\\
320.123$-$0.504&15 10 00.17& $-$58 40 18.0&$-$12.0&$-$9.0&$-$10.1&3.58&$-$10.1&3.67&C2009 (W98)\\
320.231$-$0.284&15 09 51.94& $-$58 25 38.5&$-$71.5&$-$58.0&$-$62.5&53.72&$-$62.5&31.87&C2009 (W98)\\
320.244$-$0.562&15 11 01.61& $-$58 39 37.7&$-$50.5&$-$49.0&$-$49.7&1.46&$-$49.7&1.62&2008oct20 (2006dec03)\\
320.285$-$0.308&15 10 19.00& $-$58 25 16.7&$-$69.5&$-$68.5&$-$69.0&0.84&$-$69.0&0.80&2006dec05\\
320.414+0.109&15 09 33.34& $-$57 59 46.6&$-$14.5&$-$13.0&$-$13.4&2.56&$-$13.4&3.19&2006dec03\\
320.424+0.089&15 09 41.56& $-$58 00 29.2&$-$10.0&$-$4.0&$-$8.0&2.00&$-$8.0&1.50&2006dec03\\
320.625+0.098&15 10 58.12& $-$57 53 53.4&$-$8.0&$-$6.5&$-$7.6&0.54&$-$7.6&0.80&2007nov25 \\
320.780+0.248&15 11 23.48& $-$57 41 25.1&$-$12.5&$-$2.5&$-$5.1&34.92&$-$5.1&40.03&2006dec03\\
321.030$-$0.485&15 15 51.79& $-$58 11 18.0&$-$68.0&$-$56.0&$-$66.5&30.00&$-$66.5&28.00&C2009 (W98)\\
321.033$-$0.483&15 15 52.63& $-$58 11 07.7&$-$69.0&$-$54.0&$-$61.2&76.41&$-$61.2&142.20&C2009 (W98)\\
321.148$-$0.529&15 16 48.39& $-$58 09 50.2&$-$66.5&$-$65.0&$-$66.1&9.05&$-$66.1&8.50&C2009\\
321.704+1.168&15 13 44.72& $-$56 25 34.4&$-$46.0&$-$43.0&$-$44.3&$<$0.10&$-$44.3&2.09&2007feb05 (C+95)\\
322.158+0.636&15 18 34.64& $-$56 38 25.3&$-$66.0&$-$51.0&$-$64.0&330.00&$-$63.0&253.90&C2009 (CVF95)\\
322.705$-$0.331&15 25 47.52& $-$57 09 15.5&$-$24.5&$-$20.5&$-$21.6&2.40&$-$21.6&2.02&2007feb05 (2006dec03) \\
323.459$-$0.079&15 29 19.33& $-$56 31 22.8&$-$68.5&$-$66.0&$-$66.9&17.81&$-$66.9&15.21&C97 (W98)\\
323.740$-$0.263&15 31 45.45& $-$56 30 50.1&$-$59.0&$-$42.0&$-$50.5&3114.39&$-$50.5&3231.00&CVF95 (W98)\\
323.766$-$1.370&15 36 37.69& $-$57 24 01.6&45.5&51.5&46.6&2.47&46.6&2.69&2006dec03\\
323.793$-$0.397&15 32 38.09& $-$56 35 36.5&$-$26.0&$-$22.0&$-$22.7&2.36&$-$22.7&2.06&2006dec03\\
323.799+0.017&15 30 57.12& $-$56 15 01.5&$-$60.0&$-$55.5&$-$56.2&6.67&$-$56.2&4.34&2006dec03\\
324.716+0.342&15 34 57.47& $-$55 27 23.6&$-$51.0&$-$45.0&$-$45.9&10.29&$-$45.9&10.83&C2009\\
324.789$-$0.378&15 38 22.85& $-$55 59 47.0&11.5&12.5&11.8&1.15&11.8&1.74&2006mar30\\
324.915+0.158&15 36 51.17& $-$55 29 22.9&$-$6.0&0.0&$-$2.3&8.99&$-$2.3&12.10&2006mar30\\
324.923$-$0.568&15 39 57.64& $-$56 04 08.3&$-$80.0&$-$77.0&$-$78.3&3.19&$-$78.3&4.92&2006apr02 (W98)\\
325.659$-$0.022&15 41 48.70& $-$55 11 27.0&28.6&30.5&29.4&0.57&29.4&2.25&2006mar30\\
326.323$-$0.393&15 47 04.58& $-$55 04 50.4&$-$77.5&$-$69.0&$-$69.4&3.25&$-$69.4&2.02&2006mar30\\
326.448$-$0.748&15 49 18.63& $-$55 16 51.6&$-$73.0&$-$57.5&$-$68.6&4.50&$-$71.7&9.04&2006mar30\\
326.475+0.703&15 43 16.64& $-$54 07 14.6&$-$51.0&$-$37.0&$-$38.5&122.27&$-$38.5&135.00&C2009\\
326.476+0.695&15 43 18.90& $-$54 07 35.5&$-$51.0&$-$37.0&$-$43.6&2.90&$-$43.6&3.50&C2009\\
326.608+0.799&15 43 35.99& $-$53 57 47.3&$-$45.5&$-$44.5&$-$45.2&1.40&$-$45.2&1.51&2006mar31\\
326.641+0.611&15 44 33.33& $-$54 05 31.5&$-$45.0&$-$34.5&$-$42.6&30.79&$-$42.6&26.36&C2009 (W98)\\
326.662+0.520&15 45 02.95& $-$54 09 03.1&$-$43.0&$-$38.0&$-$38.6&15.93&$-$38.6&29.09&C2009\\
326.859$-$0.677&15 51 14.19& $-$54 58 04.8&$-$63.0&$-$57.0&$-$58.1&15.26&$-$58.1&22.36&C2009\\
326.986$-$0.031&15 49 07.67& $-$54 23 04.5&$-$61.0&$-$56.5&$-$60.7&2.31&$-$60.7&1.74&2006apr02\\
327.120+0.511&15 47 32.73& $-$53 52 38.4&$-$92.0&$-$83.0&$-$87.1&54.80&$-$87.1&60.04&C2009 (W98)\\
327.282$-$0.469&15 52 36.03& $-$54 32 24.0&$-$1.5&1.5&0.0&5.40&0.0&4.05&2006mar30\\
327.291$-$0.578&15 53 07.65& $-$54 37 06.1&$-$49.0&$-$36.0&$-$37.0&2.60&$-$37.0&3.37&C2009 (2006apr02)\\
327.392+0.199&15 50 18.48& $-$53 57 06.3&$-$87.0&$-$79.0&$-$84.5&8.54&$-$84.5&11.36&C2009\\
327.395+0.197&15 50 20.06& $-$53 57 07.5&$-$90.0&$-$88.0&$-$89.5&9.00&$-$89.5&9.00&C2009\\
327.402+0.445&15 49 19.50& $-$53 45 13.9&$-$84.5&$-$72.0&$-$82.6&82.17&$-$82.6&157.80&C2009\\
327.566$-$0.850&15 55 47.61& $-$54 39 11.4&$-$30.5&$-$21.5&$-$29.0&12.50&$-$29.7&14.98&2006mar30\\
327.590$-$0.094&15 52 36.82& $-$54 03 18.7&$-$87.0&$-$85.0&$-$86.3&4.05&$-$86.3&5.15&C2009\\
327.618$-$0.111&15 52 50.22& $-$54 03 00.8&$-$99.5&$-$94.0&$-$98.0&2.10&$-$97.6&2.61&C2009 (2006apr02)\\
327.710$-$0.394&15 54 32.98& $-$54 12 36.4&$-$80.0&$-$72.0&$-$78.0&4.18&$-$78.0&4.23&2006mar30\\
327.808$-$0.634&15 56 06.91& $-$54 19 54.4&$-$43.5&$-$41.0&$-$42.3&2.65&$-$42.3&3.24&2006mar31\\
327.863+0.098&15 53 13.44& $-$53 44 01.8&$-$46.5&$-$45.0&$-$46.3&1.58&$-$46.3&1.48&2006mar31\\
327.945$-$0.115&15 54 33.91& $-$53 50 44.3&$-$52.5&$-$50.0&$-$51.6&7.75&$-$51.6&8.75&C2009\\

\hline
\end{tabular} 
\label{resotable}
\end{table*}

\begin{table*} \addtocounter{table}{-1} \centering \caption{\small cont.} 
\begin{tabular}{lcrcccrcrl}
\hline
\multicolumn{1}{c}{Source Name} & \multicolumn{2}{c}{Equatorial
Coordinates} & \multicolumn{2}{c}{Velocity range} & \multicolumn{2}{c}{MX
data} & \multicolumn{2}{c}{Survey Cube data} & \multicolumn{1}{l}{Position}\\
\ (~~~l,~~~~~~~b~~~)    &       RA(2000)        &       Dec(2000)       &
$\rm V_{L}$&$\rm V_{H}$ &  $\rm V_{pk}$(MX)     &  $\rm S_{pk}$(MX)  &
$\rm V_{pk}$(SC) & $\rm S_{pk}$(SC)     &   Refs,
epoch    \\
\ (~~~$^\circ$~~~~~~~$^\circ$~~~) & (h~~m~~~s) & (~$^\circ$~~ '~~~~") &
\multicolumn{2}{c}{(km\,s$^{-1}$ )} & (km\,s$^{-1}$ ) &  (Jy) & (km\,s$^{-1}$ ) & (Jy) & \\
\hline
328.140$-$0.432&15 56 57.75& $-$53 57 48.7&$-$40.5&$-$36.0&$-$39.2&15.39&$-$39.2&4.20&2006mar30\\
328.164+0.587&15 52 42.37& $-$53 09 50.5&$-$92.5&$-$90.0&$-$91.9&1.55&$-$91.9&2.22&2006mar30\\
328.237$-$0.547&15 57 58.28& $-$53 59 22.7&$-$47.0&$-$31.5&$-$44.7&1340.24&$-$44.7&1481.00&CVF95 (W98)\\
328.254$-$0.532&15 57 59.75& $-$53 58 00.4&$-$51.0&$-$36.0&$-$37.5&360.81&$-$37.5&440.00&CVF95 (W98)\\
328.385+0.131&15 55 47.33& $-$53 22 30.8&28.0&30.0&29.0&1.60&29.0&1.88&2006mar30\\
328.808+0.633&15 55 48.45& $-$52 43 06.6&$-$47.5&$-$42.0&$-$44.0&300.00&$-$44.0&390.00&C97 (W98)\\
328.809+0.633&15 55 48.70& $-$52 43 05.5&$-$45.0&$-$44.0&$-$44.4&351.70&$-$44.4&425.00&C97\\
328.819+1.704&15 51 24.42& $-$51 53 05.4&$-$89.5&$-$71.0&$-$86.2&1.55&$-$85.3&2.41&2006mar30\\
328.940+0.558&15 56 47.43& $-$52 41 28.2&$-$99.0&$-$87.0&$-$98.8&0.80&$-$98.8&1.40&2006mar30\\
328.942+0.565&15 56 46.17& $-$52 41 04.8&$-$96.5&$-$88.5&$-$90.9&2.13&$-$90.9&3.27&2006mar30\\
329.029$-$0.205&16 00 31.80& $-$53 12 49.6&$-$41.5&$-$33.5&$-$37.1&118.00&$-$37.1&170.90&CVF95\\
329.031$-$0.198&16 00 30.32& $-$53 12 27.3&$-$48.5&$-$41.5&$-$45.8&26.00&$-$45.8&45.00&CVF95\\
329.066$-$0.308&16 01 09.93& $-$53 16 02.6&$-$49.0&$-$40.0&$-$43.8&14.41&$-$43.8&21.89&CVF95\\
329.183$-$0.314&16 01 47.01& $-$53 11 43.3&$-$60.0&$-$50.0&$-$55.6&5.14&$-$55.6&10.70&C2009\\
329.272+0.115&16 00 21.77& $-$52 48 48.1&$-$72.5&$-$71.0&$-$72.0&1.35&$-$72.0&1.99&2006mar30\\
329.339+0.148&16 00 33.13& $-$52 44 39.8&$-$108.0&$-$105.0&$-$106.4&23.84&$-$106.4&27.16&C2009 (W98)\\
329.341$-$0.644&16 04 01.70& $-$53 20 21.1&$-$82.0&$-$81.0&$-$81.4&3.04&$-$81.4&2.98&2006mar30\\
329.405$-$0.459&16 03 32.16& $-$53 09 30.5&$-$73.0&$-$63.5&$-$70.5&50.00&$-$70.5&60.00&C2009\\
329.407$-$0.459&16 03 32.65& $-$53 09 26.9&$-$73.5&$-$66.0&$-$66.8&95.70&$-$66.8&130.30&C2009 (W98)\\
329.469+0.503&15 59 40.71& $-$52 23 27.3&$-$74.5&$-$58.5&$-$72.0&18.24&$-$72.0&21.58&C2009 (W98)\\
329.526+0.216&16 01 11.70& $-$52 34 14.5&$-$94.0&$-$92.0&$-$92.8&1.81&$-$92.8&2.73&2006mar30\\
329.556+0.181&16 01 29.44& $-$52 34 39.9&$-$109.5&$-$104.5&$-$106.5&1.55&$-$109.0&1.34&2006mar30\\
329.610+0.114&16 02 03.14& $-$52 35 33.5&$-$69.0&$-$58.5&$-$60.1&44.31&$-$60.1&49.85&C2009\\
329.622+0.138&16 02 00.30& $-$52 33 59.4&$-$86.0&$-$83.0&$-$85.0&1.98&$-$85.0&1.63&C2009 (2006apr02)\\
329.719+1.164&15 58 07.09& $-$51 43 32.6&$-$83.0&$-$72.5&$-$75.8&7.69&$-$75.8&24.44&2006mar30\\
\hline
\end{tabular} 
\label{resotable}
\end{table*}

\subparagraph{294.511--1.621} This source has had a history of variability. After being discovered in 1992 with a dominant peak of 18\,Jy at $-$10.2\,km\,s$^{-1}$ \citep{caswell95a}, it was then observed in 1995 with this feature reduced to 12\,Jy and the secondary feature at $-$11.9\,km\,s$^{-1}$, originally 7\,Jy, brightening to a comparable peak flux density of 12\,Jy \citep{caswell97}. In our survey cube observations (2006 August and 2007 January) the feature at $-$10.2\,km\,s$^{-1}$ had diminished further to $<$5\,Jy and the feature at $-$11.9\,km\,s$^{-1}$ had diminished slightly to  9\,Jy. An MX observation in 2008 March found these two features further diminished, but other features within the spectrum brightened including a feature at $-$8.7\,km\,s$^{-1}$ which doubled in brightness to 4\,Jy. In a later MX observation, in 2009 March, the feature at $-$8.7\,km\,s$^{-1}$ had further brightened to 5.5\,Jy.

\subparagraph{294.977--1.734} This new source was initially detected in the survey cube observations offset from its true position, with two spectral peaks closely spaced in velocity with comparable flux densities of $\sim$1.7 and $\sim$1.9 Jy. The peak flux density was 3.5\,Jy in the 2007 ATCA positioning observations, but then faded below 0.1\,Jy for the 2009 March MX observation. As such the survey cube spectrum is presented in Figure\,\ref{spectra}. The feature at $-$12.3\,km\,s$^{-1}$ arises from 294.990--1.719 (offset 71 arcsec), as is clear from our alignment of the spectra.

\subparagraph{298.177--0.795 and 298.213--0.343} Two masers (the first a new source) associated with H{\sc ii} regions attributed by \citet{rahman11} to the `Dragonfish Nebula', which, they claim is powered by the most luminous OB association in the Galaxy, at a distance of $\sim$9.7\,kpc. This distance is comparable to the kinematic distance for 298.213--0.343 which is $\sim$10\,kpc, However, \citet{moises11} allocate a distance of 4.7\,kpc based on photometric techniques. The second source has demonstrated variability, switching between two velocities for the brightest feature. 

\subparagraph{298.262+0.739} This source was observed in 2005 by J. L. Caswell, but not included in the 2009 compilation \citep{caswell09a}, as the position could not be confirmed. We have now been able to successfully position the source. This source was allocated to the far kinematic distance by \citet{green11b}.

\subparagraph{298.723--0.086} The weak feature at +15 \,km\,s$^{-1}$ of this new source has been positioned with the ATCA to be at the same location as the main features.

\subparagraph{300.504--0.176} This source has shown a new feature near 2.5\,km\,s$^{-1}$, expanding the velocity spread from the previous observations. This source was allocated to the near kinematic distance by \citet{green11b}.

\subparagraph{300.969+1.148} This bright source was used in the survey for system verification, observed frequently at the start of an observing run, and its past record of stability was corroborated (with several MX spectra from the survey shown in \citet{green09a}).

\subparagraph{301.136--0.226} This known source, first detected by \citet{caswell95a}, and positioned with the ATCA by \citet{caswell09a}, was allocated to the far kinematic distance by \citet{green11b}.

\begin{figure*}
 \begin{center}
 \renewcommand{\baselinestretch}{1.1}
\includegraphics[width=16.5cm]{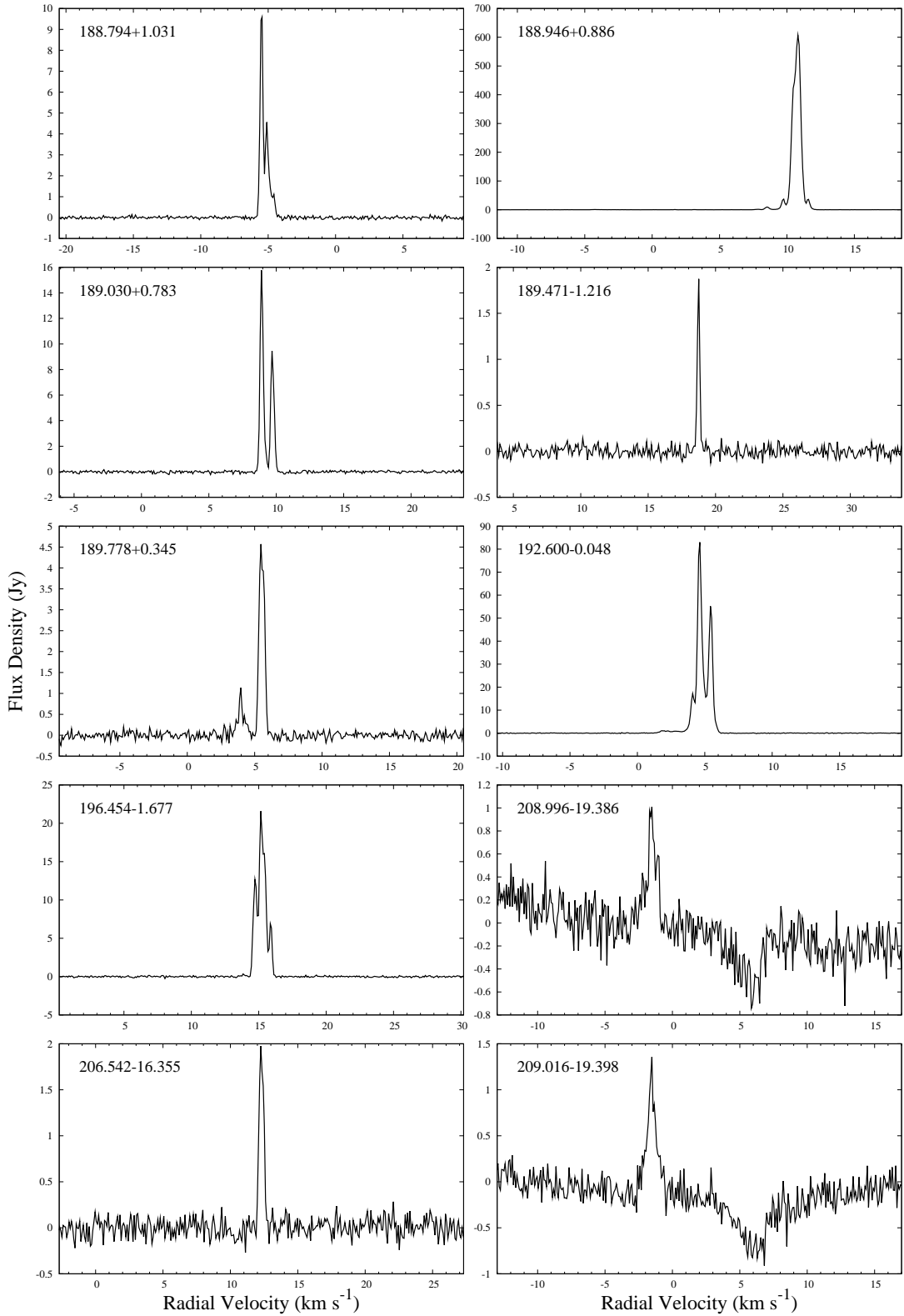}
\caption{\small Spectra of 6668-MHz methanol masers. Asterisk (*) denotes survey cube spectrum.}
\label{spectra}
\end{center}
\end{figure*}

\begin{figure*}
\addtocounter{figure}{-1}
 \begin{center}
 \renewcommand{\baselinestretch}{1.1}
\includegraphics[width=16.5cm]{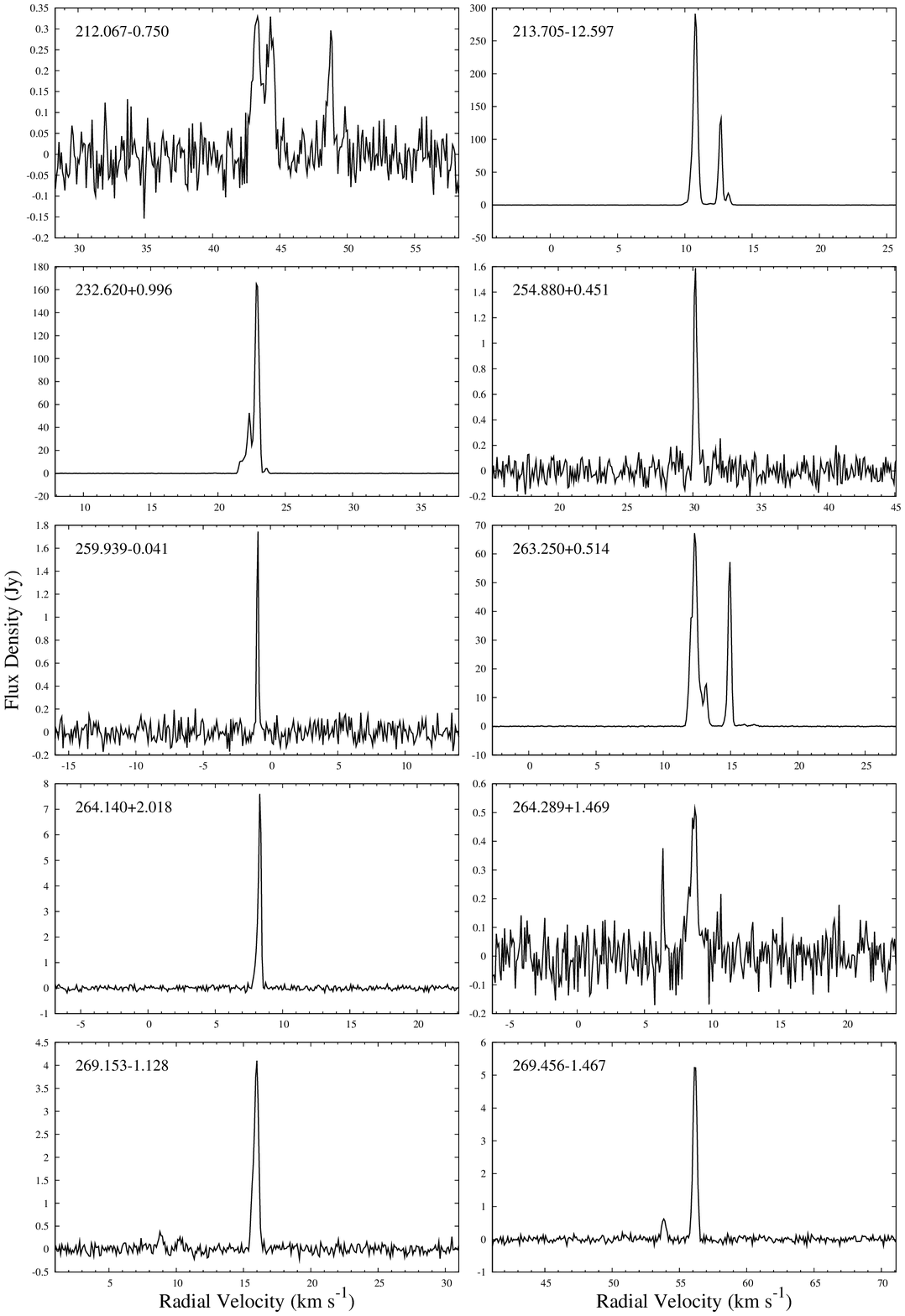}
\caption{\small -- {\it continued}}
\label{spectra}
\end{center}
\end{figure*}

\begin{figure*}
\addtocounter{figure}{-1}
 \begin{center}
 \renewcommand{\baselinestretch}{1.1}
\includegraphics[width=16.5cm]{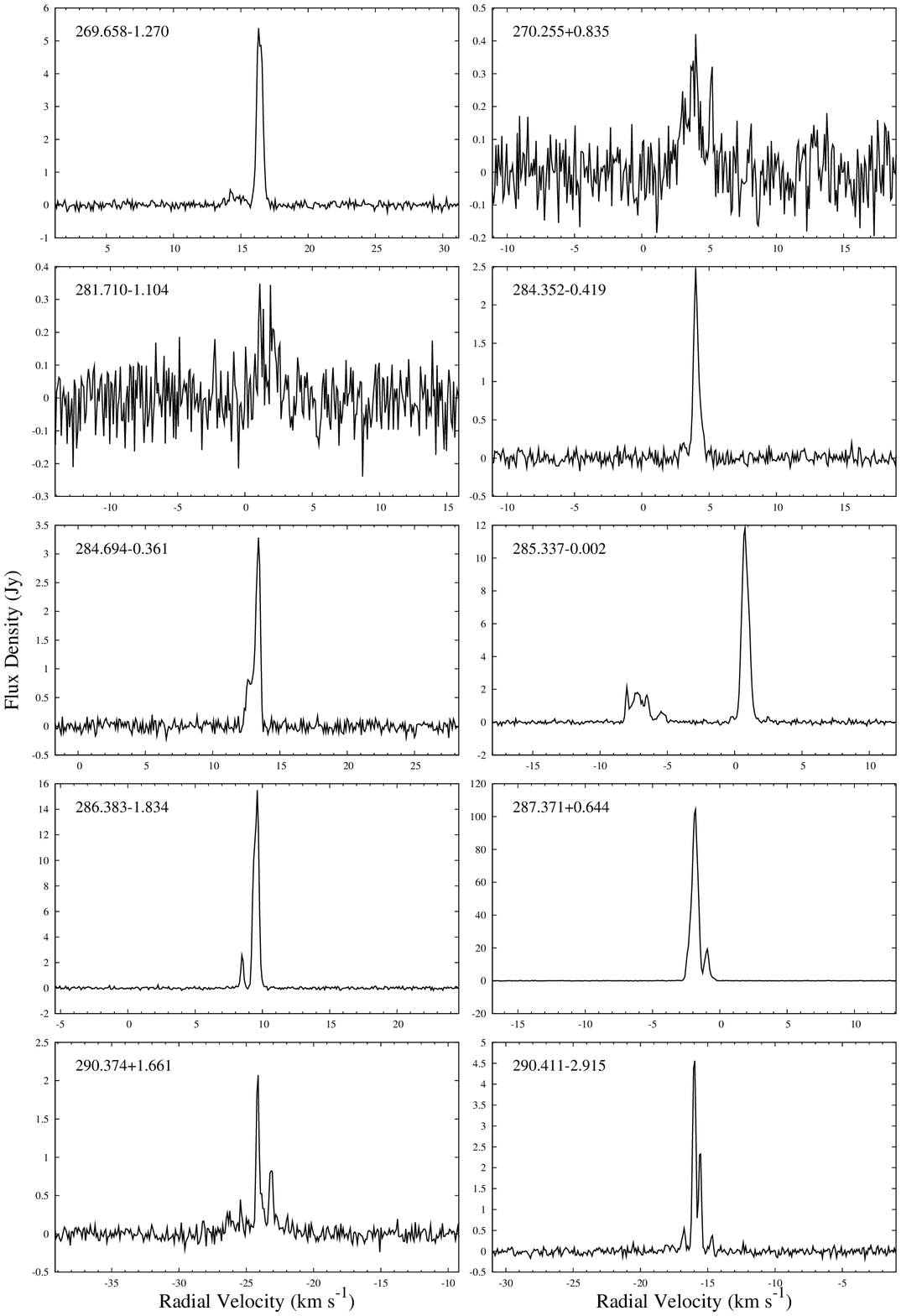}
\caption{\small -- {\it continued}}
\label{spectra}
\end{center}
\end{figure*}

\begin{figure*}
\addtocounter{figure}{-1}
 \begin{center}
 \renewcommand{\baselinestretch}{1.1}
\includegraphics[width=16.5cm]{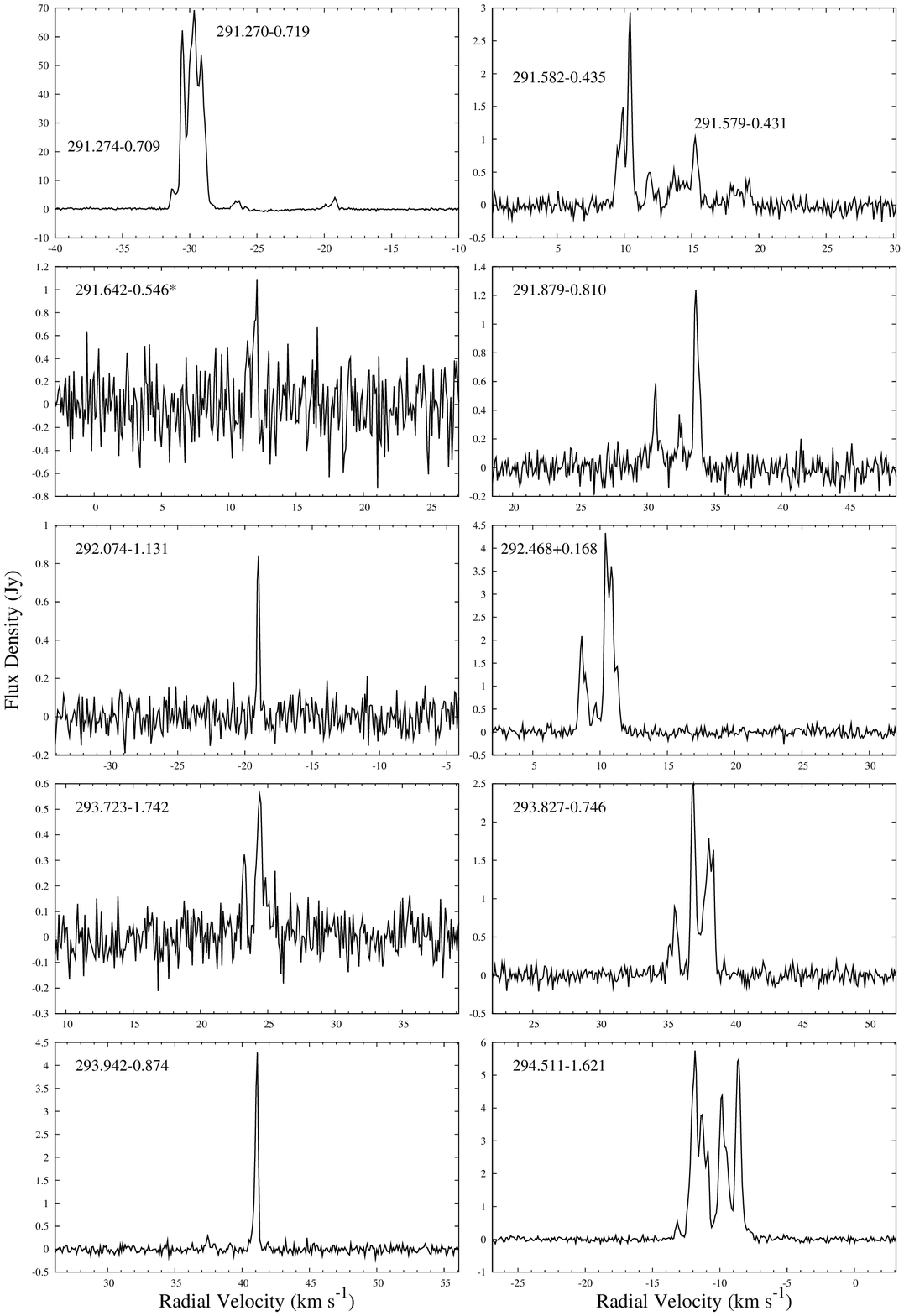}
\caption{\small }
\label{spectra}
\end{center}
\end{figure*}

\begin{figure*}
\addtocounter{figure}{-1}
 \begin{center}
 \renewcommand{\baselinestretch}{1.1}
\includegraphics[width=16.5cm]{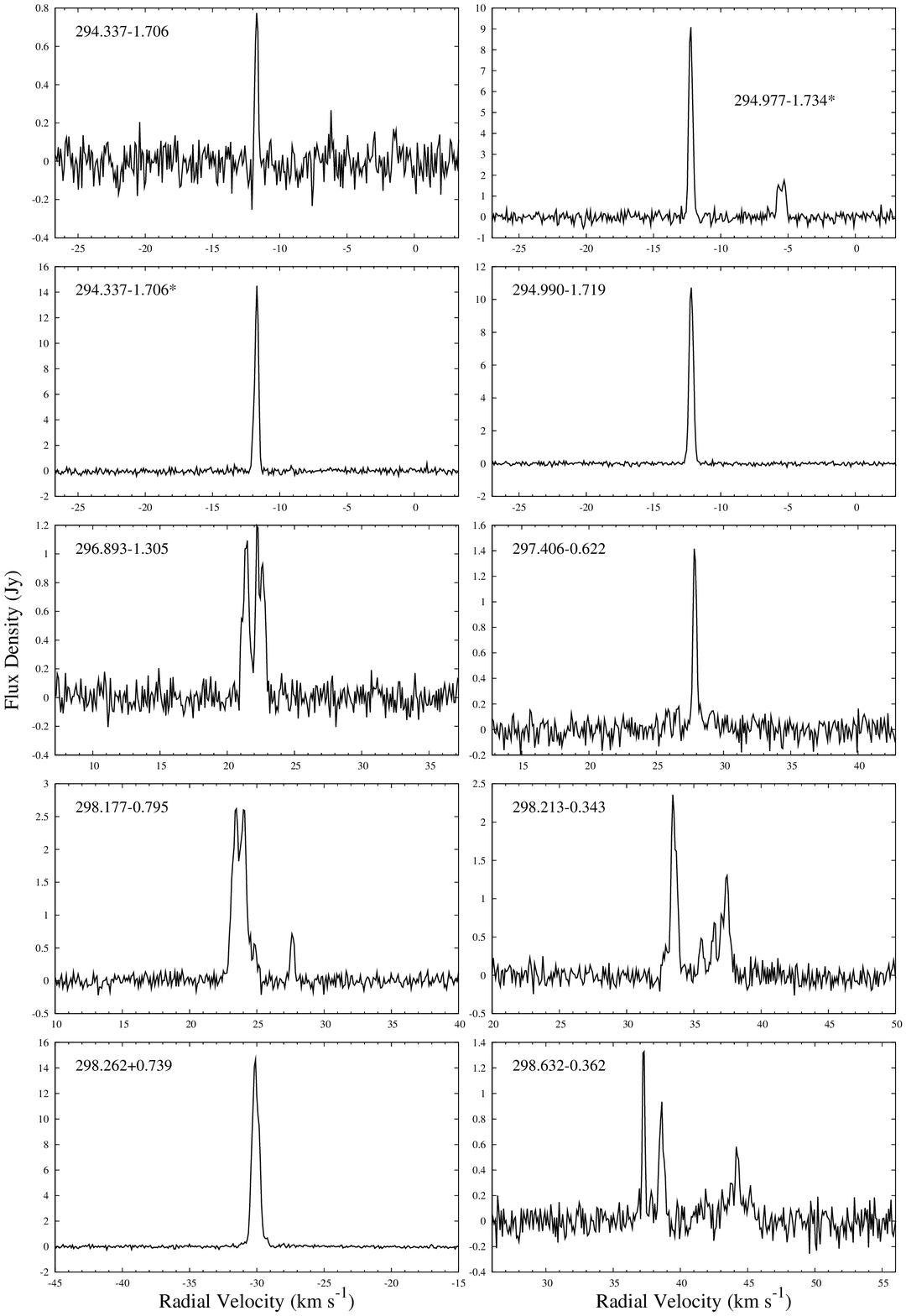}
\caption{\small -- {\it continued}}
\label{spectra}
\end{center}
\end{figure*}

\clearpage

\begin{figure*}
\addtocounter{figure}{-1}
 \begin{center}
 \renewcommand{\baselinestretch}{1.1}
\includegraphics[width=16.5cm]{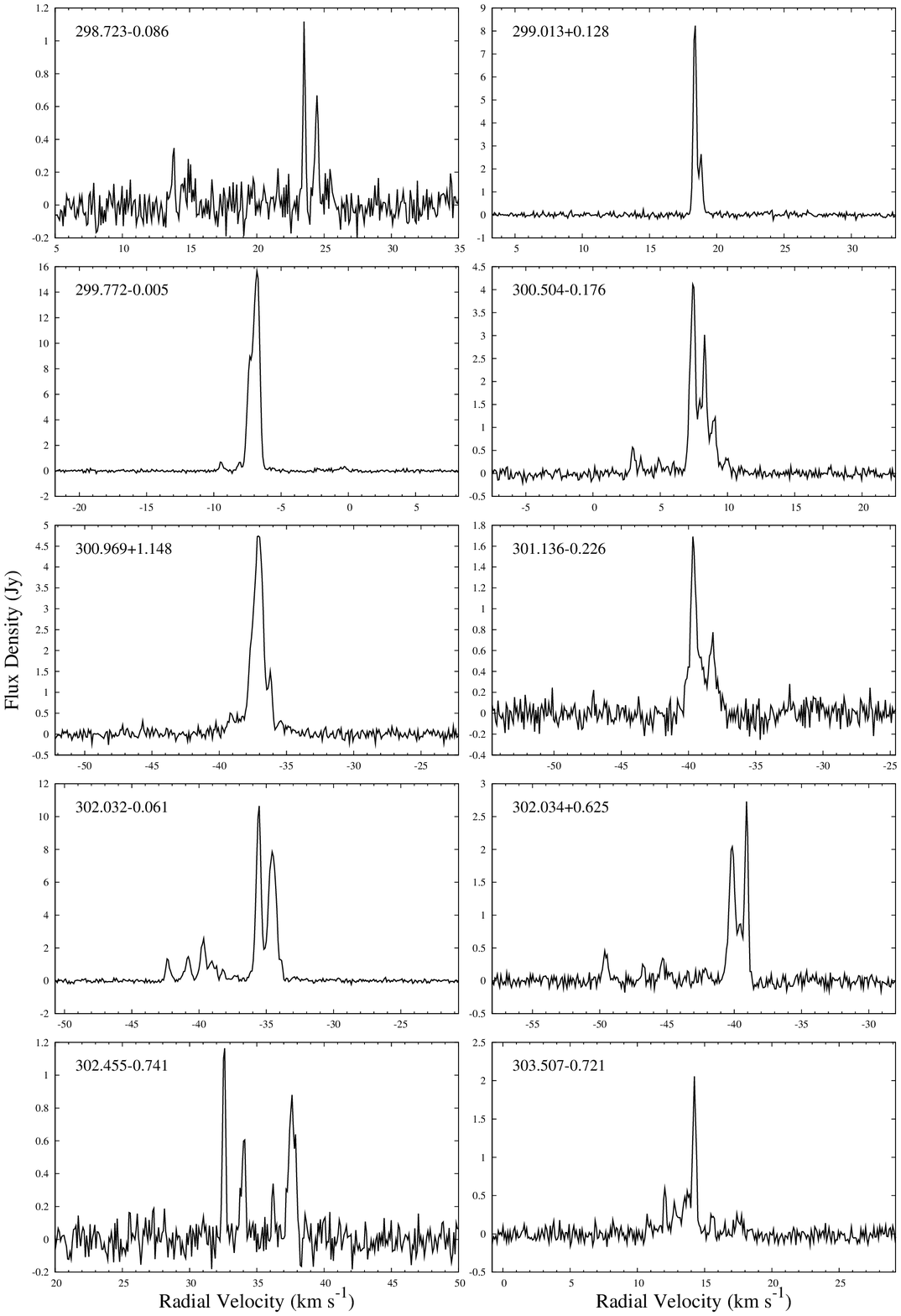}
\caption{\small -- {\it continued}}
\label{spectra}
\end{center}
\end{figure*}

\begin{figure*}
\addtocounter{figure}{-1}
 \begin{center}
 \renewcommand{\baselinestretch}{1.1}
\includegraphics[width=16.5cm]{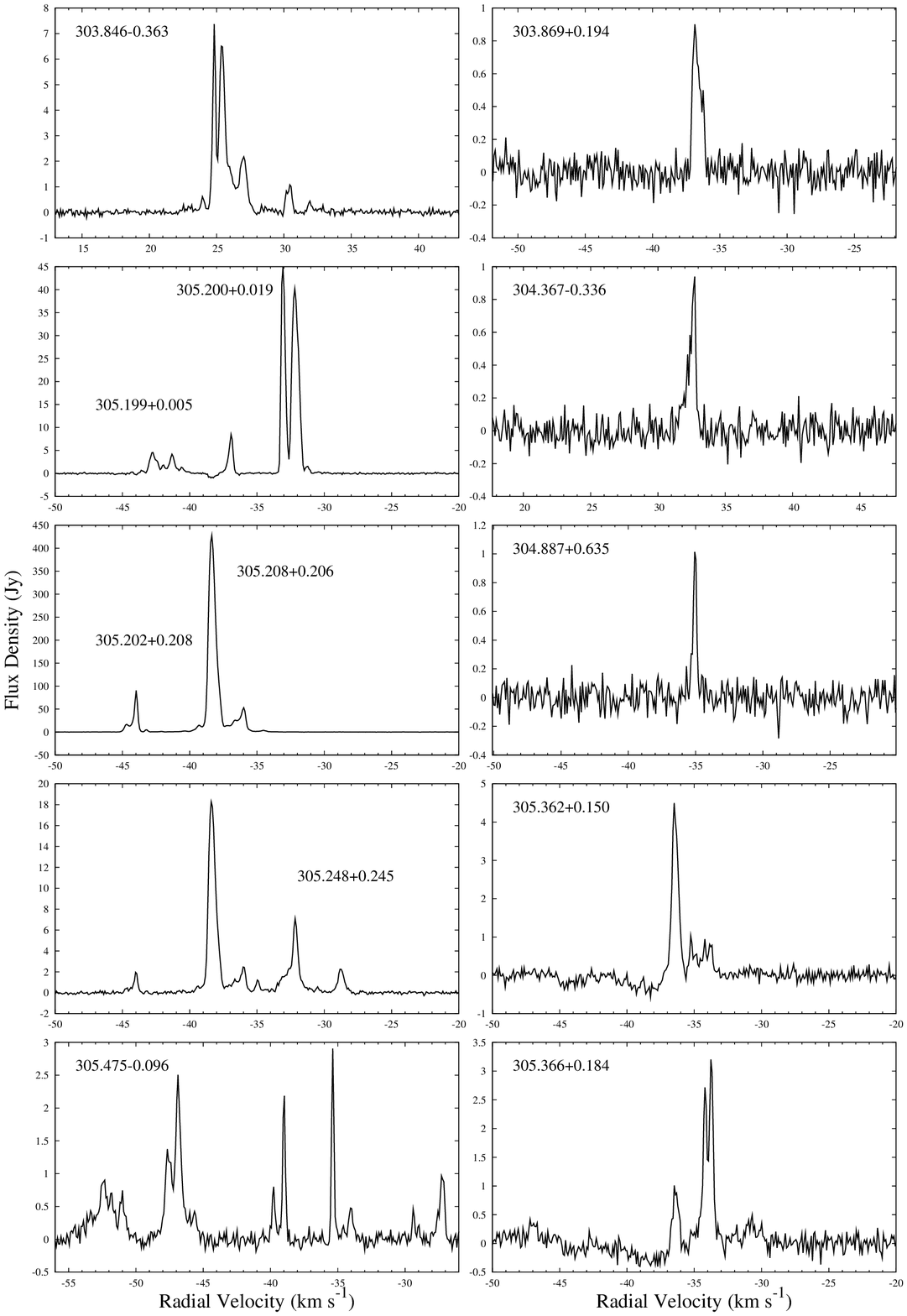}
\caption{\small -- {\it continued}}
\label{spectra}
\end{center}
\end{figure*}

\begin{figure*}
\addtocounter{figure}{-1}
 \begin{center}
 \renewcommand{\baselinestretch}{1.1}
\includegraphics[width=16.5cm]{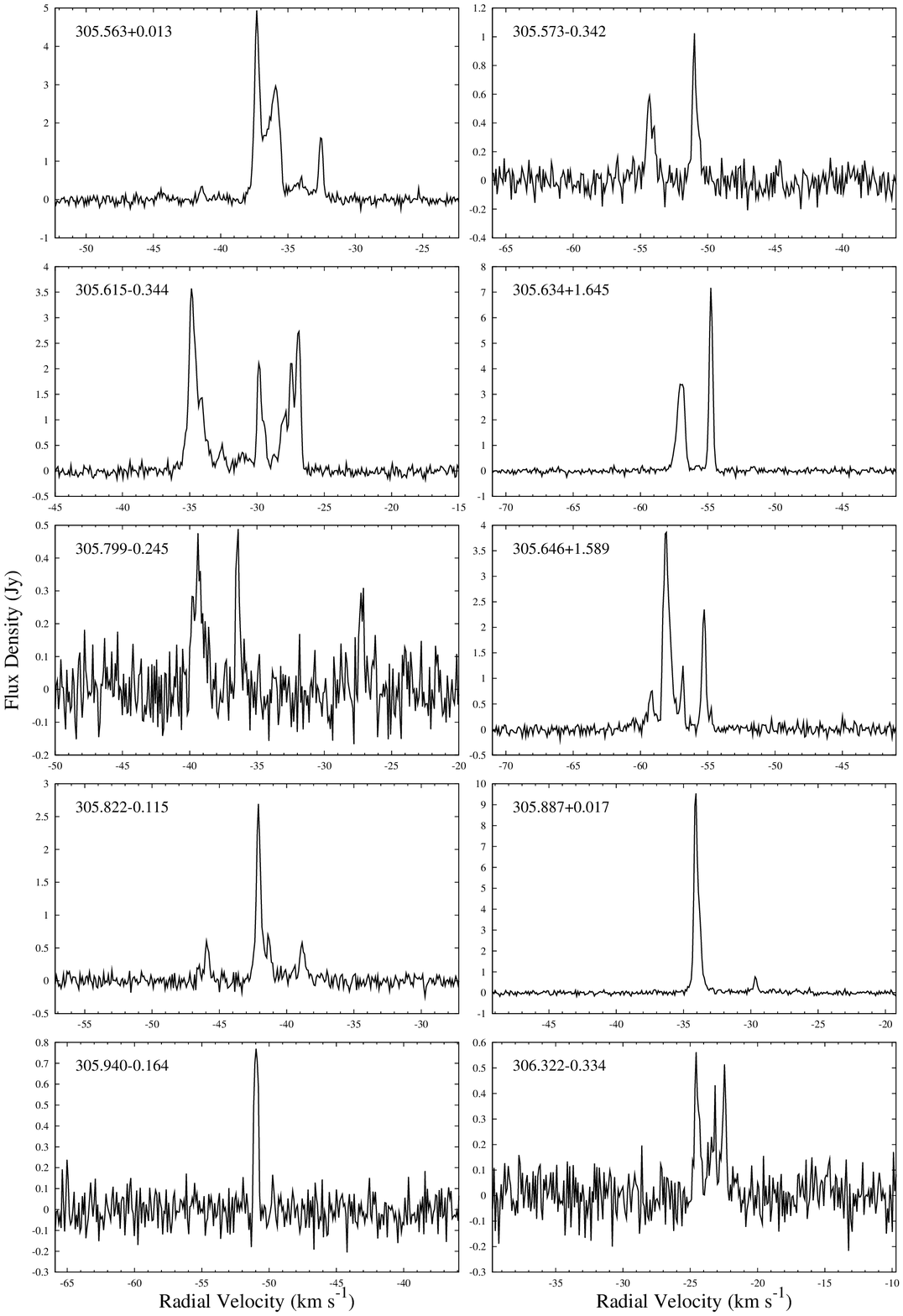}
\caption{\small -- {\it continued}}
\label{spectra}
\end{center}
\end{figure*}

\begin{figure*}
\addtocounter{figure}{-1}
 \begin{center}
 \renewcommand{\baselinestretch}{1.1}
\includegraphics[width=16.5cm]{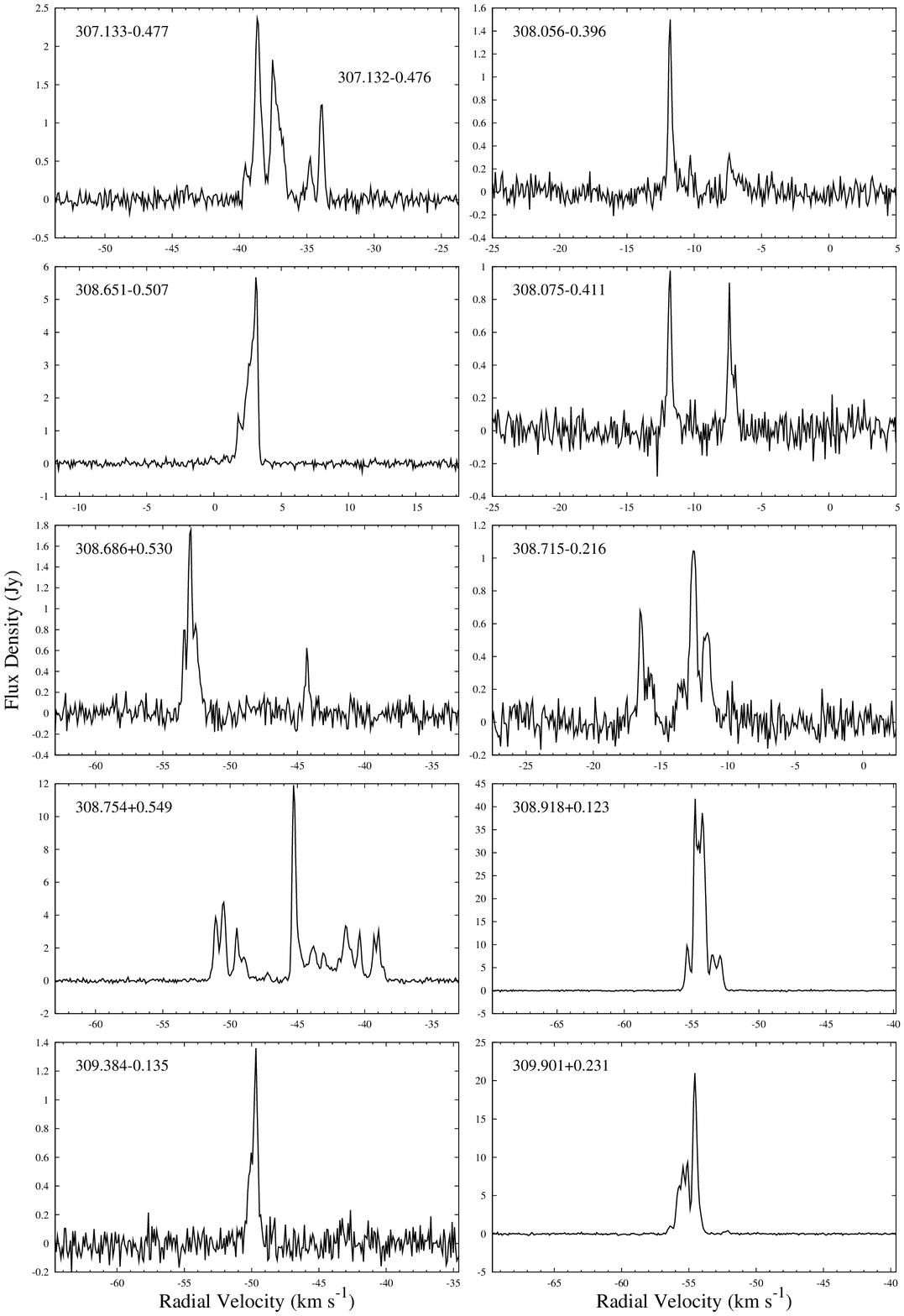}
\caption{\small -- {\it continued}}
\label{spectra}
\end{center}
\end{figure*}

\begin{figure*}
\addtocounter{figure}{-1}
 \begin{center}
 \renewcommand{\baselinestretch}{1.1}
\includegraphics[width=16.5cm]{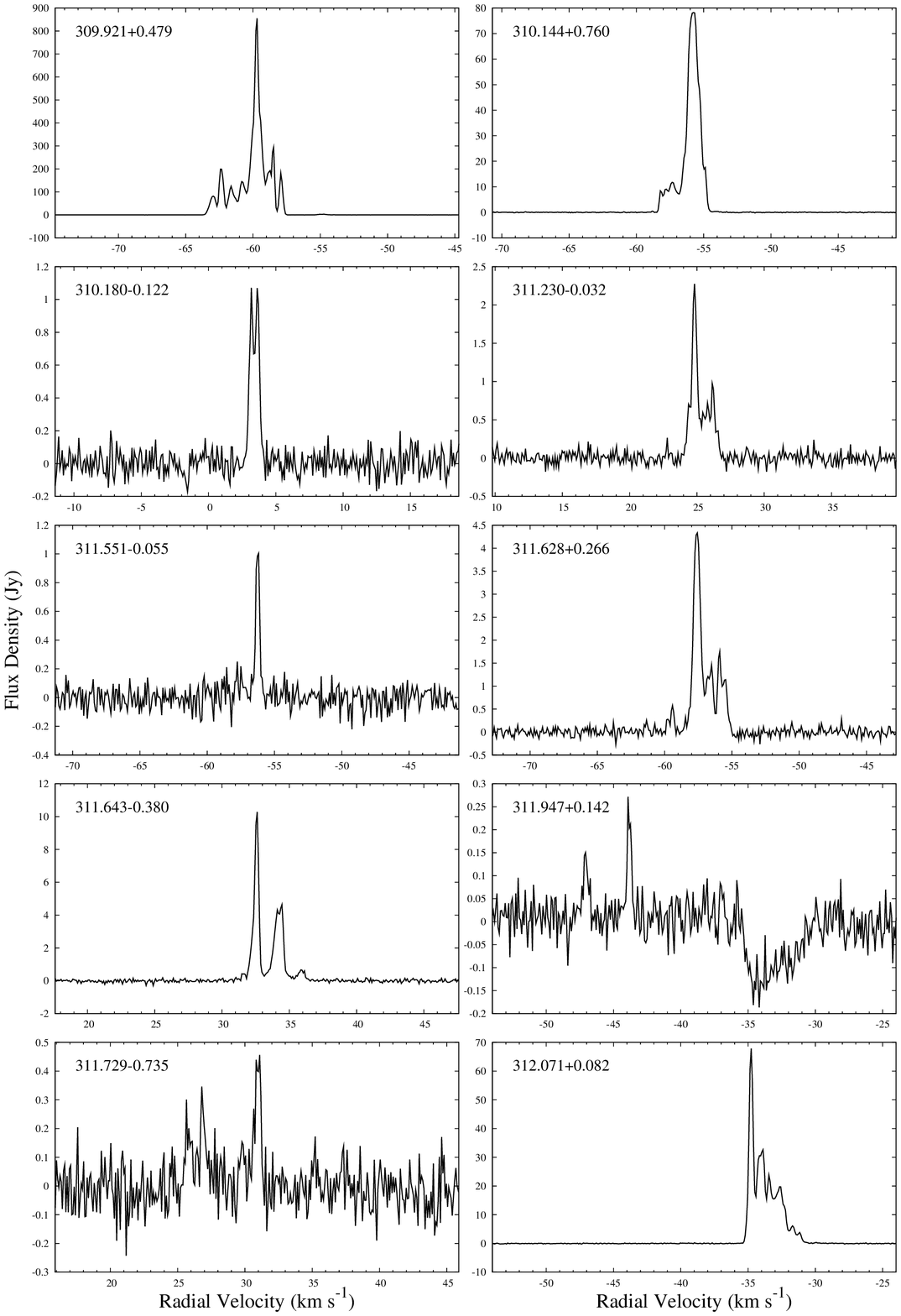}
\caption{\small -- {\it continued}}
\label{spectra}
\end{center}
\end{figure*}

\begin{figure*}
\addtocounter{figure}{-1}
 \begin{center}
 \renewcommand{\baselinestretch}{1.1}
\includegraphics[width=16.5cm]{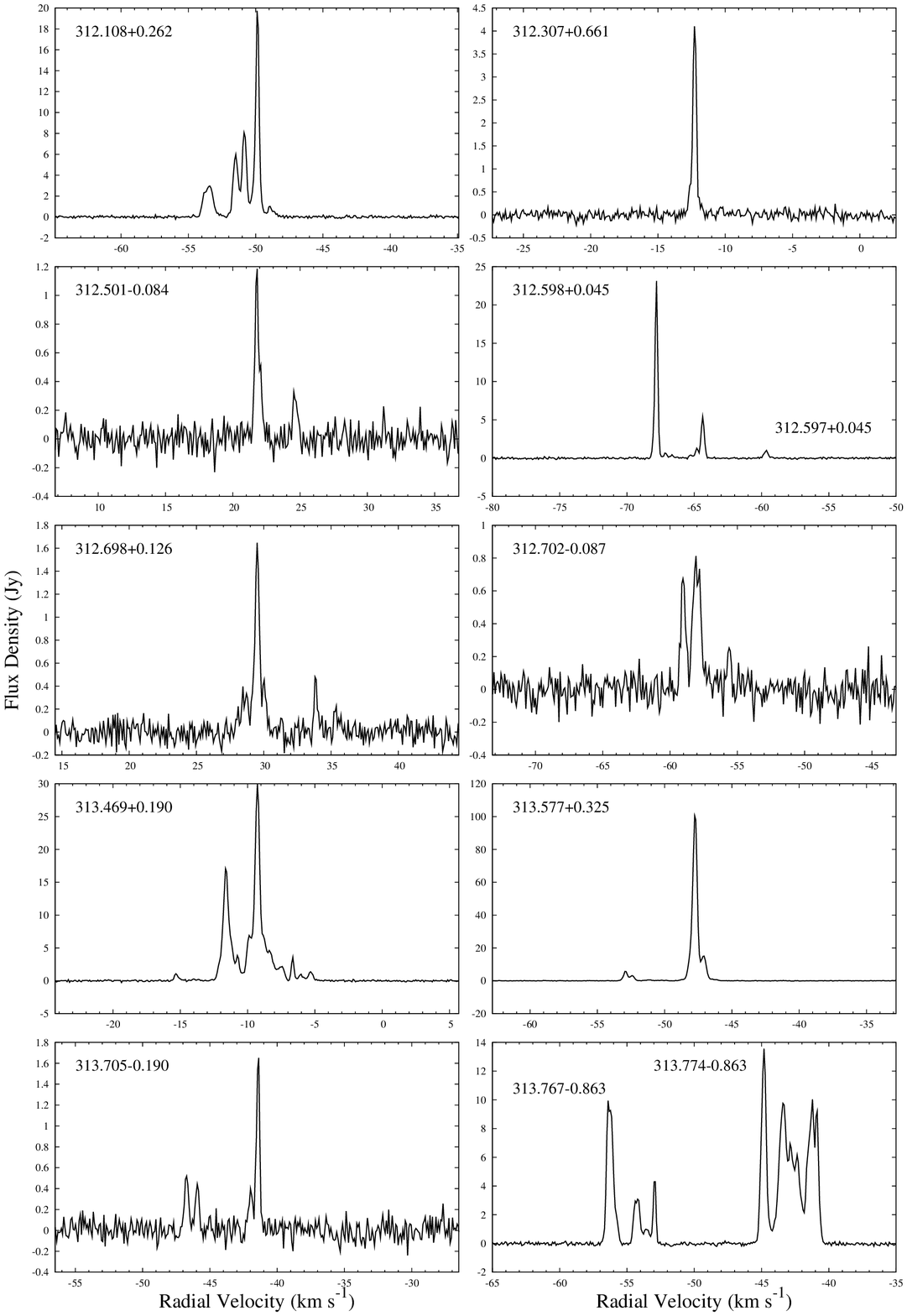}
\caption{\small -- {\it continued}}
\label{spectra}
\end{center}
\end{figure*}

\begin{figure*}
\addtocounter{figure}{-1}
 \begin{center}
 \renewcommand{\baselinestretch}{1.1}
\includegraphics[width=16.5cm]{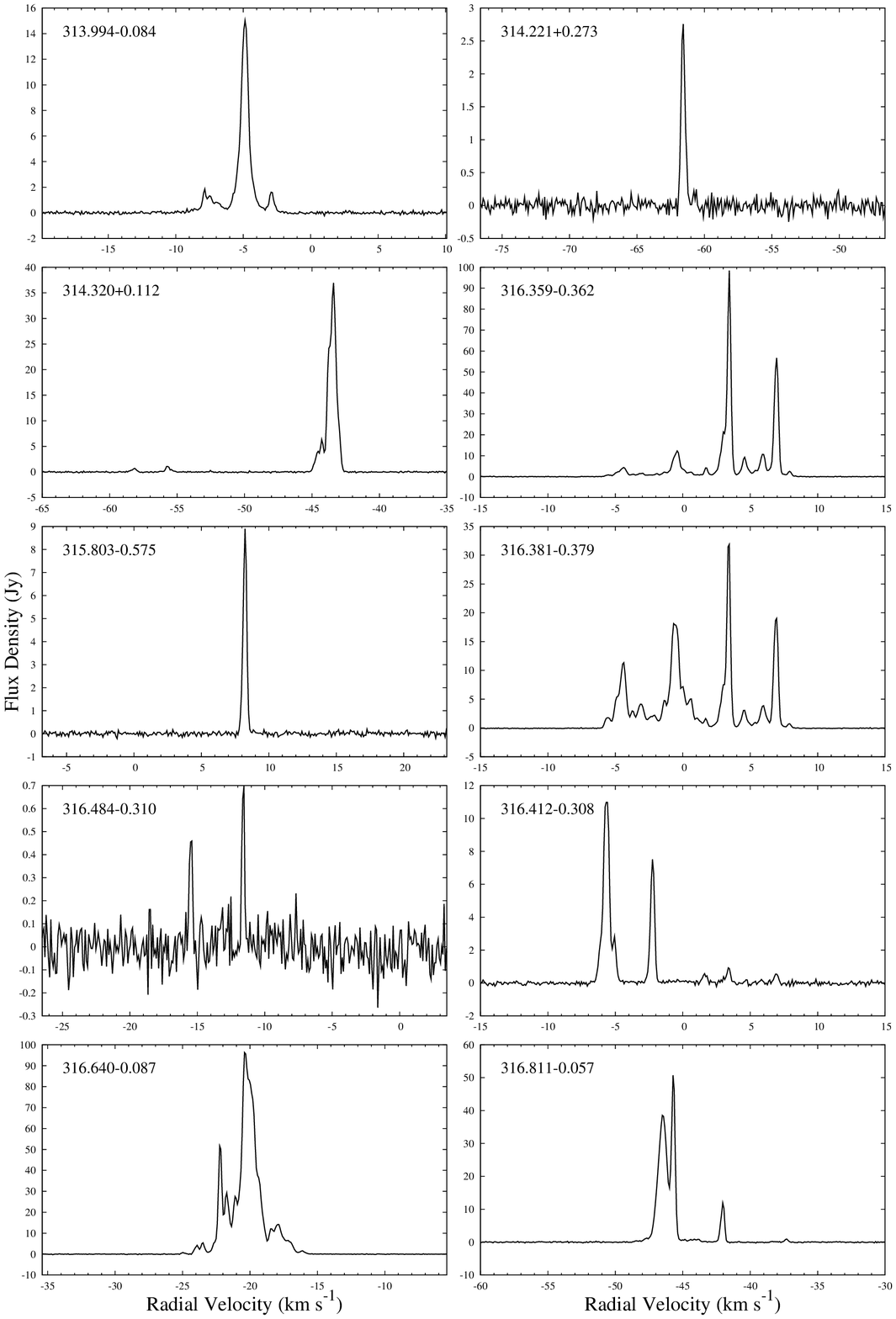}
\caption{\small -- {\it continued}}
\label{spectra}
\end{center}
\end{figure*}

\begin{figure*}
\addtocounter{figure}{-1}
 \begin{center}
 \renewcommand{\baselinestretch}{1.1}
\includegraphics[width=16.5cm]{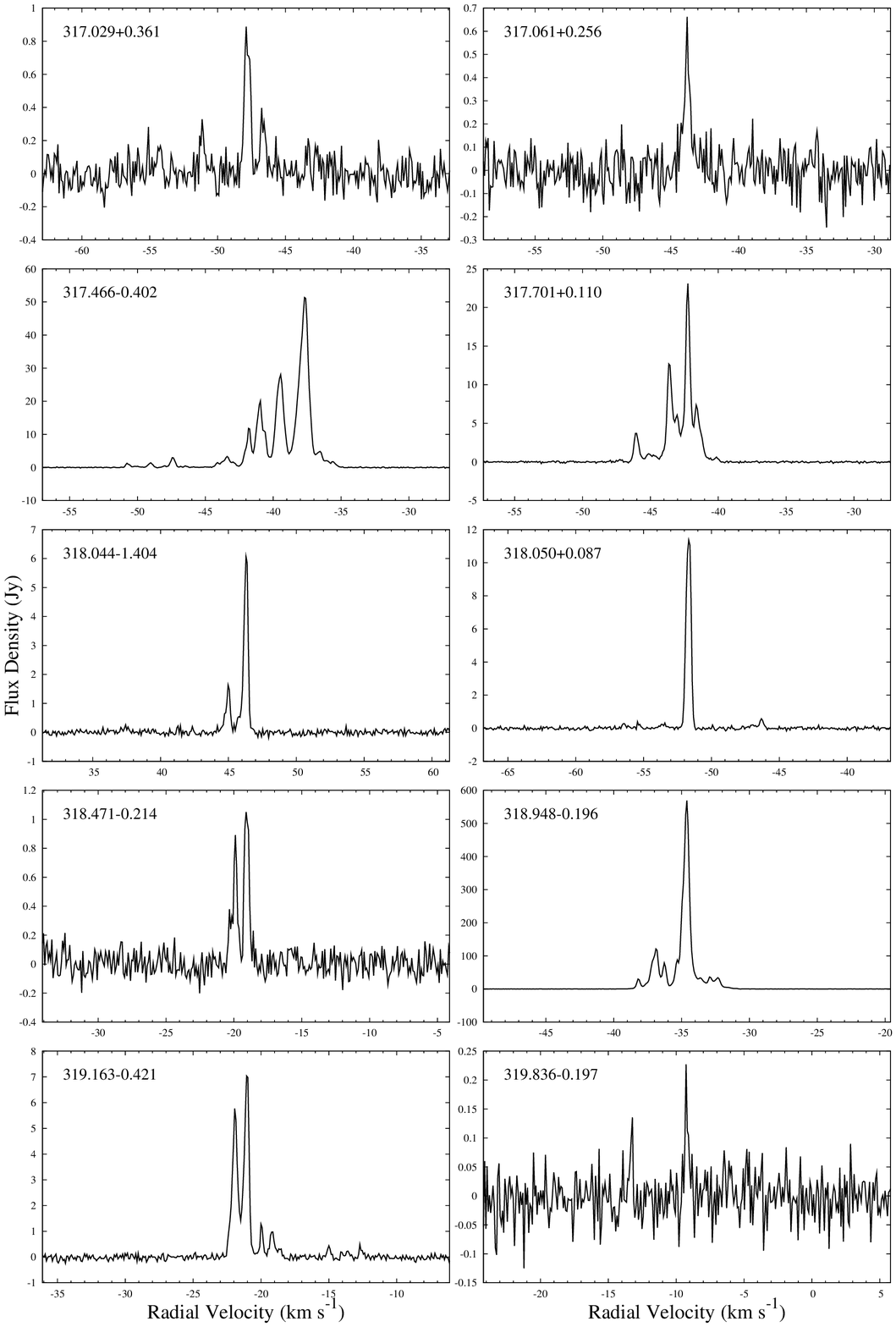}
\caption{\small -- {\it continued}}
\label{spectra}
\end{center}
\end{figure*}

\begin{figure*}
\addtocounter{figure}{-1}
 \begin{center}
 \renewcommand{\baselinestretch}{1.1}
\includegraphics[width=16.5cm]{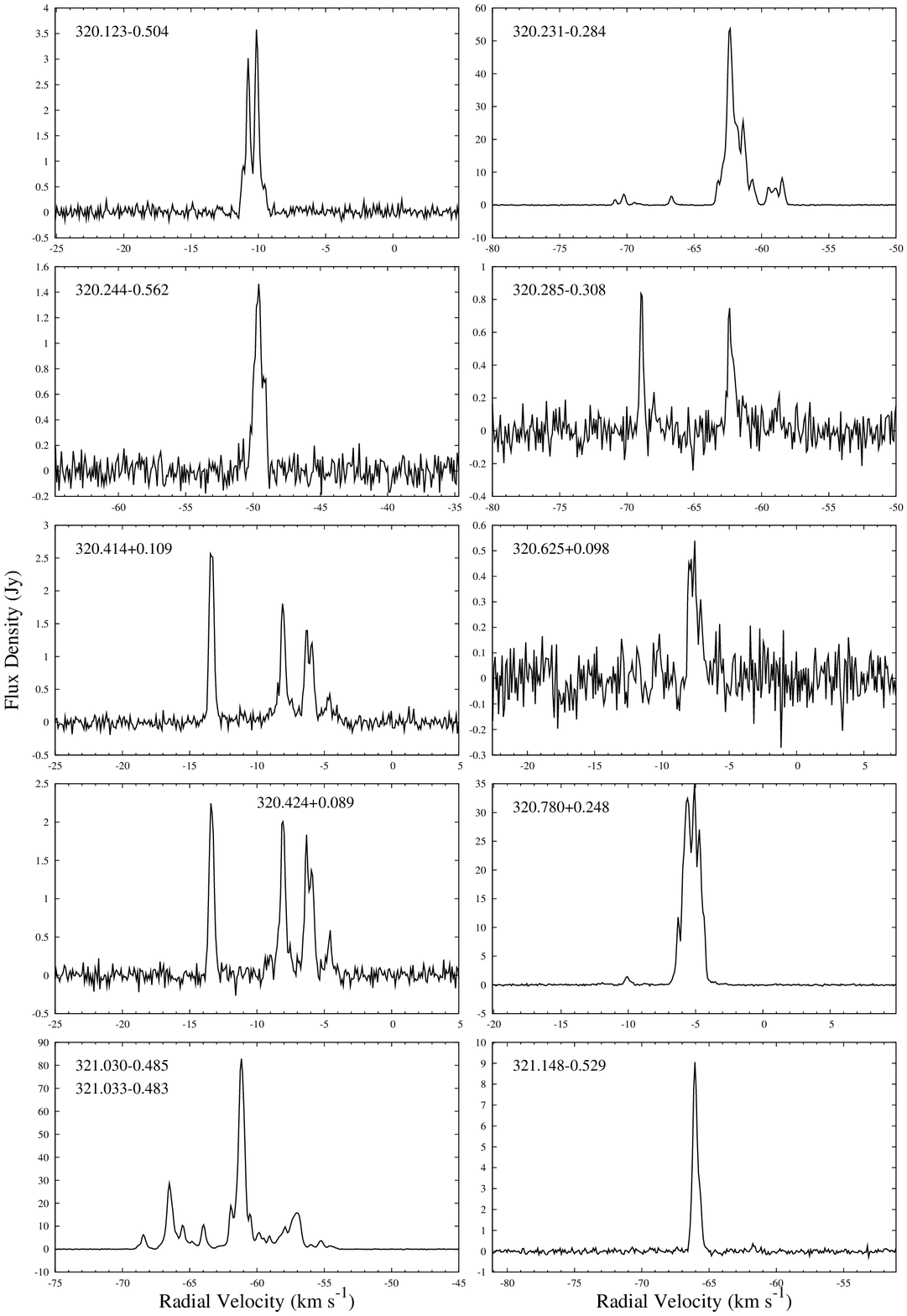}
\caption{\small -- {\it continued}}
\label{spectra}
\end{center}
\end{figure*}

\begin{figure*}
\addtocounter{figure}{-1}
 \begin{center}
 \renewcommand{\baselinestretch}{1.1}
\includegraphics[width=16.5cm]{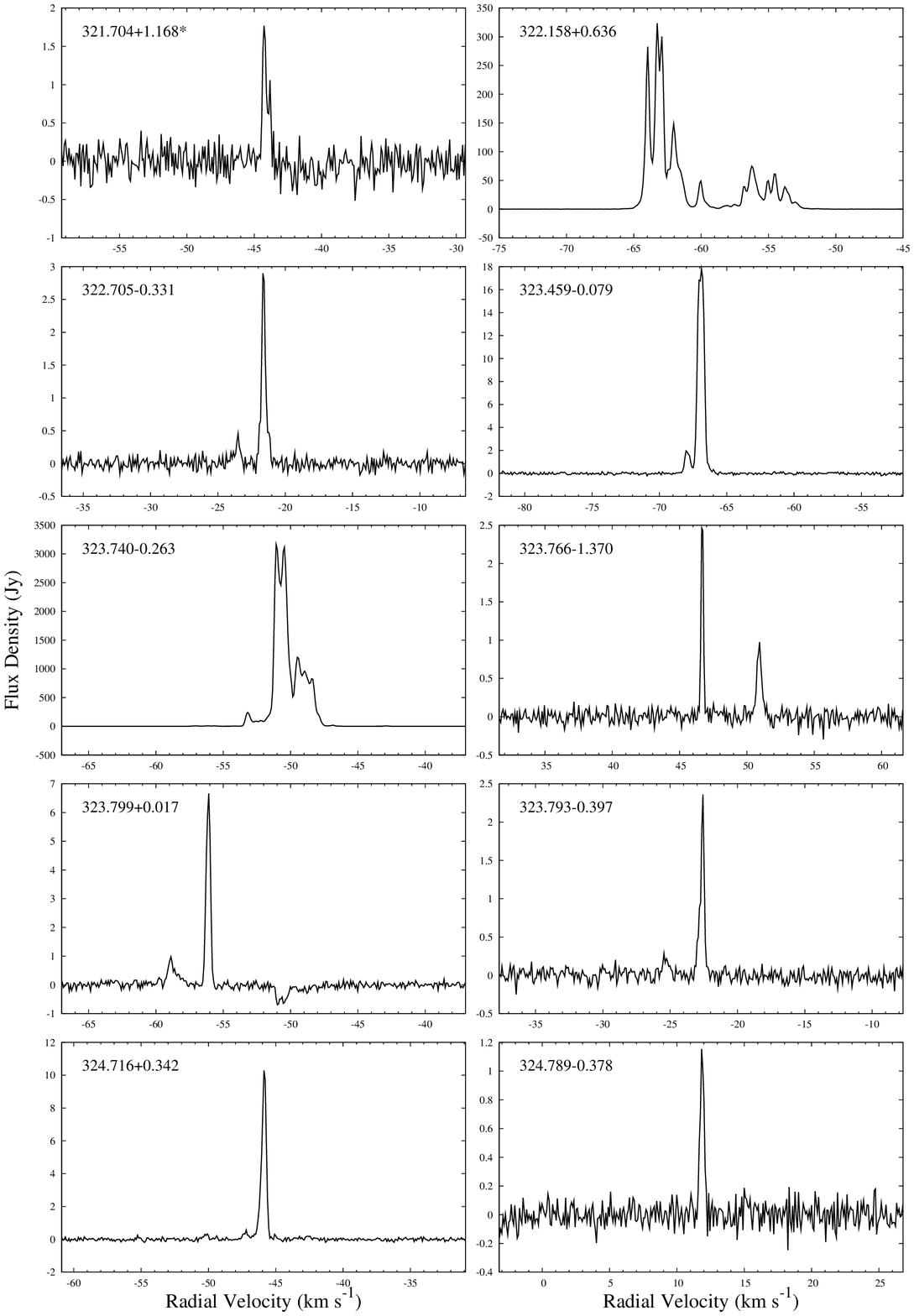}
\caption{\small -- {\it continued}}
\label{spectra}
\end{center}
\end{figure*}

\clearpage

\begin{figure*}
\addtocounter{figure}{-1}
 \begin{center}
 \renewcommand{\baselinestretch}{1.1}
\includegraphics[width=16.5cm]{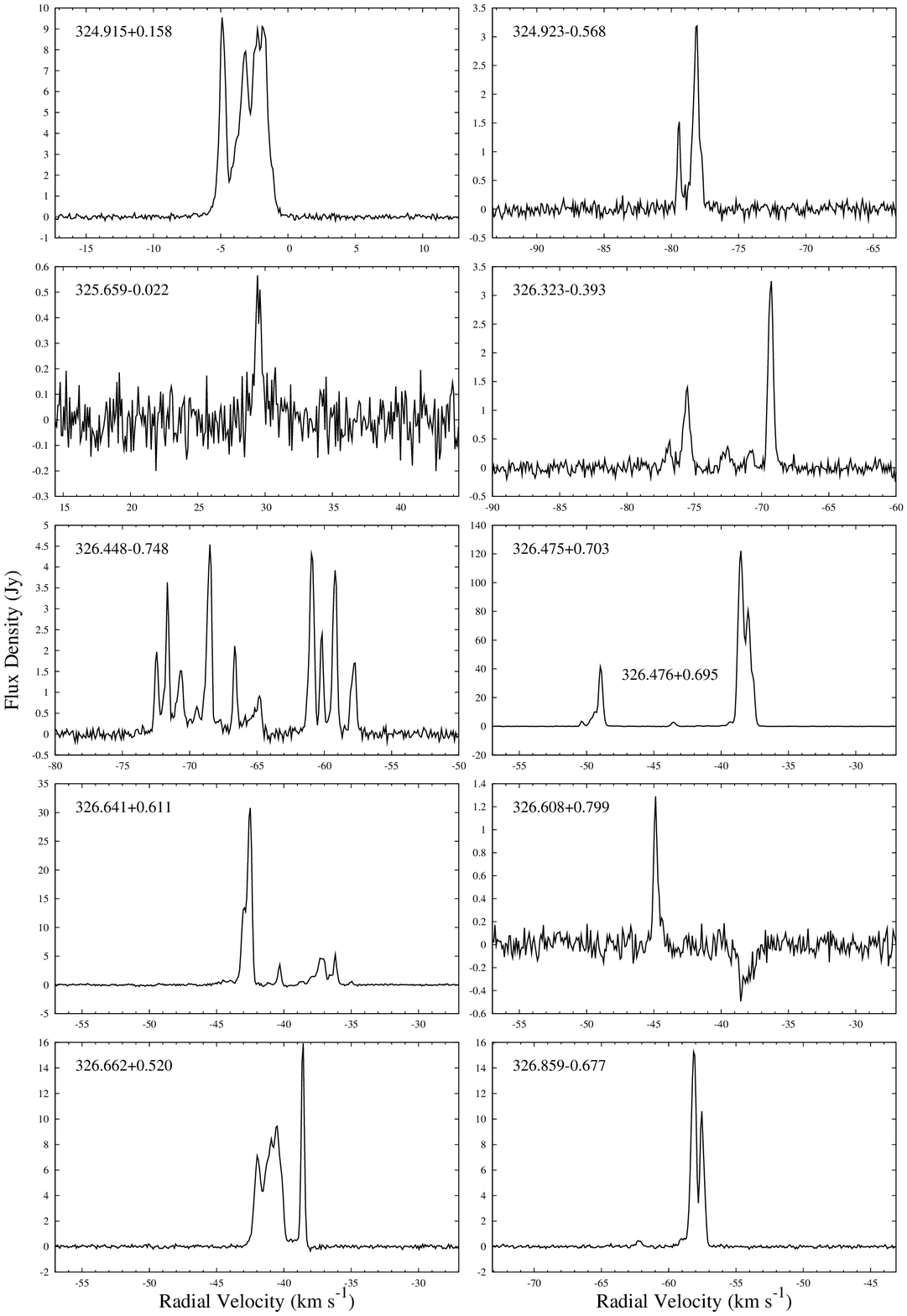}
\caption{\small -- {\it continued}}
\label{spectra}
\end{center}
\end{figure*}

\begin{figure*}
\addtocounter{figure}{-1}
 \begin{center}
 \renewcommand{\baselinestretch}{1.1}
\includegraphics[width=16.5cm]{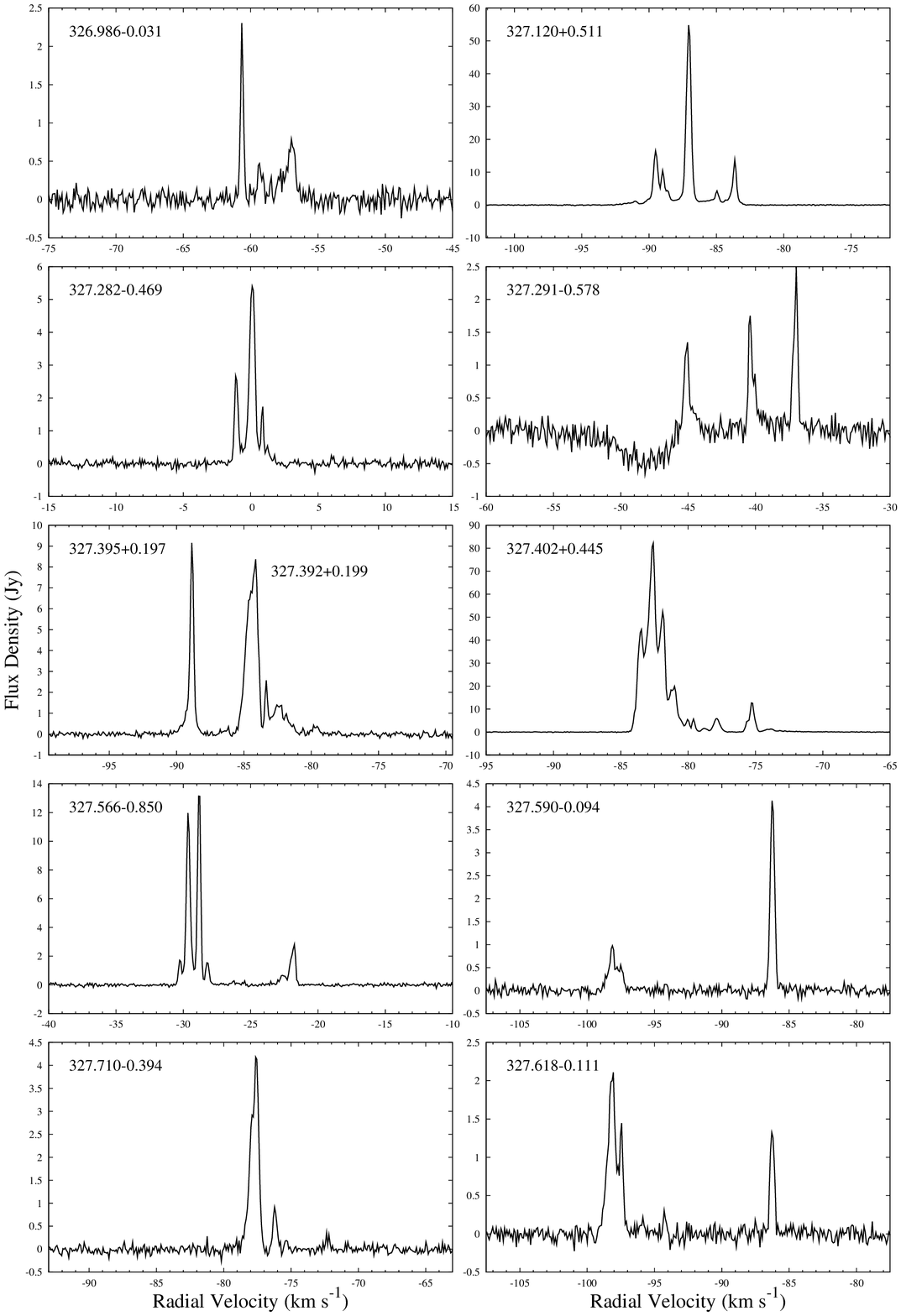}
\caption{\small -- {\it continued}}
\label{spectra}
\end{center}
\end{figure*}

\begin{figure*}
\addtocounter{figure}{-1}
 \begin{center}
 \renewcommand{\baselinestretch}{1.1}
\includegraphics[width=16.5cm]{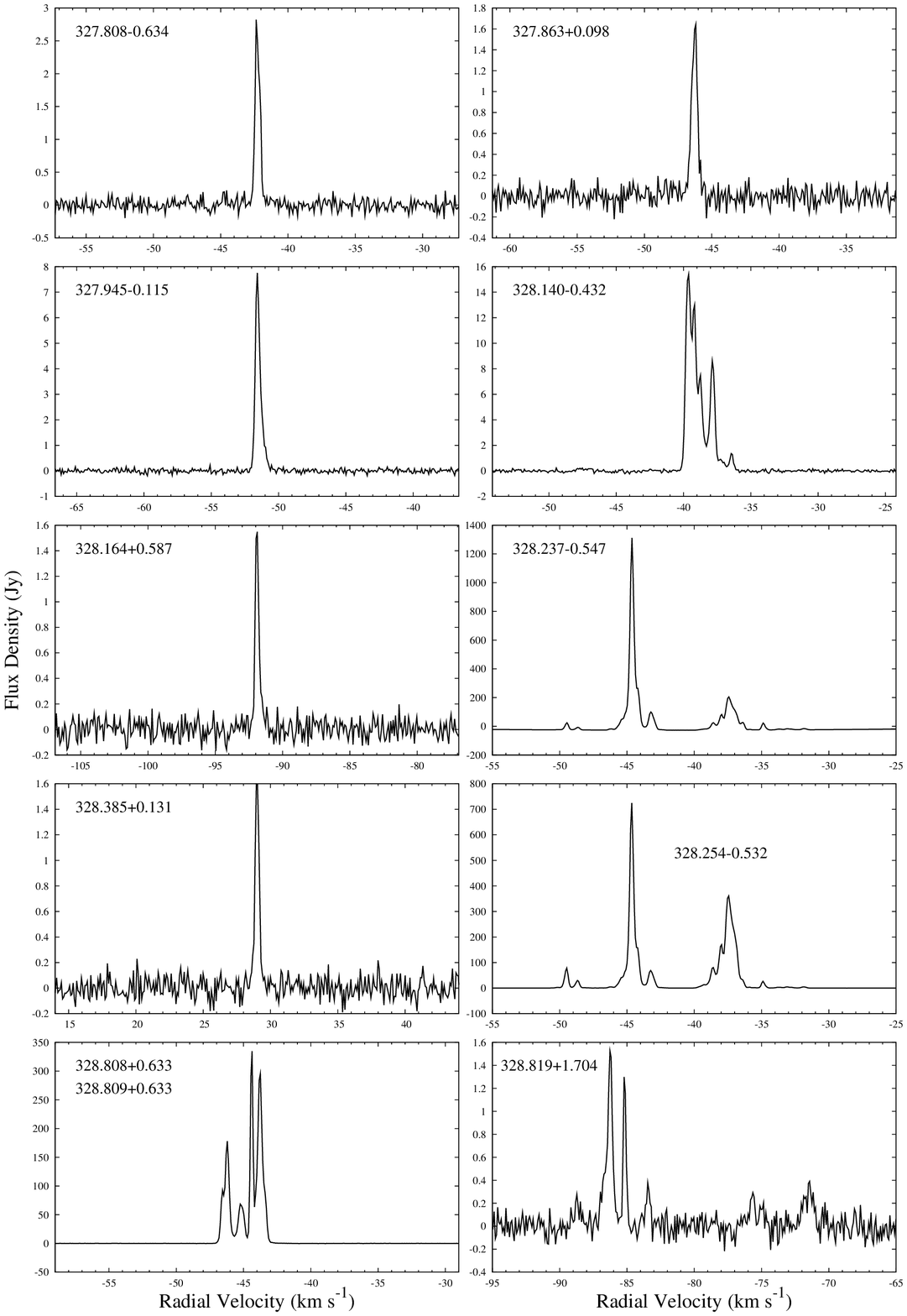}
\caption{\small -- {\it continued}}
\label{spectra}
\end{center}
\end{figure*}

\begin{figure*}
\addtocounter{figure}{-1}
 \begin{center}
 \renewcommand{\baselinestretch}{1.1}
\includegraphics[width=16.5cm]{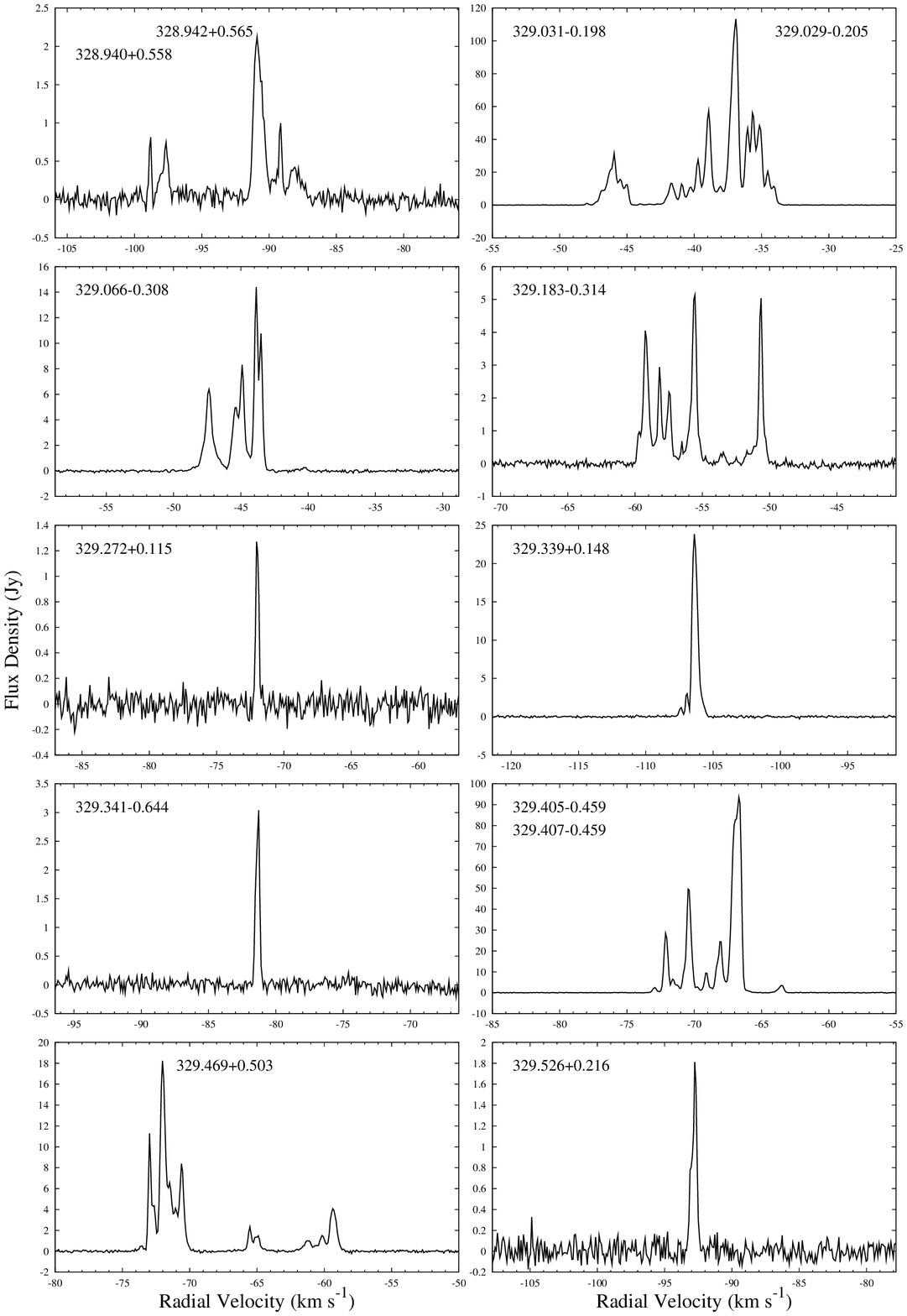}
\caption{\small -- {\it continued}}
\label{spectra}
\end{center}
\end{figure*}

\begin{figure*}
\addtocounter{figure}{-1}
 \begin{center}
 \renewcommand{\baselinestretch}{1.1}
\includegraphics[width=16.5cm]{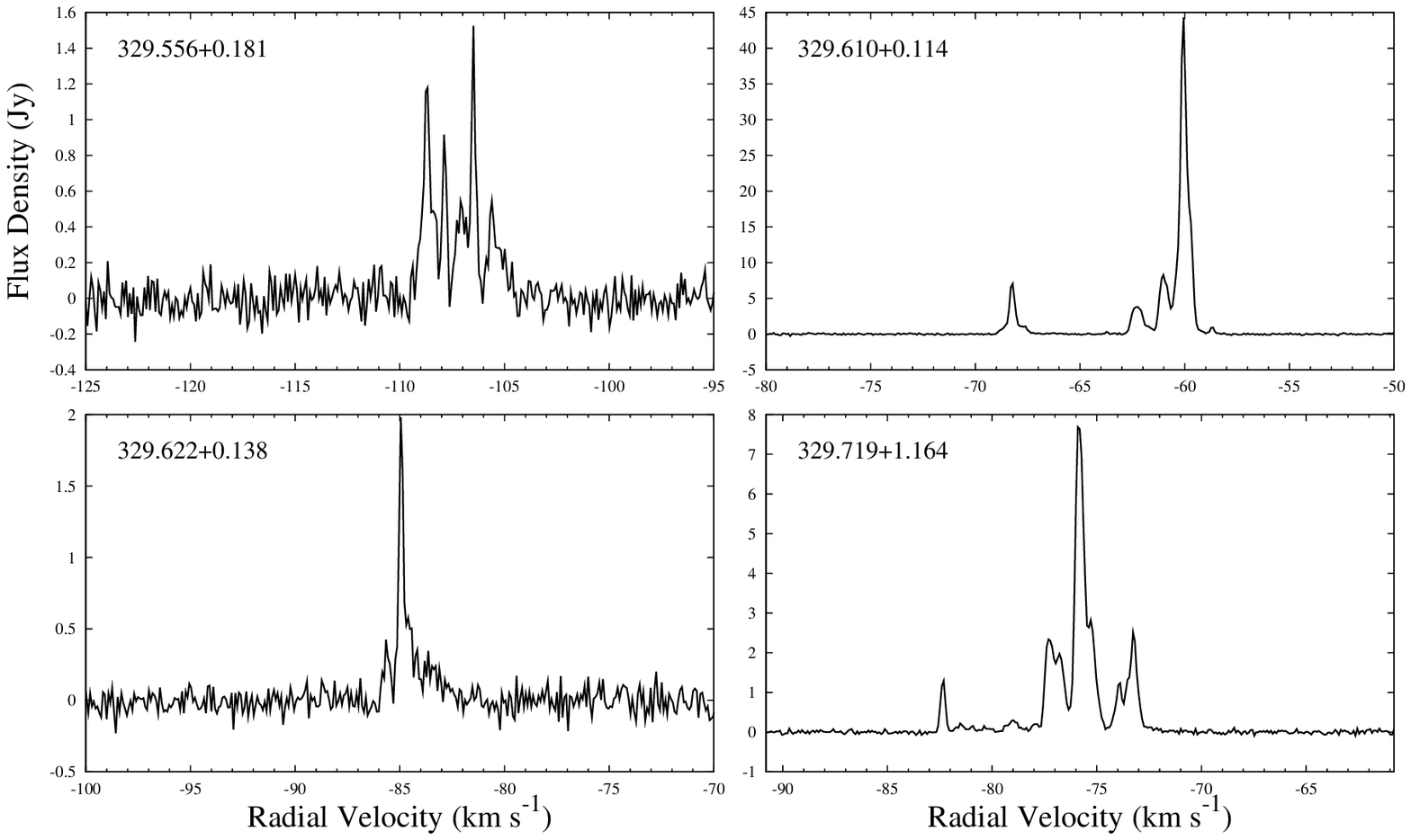}
\caption{\small -- {\it continued}}
\label{spectra}
\end{center}
\end{figure*}

\clearpage

\subparagraph{302.032--0.061} The MX spectrum shown in Figure\,\ref{spectra} for this source was taken at a position $\sim$30 arcsec offset from the actual position (requiring a flux density correction by a factor of 1.07). This source was allocated to the near kinematic distance by \citet{green11b}.

\subparagraph{302.034+0.625} This is a slightly variable source. It was initially detected with a flux density of 13\,Jy \citep{schutte93}, then seen with a peak of 10\,Jy in 2000 \citep{caswell09a}. In our  observations, we found a survey flux density of  8\,Jy, then an MX flux density (2008 March) of 11\,Jy (although this was taken at a position $\sim$40 arcsec offset from the actual position).  

\subparagraph{305.199+0.005, 305.200+0.019, 305.202+0.208, 305.208+0.206, 305.248+0.245, 305.362+0.150, 305.366+0.184, 305.475$-$0.096, 305.563+0.013, 305.573$-$0.342, 305.615$-$0.344, 305.799$-$0.245, 305.822$-$0.115 and 305.887+0.017} These sources are all loosely associated with the giant H{\sc ii} region complex, G305 \citep{clark04,walsh08,hindson10, davies11}, assigned to the far kinematic distance by  \citet{caswell75,urquhart11b,green11b}. The closely spaced quartet of sources, 305.199+0.005, 305.200+0.019, 305.202+0.208 and 305.208+0.206, consists of two pairs, shown as two aligned spectra in Figure\,\ref{spectra}. \citet{caswell97} provides a full description of the features associated with each of the sources and associated maser species. The spectrum of 305.248+0.245 has also been aligned in Figure\,\ref{spectra} with those of the previous four sources to highlight which features are sidelobes. 305.475--0.096 is a new source and has the widest velocity range of any 6.7-GHz methanol maser detected to date, with features spread across 28.5\,km\,s$^{-1}$.

\subparagraph{306.322--0.334} This source had a peak flux density of 0.4\,Jy in its initial detection \citep{caswell95a}, then brightened to 1\,Jy in 1999 \citep{caswell09a}, but returned to $\sim$0.5\,Jy in our survey and MX observations. This source was allocated to the near kinematic distance by \citet{busfield06} and \citet{green11b}.

\subparagraph{307.132--0.476 and 307.133--0.477} The first of this pair of new sources consists of the less negative velocity features, between $-$33.5\,km\,s$^{-1}$ and $-$36\,km\,s$^{-1}$; the second, the brighter of the two sources, consists of the features between $-$36\,km\,s$^{-1}$ and $-$40\,km\,s$^{-1}$.

\subparagraph{308.754+0.549} This source was allocated to the far kinematic distance by both \citet{green11b} and \citet{urquhart11b}.

\subparagraph{308.056--0.396 and 308.075--0.411} This is a pair of new sources, shown aligned in Figure\,\ref{spectra}; the first peaks at $-$11.8\,km\,s$^{-1}$, the second at $-$7.5\,km\,s$^{-1}$. The spectrum presented in Figure\,\ref{spectra} for 308.056--0.396 was obtained at an offset position (by 36 arcsec), therefore it underestimates the peak flux density of the source by a factor of 1.10. 

\subparagraph{308.686+0.530} Emission at 44\,km\,s$^{-1}$ is a feature of this new source and not a sidelobe of the nearby, stronger known source  308.754+0.549, as is demonstrated through the aligned spectra in Figure \ref{spectra}.

\subparagraph{308.918+0.123, 309.384--0.135 and 309.921+0.479} All three of these known sources, originally detected by \citet{caswell95a} were allocated to the near kinematic distance by \citet{green11b}.

\subparagraph{310.144+0.760} This source has varied greatly in intensity. It was originally two bright features with peak flux densities of $\sim$140\,Jy  \citep{walt95}. Both features have since faded, with the $-$55.7\,km\,s$^{-1}$ feature having an MX peak flux density of $\sim$80\,Jy and the $-$57.5\,km\,s$^{-1}$ feature an MX peak flux density of just 10\,Jy. Intermediate epoch positioning measurements \citep{caswell96a,caswell09a} found peak flux densities of 60\,Jy and 120\,Jy respectively.

\subparagraph{310.180--0.122 and 311.628+0.266} These two known sources, first detected by \citet{walsh97}, were allocated respectively to the far and near kinematic distances by \citet{green11b}.

\subparagraph{311.947+0.472 and 312.071+0.082} The first of these sources is a weak known source, at a star formation site allocated to the far kinematic distance by \citet{caswell75}, the second is the brightest new source, and is also responsible for the negative feature seen in the (aligned) spectrum of 311.947+0.472 in Figure\,\ref{spectra}. The known source was originally positioned \citep{caswell09a} with a feature at $-$38.3\,km\,s$^{-1}$, but that feature has now faded and features at $-$44\,km\,s$^{-1}$ and $-$47\,km\,s$^{-1}$ dominate (the original feature is below the noise in the MX observation of 2009 March, but was detected in the survey cube observations taken 2007 June-September).

\subparagraph{312.597+0.045 and 312.598+0.045} Both sources of this close pair are shown on one spectrum in Figure\,\ref{spectra}. The first source has a single narrow weak feature near $-$60\,km\,s$^{-1}$, whilst the second source spans $-$68.5\,km\,s$^{-1}$ to $-$64\,km\,s$^{-1}$. The peak feature of the second source at $-$67.9\,km\,s$^{-1}$, was found to have a flux density of 10.4\,Jy in the survey observation (2007 June-September), but doubled for the MX observation (2008 March) to 21.4\,Jy. Both sources were allocated to the near kinematic distance by \citet{green11b}.

\subparagraph{313.469+0.190} This source has almost doubled in brightness from its original observation in 1992 \citep{caswell95a} to $\sim$30\,Jy in the 2008 March MX observation. This source was allocated to the far kinematic distance by \citet{green11b} and by \citet{urquhart11b}.

\subparagraph{313.577+0.325} This known source, positioned with the ATCA by \citet{caswell09a}, was allocated to the near kinematic distance by \citet{green11b}.

\subparagraph{313.767--0.863 and 313.774--0.863} Both sources of this close pair (separation 25 arcsec) are variable. 313.767--0.863 is the more negative velocity source, with a shift in the brightest feature from 28\,Jy at $-$54.5\,km\,s$^{-1}$  in \citet{schutte93} to 9\,Jy at $-$56.3\,km\,s$^{-1}$ (the original feature fading to 3\,Jy).  313.774--0.863 accounts for the features between $-$46\,km\,s$^{-1}$ and $-$40.5\,km\,s$^{-1}$, and had a peak of 22\,Jy at $-$44.8\,km\,s$^{-1}$ in the survey observation which faded to 14\,Jy at $-$44.9\,km\,s$^{-1}$ in the 2008 March MX, and then 7\,Jy in the 2009 March MX (to be replaced by the $-$41 feature as the brightest). The $-$44.8-km\,s$^{-1}$ feature was only 12\,Jy in the original \citet{schutte93} observations and was $\sim$7\,Jy in the \citet{caswell09a} ATCA observations.

\subparagraph{314.320+0.112} This source has a wide range of velocity over which emission is seen (16.5\,km\,s$^{-1}$). It was allocated to the far kinematic distance by \citet{green11b}.

\subparagraph{315.803--0.575} This new source has doubled in peak flux density with the survey cube (2007 June-July) and positioning observations measuring 3.9\,Jy and 3.5\,Jy respectively, and the later MX observations (2008 March) measuring 8.9\,Jy.

\subparagraph{316.359--0.362, 316.381--0.379 and 316.412--0.308} This is a closely spaced trio of sources. The first is the brightest with features at the most positive velocities (between +1.5\,km\,s$^{-1}$ and +8.5\,km\,s$^{-1}$), the second covers the smaller intermediate range of  $-$0.5\,km\,s$^{-1}$ to +1.5\,km\,s$^{-1}$ and the third the most negative features, between $-$5.6\,km\,s$^{-1}$ and +2\,km\,s$^{-1}$. All three sources were allocated to the near kinematic distance by \citet{green11b}.

\subparagraph{316.484--0.310} The $-$11.5-km\,s$^{-1}$ feature of this new source flared after the survey and positioning observations (when it was $<$0.2\,Jy), with the MX observation in 2008 August finding a peak flux density of 0.7\,Jy. This shifted the peak velocity from $-$15.6\,km\,s$^{-1}$ to $-$11.5\,km\,s$^{-1}$. The original peak feature at $-$15.6\,km\,s$^{-1}$ faded from 2.3\,Jy in the survey observations to $\sim$0.45\,Jy in the MX.

\subparagraph{316.640--0.097} The intensity variability of this source is known to be quasi-periodic, with a period of 321$\pm$48 days \citep{goedhart04}. Both our survey (2007 March-July) and MX observations (2008 August) detected comparable flux densities for the peak feature.

\subparagraph{316.811--0.057} This source has demonstrated significant brightening of its emission. It was originally detected with a peak flux density of 12\,Jy in 1992 \citep{caswell95a}, and positioned in 1996 with a peak flux density of 9\,Jy \citep{caswell09a}, but was detected in the MMB survey observations with a peak flux density of 75\,Jy, before slightly diminishing to 52\,Jy in the 2008 August MX observation. This source was allocated to the near kinematic distance by \citet{busfield06}.

\subparagraph{317.029+0.361} \citet{urquhart11b} allocate the associated compact H{\sc ii} region from the Red MSX Survey to the far kinematic distance.

\subparagraph{317.466--0.402} Emission from this source has a wide range of velocity (16\,km\,s$^{-1}$). The source was allocated to the far kinematic distance by \citet{green11b}.

\subparagraph{317.701+0.110} Detected by \citet{walt95} with a peak flux density of 32\,Jy, and positioned in 2000 with a peak flux density of 8\,Jy \citep{caswell09a}. Our detection found a return to similar emission as the original detection, with the survey observations measuring 26\,Jy and the MX 23\,Jy. This source was allocated to the near kinematic distance by \citet{green11b}.

\subparagraph{318.043--1.404} The positive velocity of this source, combined with its large latitude, indicate a large kinematic distance and hence large distance from the Galactic plane (`z'), possibly attributed to the flared/warped outer Galaxy (the source is kinematically associated with the outer Galaxy portion of the Carina-Sagittarius spiral arm, see section \ref{spiralarmsection}).

\subparagraph{318.050+0.087} This source had similar survey and MX peak flux densities of $\sim$12\,Jy, comparable to the positioning observation in 1999 by \citet{caswell09a}. However, it was originally detected in 1992 \citep{caswell95a} with a peak flux density of only 3.8\,Jy. This source was allocated to the near kinematic distance by \citet{green11b}.

\subparagraph{318.948--0.196} This very bright source, observed with the ATCA by \citet{norris93} and \citet{caswell95ab}, was allocated to the near kinematic distance by \citet{green11b}.

\subparagraph{319.163--0.421} \citet{urquhart11b} allocate the associated compact H{\sc ii} region from the Red MSX Survey to the far kinematic distance.

\subparagraph{319.836--0.197} This is a consistently weak source with our observations (survey peak flux density of 0.65\,Jy, MX peak flux density of 0.23\,Jy) finding comparable flux density to the original 1992 detection (peak flux density of 0.4\,Jy) of \citet{caswell95a}. This source was allocated to the near kinematic distance by \citet{green11b}.

\subparagraph{320.123--0.504} This known source, positioned with the ATCA by \citet{caswell09a}, was allocated to the far kinematic distance by \citet{green11b}.

\subparagraph{320.231--0.284} This source has demonstrated significant variability. It was originally detected in 1992 with a peak flux density of 27\,Jy \citep{caswell95a}, before being positioned in 1999 with a peak flux density of 20\,Jy \citep{caswell09a}. For our observations it was detected in the survey cube (2006 April-May) with a peak flux density of 32\,Jy, was positioned (2007 January) with a peak flux density of 16\,Jy, and in the subsequent MX observation (2008 August) was found to peak at 54\,Jy. This source was allocated to the far kinematic distance by \citet{urquhart11b}.

\subparagraph{320.285--0.308} The spectrum of this weak new source has been aligned with the spectrum of the known bright source 320.231--0.284 in Figure\,\ref{spectra} to reveal that the feature seen at $-$62.5\,km\,s$^{-1}$ is  a sidelobe of the nearby brighter source.

\subparagraph{320.414+0.109 and 320.424+0.089} This is a pair of new sources. The spectrum presented in Figure\,\ref{spectra} for 320.414+0.109 was obtained at an offset position (by 52 arcsec) and therefore underestimates the peak flux density of the source by a factor of 1.25. The aligned spectra of the two sources clearly show that the first source is the brighter feature at $-$13.4\,km\,s$^{-1}$, whilst the second source consists of features between $-$4\,km\,s$^{-1}$ and $-$10\,km\,s$^{-1}$, peaking at $-$8\,km\,s$^{-1}$. 

\subparagraph{320.780+0.248} This is the second brightest (40\,Jy) new source.

\subparagraph{321.030--0.485 and 321.033--0.483} This is a close pair of sources originally detected by \citet{schutte93} and further observed and discussed by \citet{walt95,walsh98,caswell09a}. The second source has varied, with a peak flux density of 142\,Jy in the survey cube observation (2006 April-May), then 76\,Jy in the 2008 August MX and 58\,Jy in the 2009 March MX.

\subparagraph{321.148--0.529} This known source, positioned with the ATCA by \citet{caswell09a}, was allocated to the near kinematic distance by \citet{green11b}.

\subparagraph{321.704+1.168} This source has a history of variability. It was  originally discovered in 1992 with a peak flux density of 1.4\,Jy, but `disappeared' in 1993 \citep{caswell95a}. In our survey cube observation (2006 April-May) we found a peak flux density of 2.1\,Jy, which faded to 1.2\,Jy in the 2007 ATCA positioning measurement, and then was not present in either of two MX observations (2008 March and 2008 August). As such the survey cube spectrum is shown in Figure\,\ref{spectra}.

\subparagraph{322.158+0.636} The feature at $-$64.2\,km\,s$^{-1}$ flared by almost a factor of two between the initial survey cube observation ($\sim$150\,Jy) and the follow up MX ($\sim$280\,Jy). This feature was only 65\,Jy in the original 1992 observation of \citet{caswell95a}. The H{\sc ii} region in which the maser appears to be embedded was allocated to the near kinematic distance by \citet{caswell75} and \citet{urquhart11b}.

\subparagraph{323.459--0.079} This known source, positioned with the ATCA by \citet{caswell97},  was allocated to the near kinematic distance by \citet{green11b}.

\subparagraph{323.740--0.263} The wide velocity range (17\,km\,s$^{-1}$) of this bright source is not apparent from the spectrum presented in Figure\,\ref{spectra}, but can be seen in the spectrum previously published by \citet{caswell95a}. This source has been allocated to the near kinematic distance \citep{busfield06, green11b}.

\subparagraph{323.766--1.370} The positive velocity of this new source, combined with its large latitude, indicate a large kinematic distance and large distance from the Galactic plane (`z'), possibly attributed to the flared/warped outer Galaxy (the source is kinematically associated with the outer Galaxy portion of the Carina-Sagittarius spiral arm, see section \ref{spiralarmsection}).

\subparagraph{323.799+0.017} The spectrum of this new source has been aligned with the spectrum of 323.740--0.263 in Figure\,\ref{spectra} to reveal that the negative feature at $-$51\,km\,s$^{-1}$ is a response to the bright source 323.740--0.263 in one of the reference beams.

\subparagraph{324.716+0.342} This source has a history of variability: it had a peak flux density of 8\,Jy at 47.5\,km\,s$^{-1}$ in the original 1992 observation \citep{caswell95a}, but was then only 3.5\,Jy in 1999 \citep{caswell09a}. The secondary feature was stable though, approximately 6\,Jy in both measurements. Our MX and survey observations were both similar with $\sim$10\,Jy for the 47.5\,km\,s$^{-1}$ feature. The secondary feature was $\sim$1.5\,Jy in the survey cube, but about 0.5\,Jy in the 2008 March MX. This source was allocated to the far kinematic distance by \citet{green11b}.

\subparagraph{324.923--0.568} This known source, first detected by \citet{walsh98}, was allocated to the near kinematic distance by \citet{green11b}.

\subparagraph{325.659--0.022} This new source faded from a peak flux density of 2.3\,Jy in the survey observation (2006 January-February) to just 0.6\,Jy in the MX (2008 March). The ATCA positioning measurement detected a similar flux density to the MX observation.

\subparagraph{326.323--0.393} The brightest feature of this new source flared from a peak flux density of 2\,Jy in the survey observation (2006 January) to 3.3\,Jy in the MX observation (2008 March). The secondary feature was even more variable, flaring from 0.5\,Jy to 1.5\,Jy.

\subparagraph{326.448--0.748}  This new source has a wide velocity range of multiple spectral features, several of which dimmed between the survey observation and MX observation, shifting the peak velocity from $-$71.7\,km\,s$^{-1}$ to $-$68.6\,km\,s$^{-1}$ (the feature at $-$71.7\,km\,s$^{-1}$ also fading by a factor of $\sim$0.5).

\subparagraph{326.475+0.703 and 326.476+0.695} The sources of this close pair (separated by 29 arcsec) are shown as one spectrum in Figure\,\ref{spectra}. The two sources share the same range of velocities, but the first peaks at $-$38.5\,km\,s$^{-1}$, whilst the second peaks at $-$43.6\,km\,s$^{-1}$. These sources were allocated to the far kinematic distance by \citet{green11b}, but the near kinematic distance by \citet{busfield06}.

\subparagraph{326.608+0.799} This is a weak new source. The spectrum presented in Figure\,\ref{spectra} is aligned with the spectra of 326.475+0.703 and 326.476+0.695 to highlight that the negative feature at $-$38.5\,km\,s$^{-1}$ is a result of the bright source 326.475+0.703 falling within one of the reference beams.

\subparagraph{326.641+0.611}  This source, known since 1992 \citep[reported in][]{caswell95a}, exhibited minor variation in its intensity, with the survey observations finding a peak flux density of 26\,Jy and the MX observations 31\,Jy. This source was allocated to the near kinematic distance by \citet{urquhart11b} and has a similarly small photometric distance \citep{moises11}.

\subparagraph{326.859--0.677 and 327.120+0.511} Both of these known sources were allocated to the near kinematic distance by \citet{green11b}.

\subparagraph{327.291--0.578} This source, although varying greatly since 1992, had only minor variation between the survey observation (2006 January) and the later MX observation (2008 March), with peak flux densities of 2.6\,Jy and 3.4\,Jy respectively. The prominent absorption, centred at approximately $-$48\,km\,s$^{-1}$, is discussed in \citet{caswell95a}.

\subparagraph{327.392+0.199 and 327.395+0.197} The sources of this close pair are separated by only 14 arcsec and have almost contiguous velocity ranges. Both sources were allocated to the near kinematic distance by \citet{green11b}.

\subparagraph{327.402+0.445} This known source, first detected by \citet{caswell95a}, was allocated to the near kinematic distance by \citet{green11b}.

\subparagraph{327.590--0.094 and 327.618--0.111} The sources of this close pair (offset by $\sim$2 arcmin in Right Ascension) are shown aligned in Figure\,\ref{spectra}. Each spectrum shows a weak sidelobe response of the other source (attenuated by a factor of $\sim$3). These sources were allocated to the far kinematic distance by \citet{green11b}.

\subparagraph{327.618--0.111} This known source, positioned with the ATCA by \citet{caswell09a}, was allocated to the far kinematic distance by \citet{green11b}.

\subparagraph{327.945--0.115} This simple single-feature source has remained remarkably stable over its observational history of 15 years, with a peak flux density of $\sim$8\,Jy (it was first observed in 1994 and reported in \citealt{ellingsen96}, our MX observation was taken in 2009). 

\subparagraph{328.140--0.432} This new source was more than three times brighter in the MX observation (2008 August), than in the initial survey cube observation (2006 January).

\subparagraph{328.237--0.547 and 328.254--0.532}  This is a bright pair of sources, with overlapping velocity ranges and the brighter of the two, 328.237--0.547, is known to be periodically variable \citep{goedhart04}. The period was originally estimated to be 216 days \citep{goedhart04}, but later revised to 220 days \citep{goedhart07}. 328.237--0.547 was found to have a peak flux density of 1481\,Jy in the survey cube observations (2006 January), then 1340\,Jy in the later MX observation (2008 August). 

\subparagraph{328.808+0.633 and 328.809+0.633} These very close sources (2.5 arcsec separation) have peak velocities within 0.5\,km\,s$^{-1}$. We retain the interpretation of \citet{caswell97} and \citet{caswell09a} that the first source has features across the full velocity range, but the second is a single strong feature (at $-$44.4\,km\,s$^{-1}$ in our spectrum). Long Baseline Array observations by \citet{dodson04} confirm the spatial offset of the isolated 328.809+0.633 source. Both sources were allocated to the near kinematic distance by \citet{green11b}.

\subparagraph{328.819+1.704} This source has a wide range of velocity over which emission is seen (18.5\,km\,s$^{-1}$).

\subparagraph{328.940+0.558 and 328.942+0.565} A pair of new sources separated by 26 arcsec, the first with a peak at $-$98.8\,km\,s$^{-1}$ and the second with a peak at $-$90.9\,km\,s$^{-1}$. Both sources share comparable velocity ranges of emission.

\subparagraph{329.029--0.205 and 329.031--0.198} This is a close pair of sources separated by 26 arcsec, with the first peaking at $-$37.1\,km\,s$^{-1}$ and the second at $-$45.8\,km\,s$^{-1}$. The velocity ranges of the sources are adjacent, with the first $-$33.5\,km\,s$^{-1}$ to $-$41.5\,km\,s$^{-1}$ and the second $-$41.5\,km\,s$^{-1}$ to $-$48.5\,km\,s$^{-1}$. These sources are believed to be embedded in an Infrared Dark Cloud \citep{ellingsen06}, usually indicative of a near distance assignment, however, no H{\sc i} self-absorption was seen towards these sources, indicating a far kinematic distance \citep{green11b}. 

\subparagraph{329.066--0.308} This source was first detected as a weak feature with the Parkes Radio Telescope in 1992 \citep[reported by][]{caswell95a}, but later ATCA observations \citep{caswell95ab} found the source to be considerably offset from the previous pair, and thus yielded an estimated peak flux density of $\sim$20\,Jy. We found a comparable peak flux density of 21.9\,Jy in the survey cube, falling to 14.4\,Jy in the MX observation (2008 March). This source was allocated to the far kinematic distance by \citet{green11b}.

\subparagraph{329.183--0.314} The peak feature of this source faded from the initial survey observation (2006 January), with a peak flux density of 10.7\,Jy, to the later MX observation (2008 March), with a peak flux density of 5.1\,Jy. The feature at $-$51\,km\,s$^{-1}$ flared slightly, and the feature at $-$59\,km\,s$^{-1}$ faded slightly, such that the second brightest feature switched between the survey observation and the MX observation. This source was allocated to the near kinematic distance by \citet{green11b}.

\subparagraph{329.405--0.459 and 329.407--0.459} The second of this close pair of sources is variable, with the initial survey observation finding a peak flux density of 130\,Jy and subsequent MX observations finding 96\,Jy (2008 March) and 120\,Jy (2009 March). It was originally detected with a peak flux density of 155\,Jy in 1992 \citep{caswell95a} and positioned in 2000 with a peak flux density of 72\,Jy \citep{caswell09a}. 

\subparagraph{329.469+0.503} The peak feature of this source at $-$72\,km\,s$^{-1}$ has brightened from the original \citet{schutte93} observations, and the other spectral features have varied, with the most notable changes being the brightening of the feature at $-$73\,km\,s$^{-1}$ and the dimming of the feature at $-$70.5\,km\,s$^{-1}$. The outlying feature at $-$59\,km\,s$^{-1}$ also flared in the MX observation (2008 August) compared with the initial survey observation (2006 January). This source has a wide range of velocity over which emission is seen (16\,km\,s$^{-1}$). It was allocated to the far kinematic distance by \citet{green11b}.

\subparagraph{329.556+0.181} This new source has experienced variability in both its individual features and peak flux density, changing between the survey and MX observations, with peak velocities of  $-$109\,km\,s$^{-1}$ and $-$106.5\,km\,s$^{-1}$ respectively.

\subparagraph{329.622+0.138 and 329.610+0.114} These sources are in relative close proximity to one another (offset by $\sim$100 arcsec). Their features appear in the spectra at both positions, but are clearly distinct in velocity, with almost 15\,km\,s$^{-1}$ separation between the emission boundaries in velocity. Both sources were allocated to the far kinematic distance by \citet{green11b}.

\subparagraph{329.719+1.164} This new source exhibited significant variation in its peak intensity, with the survey observation finding a peak flux density of 24.4\,Jy and subsequent MX observations taken in 2008 August finding $\sim$8\,Jy. There has also been variation in individual features of the spectrum, but the weakest feature at $-$82.2\,km\,s$^{-1}$ has remained stable at $\sim$1\,Jy.

\begin{figure*}
 \begin{center}
 \renewcommand{\baselinestretch}{1.1}
\includegraphics[width=15cm]{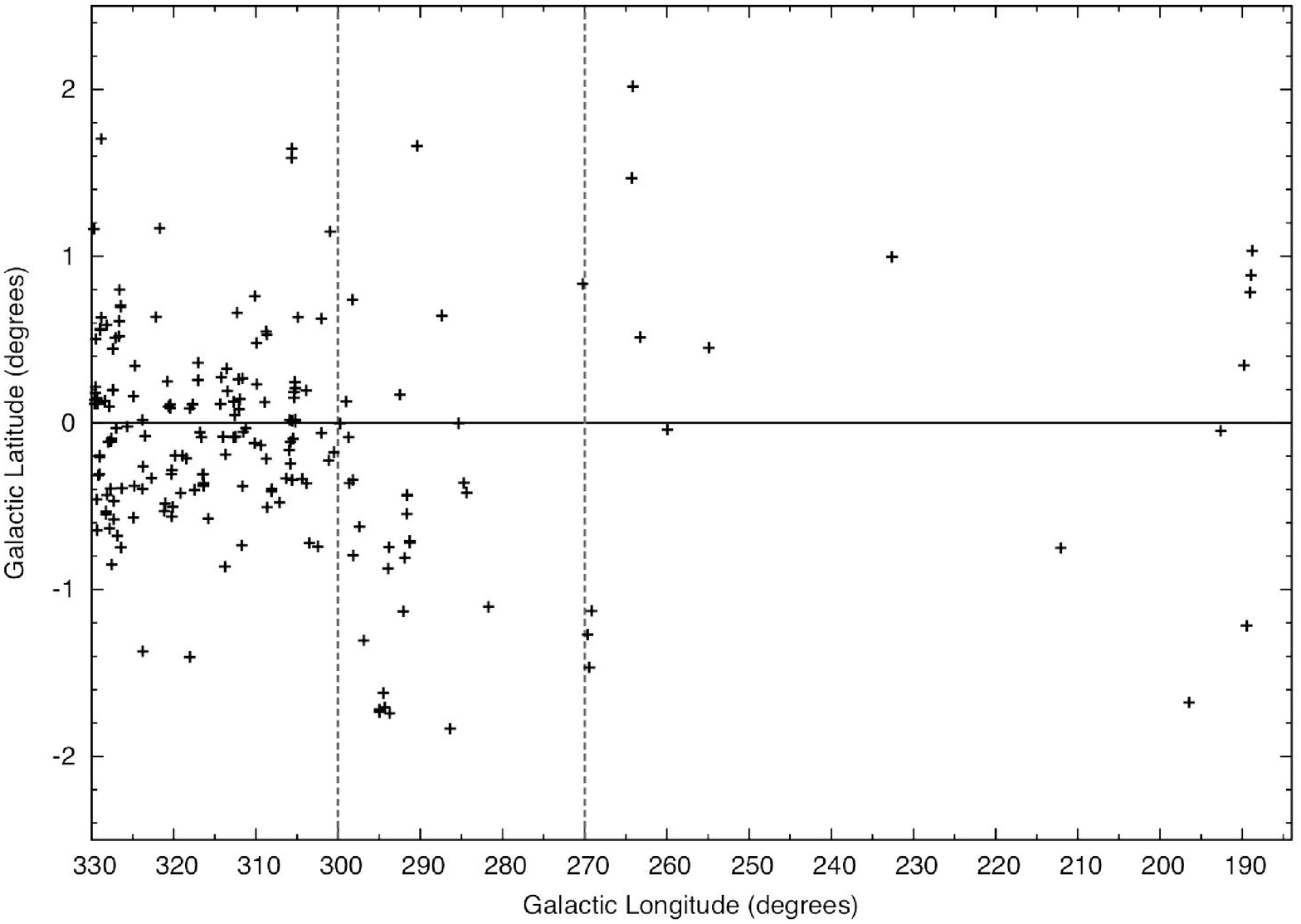}  
\caption{\small Longitude-latitude distribution of sources in the region 330$^{\circ}$ to 186$^{\circ}$. Five sources lie outside the plotted latitude range. Grey dashed lines divide longitude regions discussed in Section \ref{latdiscussion}. Note: the scale of this figure is expanded in the latitude direction for clarity.}
\label{latdist}
\end{center}
\end{figure*}

\section{Discussion}\label{discussion} 
With the region presented in the current paper, the MMB has now almost completely surveyed the half Galactic disc from longitudes 180$^{\circ}$ to 360$^{\circ}$; between 180$^{\circ}$ and 186$^{\circ}$, outside the MMB coverage, the only known source is 183.34+0.59, which was detected as a targeted observation of the IRAS source 05480+2545 \citep{slysh99,szymczak00,pestalozzi05}. Discussion of the global properties of the methanol maser population will be presented in the forthcoming survey statistics paper, analysing  the full MMB catalogue (Green et al., in prep.). Here we discuss the properties of the sources within the 186$^{\circ}$ to 330$^{\circ}$ longitude region with reference to those already published in the MMB survey 330$^{\circ}$ to 345$^{\circ}$ \citep{caswell11mmb3}; 345$^{\circ}$  through 360$^{\circ}$ to 6$^{\circ}$ \citep{caswell10mmb1}; and 6$^{\circ}$ to 20$^{\circ}$ \citep{green10mmb2}.

\subsection{Galactic Latitude distribution}\label{latdiscussion}
\begin{figure}
 \begin{center}
 \renewcommand{\baselinestretch}{1.1}
\includegraphics[width=7cm]{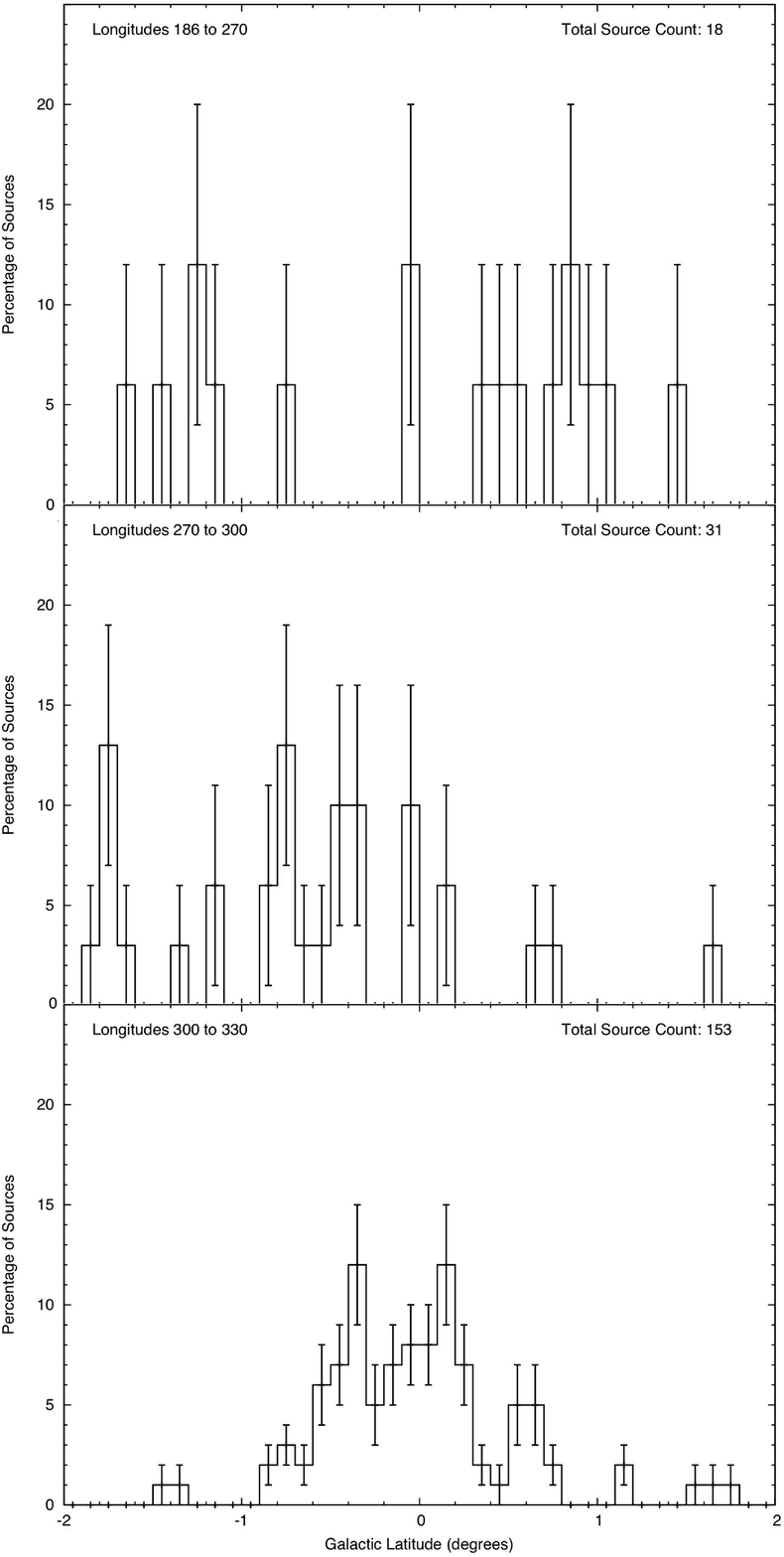}  
\caption{\small Percentage of sources in 0.1$^{\circ}$ latitude bins for the sources between longitudes 186$^{\circ}$ and 270$^{\circ}$ (top), 270$^{\circ}$ and 300$^{\circ}$ (middle) and 300$^{\circ}$ and 330$^{\circ}$ (bottom). Distributions include the statistical errors (Poisson error based on the real number of sources in each bin). The top plot shows a broad spread in latitudes characteristic of the outer Galaxy (a combination of nearby sources and Galactic flare). The middle plot shows a bias towards negative longitudes, likely due to the warp of the Galaxy between these longitudes. The bottom plot shows a distribution approaching Gaussian, but with three peaks at the 1$\sigma_{\rm Poisson}$ level. The velocities of the sources contained within these peaked bins have a significant spread and cannot be tied to any one spiral arm.}
\label{latitudehisto}
\end{center}
\end{figure}

The distribution of the latitude of sources with longitude is shown in Figure\,\ref{latdist} and as percentages of longitude regions in Figure\,\ref{latitudehisto}. We find 176 detections lie within 1$^{\circ}$ of the plane, 31 outside (including those associated with Orion and Mon R2). Longitudes 186$^{\circ}$ to 330$^{\circ}$ exhibit a much larger fraction ($\sim$15\%) of sources outside of $\pm$1$^{\circ}$ than the MMB survey results closer to the Galactic Centre (which were approximately 5\%, 3\% and 4\% for the previous catalogue releases of 345$^{\circ}$ to 6$^{\circ}$, 6$^{\circ}$ to 20$^{\circ}$ and 330$^{\circ}$ to 345$^{\circ}$ respectively). The larger spread of latitudes is most prominent in the outer Galaxy region of 186$^{\circ}$ to 270$^{\circ}$. Here, after excluding the Orion-Monoceros region, the sources are concentrated within longitude ranges 186$^{\circ}$ to 196$^{\circ}$ and 260$^{\circ}$ to 270$^{\circ}$ (with only a couple of sources at the intervening longitudes); and 12 of the 18 sources, lie at latitudes larger than 1$^{\circ}$. A combination of factors contribute to this wider distribution:  the flare of the Milky Way disk, known from H{\sc i} observations \citep[e.g.][]{kalberla07}, with a progressively larger scale height observed as one moves towards the outer Galaxy; and the presence of the local arm (see Section\,\ref{spiralarmsection}), with proportionally more nearby sources. The maser population is too sparse to reliably distinguish the two contributors. There is also a significant negative latitude bias for longitudes  270$^{\circ}$ to 300$^{\circ}$, with more than 80\% of the sources at negative latitudes (middle plot of Figure\,\ref{latitudehisto}). This is due to the long known warp of the Milky Way, seen in the H{\sc i} disk at Galactocentric distances outside the solar circle as a deviation below the plane of up to 1\,kpc at longitude 270$^{\circ}$ \citep{burke57,kerr57,henderson82,burton86}, in the radio H{\sc II} regions where more than 80\% lie at negative latitudes \citep{caswell87a}, and also seen in many other tracers such as CO and stars. Near 270$^{\circ}$, the source 269.456--1.467 stands out, with its large positive velocity corresponding to a kinematic distance of $\sim$7\,kpc ({\it z} = --180\,pc), Galactocentric distance of $\sim$11\,kpc, indicating that it lies in the warped outer Galaxy. The latitude distribution for 300$^{\circ}$ to 330$^{\circ}$ (bottom plot of Figure\,\ref{latitudehisto}) shows an approximately Gaussian distribution, but with three peaks at the 1$\sigma_{\rm Poisson}$ level. The LSR velocities of the sources contributing to these three peaks have a large spread and cannot be tied to any one spiral arm within the longitude--velocity domain (see Section \ref{spiralarmsection}). However the positive velocity sources between 300$^{\circ}$ and 330$^{\circ}$ are expected to be in the outer Galaxy (beyond the solar circle) and it is noteworthy that the majority, 15 of the 17 sources, lie at negative latitude. This is largely attributed to the Galactic warp, and these sources contribute to the histogram peak at negative latitude. The two positive latitude peaks may be dominated by clusters of sources between 325$^{\circ}$ and 330$^{\circ}$.

\subsection{Flux densities}
In this portion of the survey we detected sources with MX peak flux densities in the range of 0.23\,Jy (319.836--0.197) to 3114\,Jy (323.740--0.263). There were 18 sources which had peak flux densities below 0.7\,Jy (the 3$\sigma$ limit of the survey) in the follow-up MX observations, 10 new and 8 known, including the weaker of the two Orion sources. There were only three new sources which had peak flux densities exceeding 20\,Jy: 312.071+0.082, 320.780+0.248 and 329.719+1.164. Just over half of the new sources had peak flux densities below  2\,Jy.

\subsection{Variability}\label{varisection}
\begin{figure}
 \begin{center}
 \renewcommand{\baselinestretch}{1.1}
\includegraphics[width=8cm]{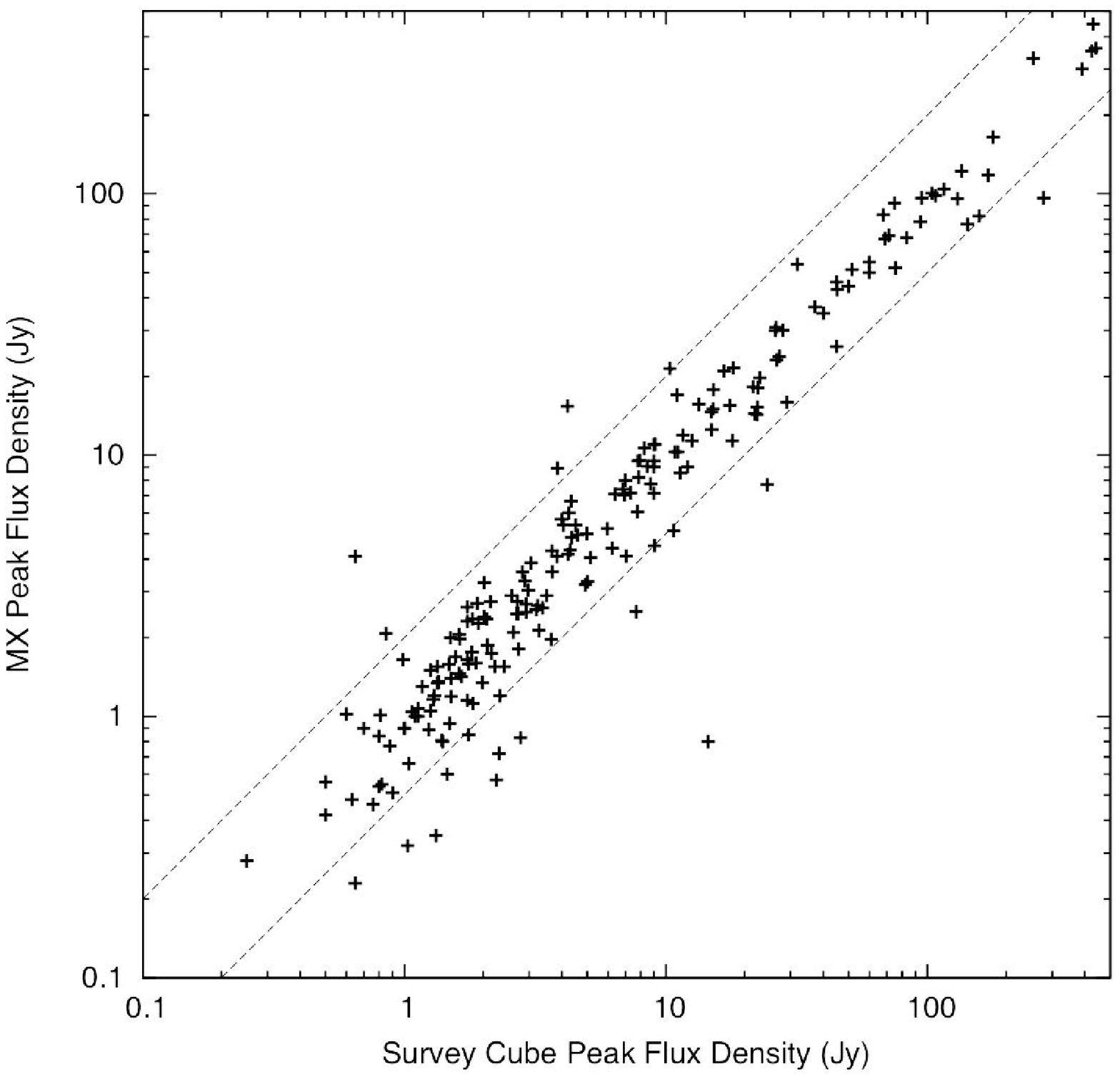}  
\caption{\small Comparison between peak flux density in the initial survey observations (`Survey') and that seen in the later targeted (`MX') observations. The dashed lines bound those sources which have varied by a factor of less than two. Five sources lay outside the MMB survey region and are therefore not plotted since there is no initial survey cube data.}
\label{fluxratio}
\end{center}
\end{figure}

Comparison of the peak flux densities measured in the initial survey observations with those of the later targeted MX observations is given in Figure\,\ref{fluxratio}. The median peak flux density ratio between the MX data and the survey cube is consistent with the findings of the previous catalogues with a value of 0.92. Again we attribute a small decrease in flux density in the subsequent MX observations to the greater impact of pointing errors with the smaller beam (3.2 arcmin, not smoothed to 4.4 arcmin as with the survey observations) and the effect of the higher noise of the survey cubes on the weaker sources. There were 14 sources (including two in the Orion-Monoceros region) which decreased by more than a factor of two, with the largest decrease, a factor of 10, that of 294.337--1.706. There were five sources which increased by more than a factor of 2, with the largest increase, a factor of 6.3, that of 269.153--1.128 (we disregard 264.140+2.018 for which the tabulated survey value is a lower limit as discussed in the source notes). As with the other regions of the survey there was a wide range of subsidiary feature variability (as discussed individually in the source notes).

\begin{figure*}
 \begin{center}
 \renewcommand{\baselinestretch}{1.1}
\includegraphics[width=15cm]{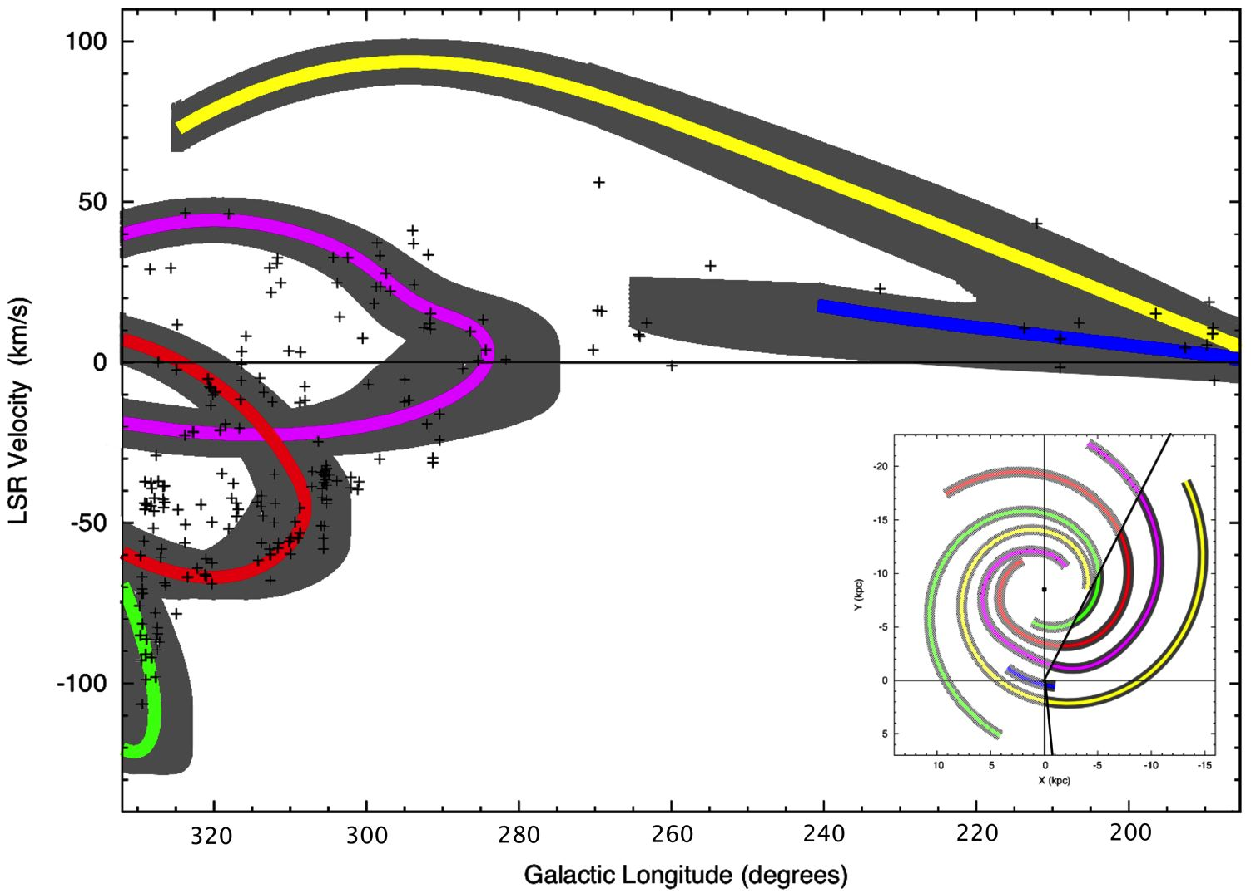}  
\caption{\small Longitude--velocity distribution of sources in the region 330$^{\circ}$ to 186$^{\circ}$, overlaid with example spiral arm loci (the spatial pattern of \citet{taylor93}). A flat rotation curve with a circular rotation of 246\,km\,s$^{-1}$ \citep{reid09, bovy09} was used to transfer the arms to the `{\it l--v}' domain. Coloured loci are the spiral arms defined by the model, grey shading incorporates an arm width of 1\,kpc and a velocity tolerance of $\pm$7\,km\,s$^{-1}$. Yellow loci represent the Perseus spiral arm; Purple - Carina-Sagittarius; Orange - Crux-Scutum; Green - Norma; Blue - Local arm (Orion-Cygnus). Crosses show 6.7--GHz methanol masers of the MMB survey. The sub-image shows the top-down view of the spiral arms within the Galaxy, with bold lines delineating the longitude region presented here. Note the longitude range presented is 2$^{\circ}$ wider than the catalogue region so as to show the sources near 330$^{\circ}$ without confusion (but masers between longitudes 330$^{\circ}$ and 332$^{\circ}$  are not shown). The location of the spiral arm loci differs slightly from the third catalogue paper owing to the adoption of a different rotation curve for this section of the Galaxy.}
\label{lvdist}
\end{center}
\end{figure*}

\subsection{Distribution of velocities} 
Within the longitude range 270$^{\circ}$ to 330$^{\circ}$ there are 40 sources with positive velocities, formally corresponding to kinematic distances outside the solar circle. Between the longitudes of 186$^{\circ}$ and 270$^{\circ}$, all the sources, with the exception of three (188.78+1.03, 209.016--19.398 and 259.939--0.041), have positive velocities as would be expected for this Galactic quadrant. The three sources with negative, formally `forbidden', velocities are within 10\,km\,s$^{-1}$ of 0\,km\,s$^{-1}$ and therefore are likely to be either nearby sources or simply consistent with the commonly recognised small scatter in peculiar motions. The most negative velocity within longitudes 186$^{\circ}$ to 330$^{\circ}$ is --106.5\,km\,s$^{-1}$ for 329.339+0.148. The most positive velocity is +56.1\,km\,s$^{-1}$ for 269.456--1.467, which corresponds to the Perseus spiral arm. 318.043--1.404 and 323.766-1.370 also have significantly high positive velocities, corresponding to either the Perseus spiral arm or the Carina-Sagittarius spiral arm (depending on the model of arms and Galactic rotation adopted) and their negative latitudes reflect the warp in this outer Galaxy region. These last two sources were found close to the boundary of the velocity coverage of the MMB (within 20\,km\,s$^{-1}$). As mentioned previously, the velocity coverage of the survey was chosen to trace that of the \citet{dame01} CO distribution, but as can be seen from the spiral arm loci presented in Figure\,\ref{lvdist} the velocities of the distant, outer Galaxy portion of the Perseus arm may be more extreme and this leaves the opportunity for future surveys to search these LSR velocities (the MMB limit approximately follows the central locus of the arm presented in Figure\,\ref{lvdist}, as can be seen in \citet{green09a}).

\subsection{Individual source velocity spreads}
As with the previous catalogue releases \citep{caswell10mmb1,green10mmb2,caswell11mmb3}, for any source observed at several epochs, we have chosen the largest velocity range measured at any epoch for the listing in Table\,\ref{resotable}. The median velocity spread for sources in this region of the survey is 5\,km\,s$^{-1}$. There are seven sources with velocity spreads equal to or wider than 16\,km\,s$^{-1}$: 188.946+0.886, 305.475--0.096, 314.320+0.112, 317.466--0.402, 323.740--0.263, 328.819+1.704,  329.469+0.503. One of these sources, the new source 305.475--0.096, has a velocity width of 28.5\,km\,s$^{-1}$, which is the widest velocity range detected in the MMB survey and is the widest velocity range of any 6.7-GHz methanol maser reported to date. 

\subsection{Structure of the spiral arms}\label{spiralarmsection}
The longitude-velocity or `{\it l--v}' plot for the longitude region of 330$^{\circ}$ to 186$^{\circ}$ is shown in Figure\,\ref{lvdist} with the loci of spiral arms. This region of the Galactic plane is particularly notable for tangents of the Crux-Scutum and Sagittarius spiral arms together with an unconfused portion of the Perseus arm and the edge of the tangent of the Norma arm. We see a large proportion of masers around both the Norma and Crux-Scutum arm tangents; the Carina-Sagittarius arm tangent is less pronounced, but its orientation with respect to the line-of-sight means that the tangent covers a larger area of the  `{\it l--v}' domain. We see a number of masers associated with the outer Perseus arm and the local arm spur, indicating high-mass star formation in these locales. We see no evidence for 6.7-GHz methanol masers in the outer spiral arm of \citet{mcclure04}, an extension of the Norma arm which has velocities in excess of the Perseus arm loci (by $\sim$40\,km\,s$^{-1}$). This is consistent with the understanding that this outer arm is primarily a H{\sc i} feature.

There are a few sources around longitude 295$^{\circ}$ with velocities higher than the loci of the Carina-Sagittarius arm (at about +40\,km\,s$^{-1}$). This indicates either unusual velocities within the arm, or that the arm itself (rather than the logarithmic spiral arm fit of the model) actually exhibits these higher velocities. This can be partially accounted for by avoiding the adjustments to the arm shape that \citet{taylor93} made. Seven sources lie between 263$^{\circ}$ and 271$^{\circ}$, four of which are associated with the local spur locus. The three others could be associated with the Perseus arm, despite somewhat irregular velocities, but are more likely associated with the local spur arm, if it were extended beyond the simple model. All seven are nominally attributed to the Vela molecular cloud complex (see source notes).

There is a cluster of sources between longitudes 325$^{\circ}$ and 329$^{\circ}$ with velocities of approximately --40\,km\,s$^{-1}$ that do not currently align with any of the spiral arms as shown in  Figure\,\ref{lvdist}. This could indicate that the velocities of the Norma arm tangent have a broader range than expected, that the Carina-Sagittarius arm locus extends to more negative velocities than is currently depicted or that Crux-Scutum arm has more positive velocities than depicted. All three options are possible depending on the choice of spiral arm shape and rotation curve parameters. Four of the sources, near longitude 326$^{\circ}$, have consistent positive latitudes and the presence of H{\sc i} Self Absorption has indicated they lie at the near kinematic distance \citep{green11b}, implying the Crux-Scutum arm adjustment may be appropriate. However the other sources in the cluster, between 327$^{\circ}$ and 329$^{\circ}$, have a range of latitudes. These sources are further examples of ideal candidates for parallax observations with Very Long Baseline Interferometry  - which would clearly define their location within the Galaxy.

\subsection{Rate of detection}
Combined with the previous catalogue releases of \citet{caswell10mmb1} and \citet{caswell11mmb3} we find that there is at least one source (and usually many more) within each one degree longitude bin at all longitudes between 290$^{\circ}$ and 360$^{\circ}$. There is a prominent gap in detections between 270$^{\circ}$ and 280$^{\circ}$, corresponding to a gap in nearby spiral arms (the only arm at these longitudes is the outer portion of the Perseus arm). There are also gaps in detections between 253$^{\circ}$ and 233$^{\circ}$ and again between 231$^{\circ}$ and 214$^{\circ}$. 

\section{SUMMARY} 
We present the 207 detections of masers in the MMB survey for the longitude region 186$^{\circ}$ to 330$^{\circ}$, including 89 new sources, one of which has the widest velocity range of any 6.7-GHz methanol maser detected to date (28.5\,km\,s$^{-1}$). This region of the survey features a significant fraction of sources at large latitudes ($>$1$^{\circ}$ from the plane). We also present the first blind survey of the Orion-Monoceros star forming region, in which we detected the four known sources, but made no new detections. We positioned the Orion-A (south) and Orion-B sources with the MERLIN array, finding positions consistent within the uncertainties of previous observations. The MMB survey has now surveyed over half of the Galactic plane. Digital versions of the spectra of all the MMB masers together with various ancillary data are available at www.astromasers.org or www.manchester.ac.uk/jodrellbank/mmb.

\section*{Acknowledgments}
The authors thank Thomas Dame for providing Orion-Monoceros $^{12}$CO data in a machine readable format and Anita Richards for assistance with the MERLIN data reduction. MERLIN is a national facility operated by the University of Manchester on behalf of STFC. We thank the staff at the Parkes Radio Telescope and Australia Telescope Compact Array for ensuring the smooth running of our observations.
The Parkes Observatory and the Australia Telescope Compact Array are part of the Australia Telescope which is funded by the Commonwealth of Australia for operation as a National Facility managed by CSIRO.

\bibliographystyle{mn2e} \bibliography{UberRef}

\label{lastpage}

\end{document}